\AddToHook{package/hyperxmp/before}{%
  \RequirePackage[bookmarksnumbered,unicode]{hyperref}}
\documentclass[acmtog,screen,nonacm]{acmart}
\usepackage{ifthen}
\usepackage{titlesec}
\usepackage{dblfloatfix}
\usepackage{cuted}
\usepackage{caption} %

\usepackage[raggedrightboxes]{ragged2e}
\usepackage{longtable}

\usepackage{scalerel}
\usepackage{afterpage}
\usepackage[defaultlines=3,all]{nowidow}

\newboolean{showcomments}
\setboolean{showcomments}{false}  %

\newboolean{ms}
\usepackage[inkscapelatex=false]{svg}

\newcommand{\commentMotive}[1]{%
  \ifthenelse{\boolean{showcomments}}%
    {\par\noindent\textcolor{red}{#1}\par}%
    {}%
}

\newcommand{\VA}{Vandermonde}
\newcommand{\SYS}{ElementForge}

\newcommand{\IPTEM}{IPTEM}

\usepackage{amsmath,amsthm,amsfonts}
\usepackage{syntax}
\usepackage{framed}
\usepackage{mdframed}
\usepackage{wrapfig}

\usepackage{graphicx}

\DeclareMathOperator{\Inv}{Inv}

\theoremstyle{definition}
\newtheorem{definition}{Definition}[section]
\newtheorem{example}{Example}[section]
\theoremstyle{plain}

\newtheorem{proposition}[definition]{Proposition}
\newtheorem{theorem}{Theorem}[definition]

\usepackage{algorithm}
\usepackage{algpseudocode}
\usepackage{cancel}
\usepackage{quiver}
\usepackage{algpseudocode}

\usepackage{color}
\definecolor{darkred}{rgb}{0.6,0.0,0.0}
\definecolor{darkgreen}{rgb}{0,0.50,0}
\definecolor{lightblue}{rgb}{0.0,0.42,0.91}
\definecolor{orange}{rgb}{0.99,0.48,0.13}
\definecolor{grass}{rgb}{0.18,0.80,0.18}
\definecolor{pink}{rgb}{0.97,0.15,0.45}

\newboolean{showpeople}
\setboolean{showpeople}{false} %

\ifthenelse{\boolean{showpeople}}{
\newcommand{\todo}[1]{\textcolor{red}{TODO: #1}}
\newcommand{\glb}[1]{\textcolor{blue}{Gilbert: #1}}
\newcommand{\saman}[1]{\textcolor{orange}{Saman: #1}}
\newcommand{\oded}[1]{\textcolor{purple}{Oded: #1}}
\newcommand{\odedtodo}[1]{\textcolor{purple}{Oded Todo: #1}}
\newcommand{\teo}[1]{\textcolor{green}{Teo: #1}}

}{
\newcommand{\todo}[1]{}
\newcommand{\glb}[1]{}
\newcommand{\saman}[1]{}
\newcommand{\oded}[1]{}
\newcommand{\odedtodo}[1]{}
\newcommand{\teo}[1]{}

}

\usepackage{subcaption}

\usepackage{listings}

\lstdefinelanguage{PythonPlus}[]{Python}{
  morekeywords=[1]{,as,assert,nonlocal,with,yield,self,True,False,None,match,case,} %
  morekeywords=[2]{,__init__,__add__,__mul__,__div__,__sub__,__call__,__getitem__,__setitem__,__eq__,__ne__,__nonzero__,__rmul__,__radd__,__repr__,__str__,__get__,__truediv__,__pow__,__name__,__future__,__all__,}, %
  morekeywords=[3]{,object,type,isinstance,copy,deepcopy,zip,enumerate,reversed,list,set,len,dict,tuple,range,xrange,append,execfile,real,imag,reduce,str,repr,}, %
  morekeywords=[4]{,Exception,NameError,IndexError,SyntaxError,TypeError,ValueError,OverflowError,ZeroDivisionError,}, %
  morekeywords=[5]{,ode,fsolve,sqrt,exp,sin,cos,arctan,arctan2,arccos,pi, array,norm,solve,dot,arange,isscalar,max,sum,flatten,shape,reshape,find,any,all,abs,plot,linspace,legend,quad,polyval,polyfit,hstack,concatenate,vstack,column_stack,empty,zeros,ones,rand,vander,grid,pcolor,eig,eigs,eigvals,svd,qr,tan,det,logspace,roll,min,mean,cumsum,cumprod,diff,vectorize,lstsq,cla,eye,xlabel,ylabel,squeeze,}, %
}
\lstdefinelanguage{PyBrIM}[]{PythonPlus}{
  emph={d,E,a,Fc28,Fy,Fu,D,des,supplier,Material,Rectangle,PyElmt},
}

\lstdefinestyle{colorEX}{
  backgroundcolor=\color{white},
  commentstyle=\color{darkgreen}\slshape,
  keywordstyle=\color{blue}\bfseries\itshape,
  keywordstyle=[2]\color{blue}\bfseries,
  keywordstyle=[3]\color{grass},
  keywordstyle=[4]\color{red},
  keywordstyle=[5]\color{orange},
  stringstyle=\color{darkred},
  emphstyle=\color{pink}\underbar,
}
\usepackage{letltxmacro}
\newcommand*{\SavedLstInline}{}
\LetLtxMacro\SavedLstInline\lstinline
\DeclareRobustCommand*{\lstinline}{%
  \ifmmode
    \let\SavedBGroup\bgroup
    \def\bgroup{%
      \let\bgroup\SavedBGroup
      \hbox\bgroup
    }%
  \fi
  \SavedLstInline
}

\newcommand{\tx}{\mathbf{tx}}
\newcommand{\nvar}{\mathbf{n}}
\newcommand{\bvar}{\mathbf{b}}
\newcommand{\fvar}{\mathbf{f}}
\newcommand{\cvar}{\mathbf{c}}

\newcommand{\s}{\mathbf{s}}
\newcommand{\szero}{\mathbf{s_{0}}}
\newcommand{\sone}{\mathbf{s_{1}}}
\newcommand{\sstar}{\mathbf{s^{\star}}}
\newcommand{\vstar}[1]{\mathbf{#1^{\star}}}
\newcommand{\Rspace}{\mathbf{R}}

\newcommand{\e}{\mathbf{e}}
\newcommand{\ezero}{\mathbf{e_{0}}}
\newcommand{\eone}{\mathbf{e_{1}}}

\newcommand{\nzero}{\mathbf{n_{0}}}
\newcommand{\nk}{\mathbf{n_{k}}}
\newcommand{\txzero}{\mathbf{tx_{0}}}

\newcommand{\boldoplus}{\mathbf{\oplus}}
\newcommand{\boldotimes}{\mathbf{\otimes}}
\newcommand{\boldplus}{\mathbf{+}}
\newcommand{\boldtimes}{\mathbf{\times}}
\newcommand{\boldstar}{\mathbf{\star}}
\newcommand{\boldsharp}{\mathbf{\sharp}}
\newcommand{\boldsharparg}[1]{\mathbf{#1}^{\mathbf{\sharp}}}
\newcommand{\esharp}{\mathbf{e}^{\boldsharp}}

\newcommand{\Kernel}{\operatorname{\mathbf{Kernel}}}
\newcommand{\Image}{\operatorname{\mathbf{Image}}}
\newcommand{\CoImage}{\operatorname{\mathbf{CoImage}}}
\newcommand{\CoKernel}{\operatorname{\mathbf{CoKernel}}}

\newcommand{\Zero}{\operatorname{\mathbf{Zero}}}
\newcommand{\Eval}{\operatorname{\mathbf{Eval}}}
\newcommand{\Abstract}{\operatorname{\mathbf{Abstract}}}
\newcommand{\Reshape}{\operatorname{\mathbf{Reshape}}}
\newcommand{\Inner}{\operatorname{\mathbf{Inner}}}
\newcommand{\Project}{\operatorname{\mathbf{Project}}}
\newcommand{\Embed}{\operatorname{\mathbf{Embed}}}
\newcommand{\Inverse}{\operatorname{\mathbf{Inverse}}}
\newcommand{\Restrict}{\operatorname{\mathbf{Restrict}}}
\newcommand{\NonLinear}{\operatorname{\mathbf{NonLinear}}}

\newcommand{\Injective}{{\operatorname{\mathbf{Injective}}}}
\newcommand{\Surjective}{\mathbf{\operatorname{\mathbf{Surjective}}}}
\newcommand{\Bijective}{\mathbf{\operatorname{\mathbf{Bijective}}}}
\newcommand{\gprop}{\mathbf{\operatorname{\mathbf{prop}}}}

\newcommand{\gV}{\mathbf{V}}
\newcommand{\gW}{\mathbf{W}}
\newcommand{\gF}{\mathbf{F}}
\newcommand{\attemptmath}[1]{\mathbf{#1}}

\newcommand{\sh}{\mathbf{sh}}
\newcommand{\av}{\mathbf{av}}
\newcommand{\en}{\mathbf{en}}

\newcommand{\rf}{\mathbf{rf}}
\newcommand{\rfzero}{\mathbf{rf_{0}}}
\newcommand{\rfone}{\mathbf{rf_{1}}}

\newcommand{\ar}{\mathbf{ar}}

\newcommand{\stmt}{\mathbf{stmt}}
\newcommand{\formula}{\mathbf{formula}}

\newcommand{\bolddiv}{\mathbf{/}}

\newcommand{\Array}{\operatorname{\mathbf{Array}}}
\newcommand{\Einsum}{\operatorname{\mathbf{Einsum}}}
\newcommand{\QR}{\operatorname{\mathbf{QR}}}
\newcommand{\SVD}{\operatorname{\mathbf{SVD}}}

\newcommand{\bzero}{\mathbf{b_{0}}}
\newcommand{\bone}{\mathbf{b_{1}}}

\newcommand{\RawBasis}{\operatorname{\mathbf{RawBasis}}}
\newcommand{\DualBasis}{\operatorname{\mathbf{DualBasis}}}
\newcommand{\PrimalBasis}{\operatorname{\mathbf{PrimalBasis}}}
\newcommand{\ProjectedBasis}{\operatorname{\mathbf{ProjectedBasis}}}
\newcommand{\ImageBasis}{\operatorname{\mathbf{ImageBasis}}}
\newcommand{\tdim}{\mathbf{tdim}}
\newcommand{\dervType}{\mathbf{dervType}}
\newcommand{\topType}{\mathbf{topType}}
\newcommand{\chartTy}{\mathbf{chartTy}}
\newcommand{\fieldTy}{\mathbf{fieldTy}}
\newcommand{\Euclidean}{\operatorname{\mathbf{Euclidean}}}
\newcommand{\Tangential}{\operatorname{\mathbf{Tangential}}}

\newcommand{\Simplex}{\operatorname{\mathbf{Simplex}}}

\newcommand{\TopChart}{\operatorname{\mathbf{TopChart}}}
\newcommand{\BoundaryChart}{\operatorname{\mathbf{BoundaryChart}}}
\newcommand{\Chart}{\operatorname{\mathbf{Chart}}}

\newcommand{\TangentSpace}{\operatorname{\mathbf{TangentSpace}}}
\newcommand{\EuclideanSpace}{\operatorname{\mathbf{EuclideanSpace}}}

\newcommand{\field}{\operatorname{\mathbf{field}}}

\newcommand{\cv}{\mathbf{cv}}
\newcommand{\fv}{\mathbf{fv}}

\newcommand{\argVar}{\mathbf{arg}}
\newcommand{\args}{\mathbf{args}}
\newcommand{\PTEM}{\mathbf{PTEM}}

\newcommand{\spaceOf}{\operatorname{\mathbf{spaceOf}}}

\newcommand{\TangentVector}{\operatorname{\mathbf{TangentVector}}}
\newcommand{\CoTangentVector}{\operatorname{\mathbf{CoTangentVector}}}
\newcommand{\NormalVector}{\operatorname{\mathbf{NormalVector}}}
\newcommand{\CoNormalVector}{\operatorname{\mathbf{CoNormalVector}}}

\newcommand{\Xop}{\operatorname{\mathbf{X}}}
\newcommand{\FieldAccess}{\operatorname{\mathbf{FieldAccess}}}
\newcommand{\Dop}{\operatorname{\mathbf{D}}}

\newcommand{\Pointwise}{\operatorname{\mathbf{Pointwise}}}

\newcommand{\bc}{\mathbf{bc}}
\newcommand{\integral}{\mathbf{integ}}
\newcommand{\op}{\mathbf{op}}
\newcommand{\overVar}{\mathbf{over}}
\newcommand{\reduction}{\mathbf{reduction}}
\newcommand{\reductions}{\mathbf{reductions}}

\newcommand{\PointEval}{\operatorname{\mathbf{PointEval}}}
\newcommand{\IntegralOp}{\operatorname{\mathbf{Integral}}}

\newcommand{\MinOp}{\operatorname{\mathbf{Min}}}
\newcommand{\MaxOp}{\operatorname{\mathbf{Max}}}
\newcommand{\SumOp}{\operatorname{\mathbf{Sum}}}

\newcommand{\Difference}{\operatorname{\mathbf{Difference}}}
\newcommand{\Avg}{\operatorname{\mathbf{Avg}}}
\newcommand{\choice}{\operatorname{\mathbf{choice}}}
\newcommand{\choiceAllEq}{\operatorname{\mathbf{choiceAllEq}}}

\newcommand{\LocalChart}{\operatorname{\mathbf{LocalChart}}}
\newcommand{\LocalField}{\operatorname{\mathbf{LocalField}}}

\newcommand{\GlobalChart}{\operatorname{\mathbf{GlobalChart}}}
\newcommand{\GlobalField}{\operatorname{\mathbf{GlobalField}}}

\newcommand{\ChartOp}{\operatorname{\mathbf{Chart}}}
\newcommand{\Basis}{\operatorname{\mathbf{Basis}}}
\newcommand{\ALL}{\operatorname{\mathbf{ALL}}}

\newcommand{\Argument}{\operatorname{\mathbf{Argument}}}
\newcommand{\Default}{\operatorname{\mathbf{Default}}}
\newcommand{\Storage}{\operatorname{\mathbf{Store}}}

\newcommand{\IPTX}{\operatorname{\mathbf{IPTX}}}

\newcommand{\degVar}{\mathbf{deg}}
\newcommand{\Psp}{\mathbf{P}}   %
\newcommand{\Deg}{\operatorname{\mathbf{Deg}}}
\newcommand{\RangeOp}{\operatorname{\mathbf{Range}}}
\newcommand{\dvar}{\mathbf{d}}
\newcommand{\IPTEMS}{\mathbf{IPTEMS}}

\newcommand{\Pure}{\operatorname{\mathbf{Pure}}}
\newcommand{\KernelOp}{\operatorname{\mathbf{Kernel}}}
\newcommand{\IKernel}{\operatorname{\mathbf{IKernel}}}
\newcommand{\BKernel}{\operatorname{\mathbf{BKernel}}}

\usepackage{tabularx}
\usepackage{longtable}
\usepackage{makecell}
\usepackage{caption}
\usepackage{adjustbox}
\usepackage{threeparttable}

\makeatletter
\lst@AddToHook{Init}{\def\@currentcounter{lstnumber}}%
\makeatother

\usepackage{cleveref}

\crefname{lstlisting}{listing}{listings}
\Crefname{lstlisting}{Listing}{Listings}
\crefname{example}{example}{examples}
\Crefname{example}{Example}{Examples}
\crefname{paragraph}{paragraph}{paragraphs}
\Crefname{paragraph}{Paragraph}{Paragraphs}

\AddToHook{env/definition/begin}{\crefalias{section}{definition}}
\AddToHook{env/example/begin}{\crefalias{section}{example}}
\AddToHook{env/proposition/begin}{\crefalias{section}{proposition}}
\AddToHook{env/theorem/begin}{\crefalias{section}{theorem}}

\setboolean{ms}{false}  %
\AtBeginDocument{%
  \providecommand\BibTeX{{%
    \normalfont B\kern-0.5em{\scshape i\kern-0.25em b}\kern-0.8em\TeX}}}

\setcopyright{none}
\begin{document}
\lstset{language=PythonPlus}

\title{Extensible Programmatic Specification of Finite Element Methods}

\author{Theadora Fields Collin}
\email{teoc@mit.edu}
\affiliation{\institution{Massachusetts Institute of Technology}
  \city{Cambridge}\state{Massachusetts}\country{USA}}

\author{Oded Stein}
\email{ostein@usc.edu}
\affiliation{\institution{University of Southern California}
  \city{Los Angeles}\state{California}\country{USA}}
\affiliation{\institution{Technion---Israel Institute of Technology}
  \city{Haifa}\country{Israel}}

\author{Gilbert Bernstein}
\email{gilbo@cs.washington.edu}
\affiliation{\institution{University of Washington}
  \city{Seattle}\state{Washington}\country{USA}}

\author{Saman Amarasinghe}
\email{saman@csail.mit.edu}
\affiliation{\institution{Massachusetts Institute of Technology}
  \city{Cambridge}\state{Massachusetts}\country{USA}}

\renewcommand{\shortauthors}{Collin et al.}

\begin{abstract}
Finite element methods (FEMs) are a powerful and ubiquitous tool for solving engineering problems.
Experimenting with different finite elements can improve the quality and efficiency of solutions.
In some cases, the wrong (but nonetheless most common) choice of finite element will produce solutions which converge to the wrong answer regardless of mesh resolution.
However, in practice, the choice of finite element is not explored due to the complexity of re-deriving and re-implementing finite element methods.
Trying a new finite element is challenging because practitioners must manually deduce formulas to use these elements and they must implement these formulas within the context of a potentially complex system.
We address this problem by introducing \SYS{}, a finite element system that is parametric over the literate mathematical specification of a finite element in a domain-specific language (DSL).
The \SYS{} compiler reasons about tensor spaces, tensors, and tensor bases from first principles to derive implementations of finite elements.
The \SYS{} compiler is able to automatically derive implementations of finite elements previously only derived by hand.
Further, \SYS{} minimally couples several key mathematical concepts, mainly tensor fields, mesh topologies, sparse tensors, and assembled finite element operators, to produce a complete finite element system that is parametric over the choice of element.
Consequently, the elements derived by the compiler can be applied parametrically to new meshes, PDEs, and boundary conditions.
We evaluate our system by implementing several simulations with different finite elements, demonstrating that our system can explore tradeoffs in generality, accuracy, speed, and representational complexity.
For example, we are able to implement the Morley, Bell, Argyris, and Hermite like elements with less than 50 lines of code and use them all in a single simulation.

\end{abstract}


\begin{teaserfigure}
  \includegraphics[width=\textwidth]{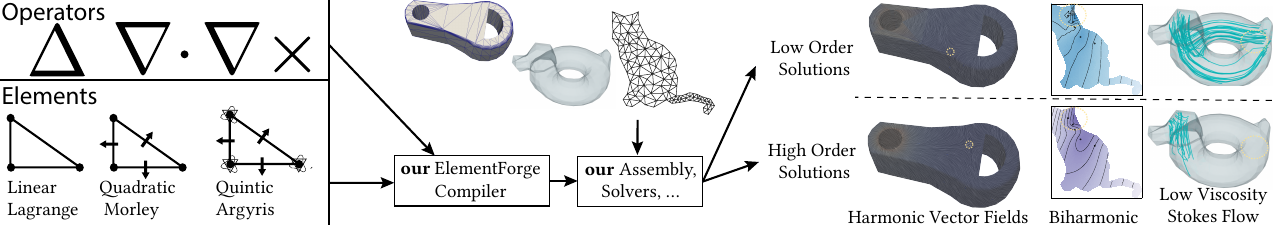}
  \caption[System Architecture]{From a high level description of a finite element and an operator, our system produces routines that can be used to deliver better higher order solutions to a variety of graphics problems. Yellow dotted circles indicate telltales that the bottom solutions are higher quality.}
  \Description{A diagram connects operator and finite-element specifications
    and input meshes to the ElementForge compiler and assembly routines.
    Paired low- and high-order results show harmonic vector fields,
    biharmonic solutions, and low-viscosity Stokes flow, with yellow circles
    marking differences in solution quality.}
  \label{fig:teaser}
\end{teaserfigure}

\maketitle

\section{Introduction}
\label{sec:intro}

Finite Element Methods (FEMs) have become an important tool in a variety of fields, including applied mathematics, scientific computing,
engineering,
and computer graphics~\cite{finiteelementmethodforellipticproblems2002,thefiniteelementmethod2000,themathematicaltheoryoffiniteelementmethods2008}.
Within computer graphics, FEMs are a fundamental tool for many simulation and geometry processing pipelines.
FEMs are powerful because one can vary the mesh, the problem formulation, the solver, and the
\emph{element}, a discrete local function space.
One can choose from more than 100 families of finite elements
\cite{defelementanencyclopediaoffiniteelementdefinitions2023}, each of which provide specific quantitative and qualitative advantages and disadvantages.
Within computer graphics specifically, practitioners have examined the trade-offs between convergence and computational cost of various higher order Lagrange elements~\cite{alargescalecomparisonoftetrahedralandhexahedralelementsforsolvingellipticpdeswiththefiniteelementmethod2022,polysplinefiniteelementmethod2019,fernandez2025symx}.

However, practitioners face three challenges if they want to determine and exploit the trade-offs that different finite elements (FEs) offer.
The first challenge is that practitioners must deduce, from the definition of the FE, various mathematical formulas that software packages require to implement an FE~\cite{finatfinatasmarterlibraryoffiniteelements,algorithm839fiatanewparadigmforcomputingfiniteelementbasisfunctions2004}.
The second challenge is that practitioners must understand global FE operations (e.g., methods to build a sparse matrix or enforce a boundary condition) at both a mathematical and software level so that they can reimplement these operations for their class of FE.
The third challenge is that for practitioners to explore these trade-offs, they must experiment with multiple FEs, encountering the first two challenges each time, and they must also experiment with their problem formulation to properly exploit some FEs.

The first challenge motivates our focus on deriving element transformations directly from mathematical specifications.
Many systems offer partial solutions to the second challenge because they internally provide an interface to FEs and use this interface to implement many
global FEM operations~\cite{codegenerationforgenerallymappedfiniteelements2019,getfemautomatedfemodelingofmultiphysicsproblemsbasedonagenericweakformlanguage2020,mfemamodularfiniteelementmethodslibrary2021}.
 In software packages that offer solutions to a set of narrow problems, interfaces to an FE might not be present (the FE is baked into the script) or the interface only serves a narrow set of problems (e.g., it only supports certain derivatives or boundary conditions (BCs)).
Finally, many systems address third problem via libraries or domain specific languages (DSLs) to vary the problem formulation~\cite{fernandez2025symx, unifiedformlanguageadomainspecificlanguageforweakformulationsofpartialdifferentialequations2014,fullyandsemiautomatedshapedifferentiationinngsolve2020,automatedsolutionofdifferentialequationsbythefiniteelementmethod2016,firedrakeautomatingthefiniteelementmethodbycomposingabstractions2016,mfemamodularfiniteelementmethodslibrary2021,dunefemageneralpurposediscretizationtoolboxforparallelandadaptivescientificcomputing2012}.
These DSLs offer a high degree of flexibility for problem specifications and can work with different FEs. Our focus is the additional step of deriving reference-to-world transformations from element definitions, rather than requiring element-specific implementations of those transformations.
As a practical consequence of the prior lack of automation, many communities stick to varying the problem to work with linear Lagrange elements, hiding a large space of trade-offs involving both higher order Lagrange elements and \emph{non-Lagrangian elements}.
Thus, we propose to significantly lower the cost of varying FEs and problems with \SYS{}, the first FE DSL that is fully parametric over a mathematical definition of an FE.
This parametricity extends through the FEM pipeline: interpolation, boundary-condition handling, and operator assembly are expressed using common primitives that adapt to the chosen element's mathematical definition (\cref{sec:GFDC}).
We believe that \SYS{} will allow many to find and take advantage of trade-offs among FEs and associated problem formulations.

To meet all three challenges, our work, \SYS{}, offers abstractions, languages, and deductive systems for fundamental aspects of FEM.
To answer the first challenge, \SYS{} contains a language for specifying FEs and a compiler that automatically deduces the formulas required to be implemented by hand in other packages, thus automating work previously found in mathematical journals.
To answer the second challenge, \SYS{} offers several core primitives that we show are sufficient for users to systematically generate global finite element operations that translate local formulas to useful code for building and manipulating sparse matrices.
To answer the third challenge, \SYS{} allows users to specify suitable problems with the same language that it provides to specify FEs.
Finally, \SYS{} agglomerates these solutions into a single package that allows users to define, reuse, and combine various FEs, problem specifications, and global routines to rapidly explore FEMs. 
The core design principle of \SYS{} is to systematically model the discrete mathematical objects of FEMs.
In contrast to prior work, our software architecture does not emphasize either the global theory of PDEs or the structure of a specific class of problems.
\SYS{} emphasizes the discrete mathematical objects at the core of FEM \emph{implementations} as these are the tools required to save labor through automation.
Consequently, \SYS{} pays special care to pointwise manipulation of tensor fields on manifolds, the definition of an FE, the construction of a discrete mesh, and the algebraic construction of global operations via the interaction of pointwise expressions and discrete meshes.
These new tools provide the means to go from a mathematically specified FE and operator to a sparse matrix, automatically, and, hopefully, understandably.

Our implementation is not without some limitations, especially as we currently do not pay attention to low level performance issues; many FEM packages have outlined techniques to optimize code generated for FEs with special structures and these techniques could eventually be employed here~\cite{automaticcodegenerationforhighperformancediscontinuousgalerkinmethodsonmodernarchitectures2020,astudyofvectorizationformatrixfreefiniteelementmethods2020,exposingandexploitingstructureoptimalcodegenerationforhighorderfiniteelementmethods2017}.
Even with these limitations, we are able to implement novel simulations of independent interest to different areas of computer graphics.
Moreover, we view these simulations and our software as part of an effort to make FEs more accessible. 
We make the following contributions:

\begin{enumerate}
    \item To the best of our knowledge, the first FE DSL that is fully parametric over a mathematical definition of an FE, enabling users to reuse and adapt global FEM routines across different elements and problems.
    \item A novel DSL and compiler for specifying FEs and operators via pointwise integrals of coordinate-free tensor algebra. This system enables the automated deduction of formulas needed to implement FEs at the local level.
    \item An interface for the systematic construction of global FE operations, enabling core and new global FE operations to be built on a wide variety of elements.
    \item An overall compiler taking as input (i) an FE description, (ii) operator specifications, and (iii) a mesh, and producing discretization operators (and interpolation functions) as output.
    \item Several novel simulations of interest to computer graphics:
    \begin{enumerate}
        \item A more efficient yet simpler data interpolation/smoothing method based on the natural BCs of the biharmonic equation, identified in ~\cite{naturalboundaryconditionsforsmoothingingeometryprocessing2018}, and the Morley element~\cite{weakgalerkinmethodsbasedmorleyelementsongeneralpolytopalpartitions2022}. Moreover, we have explored a variety of FEs for this class of problems, leading to characterization of some of the tradeoffs involved.
        \item To the best of our knowledge, the first implementation of the finite element exterior calculus (FEEC) that is derived purely and automatically from the mathematical specification, which allows us to re-implement the analogues of all the discrete exterior calculus (DEC) operators, leading to FEM versions of classic DEC applications. Our methods also expose the cheaper first kind Nédélec elements~\cite{finiteelementexteriorcalculusfromhodgetheorytonumericalstability2010}, which are not properly characterized in the graphics literature (compared to the standard second kind elements). 
        \item Several FEs for Stokes flows problems in 2D+time and 3D+time, illustrating the phenomena of locking incompressible fluids (in contrast to prior work that focused on elasticity).
        
    \end{enumerate}
    \ifthenelse{\boolean{ms}}{
        }
        {}
\end{enumerate}

\subsection{Paper Outline}
\begin{figure}
    \centering
    \includegraphics[width=\linewidth]{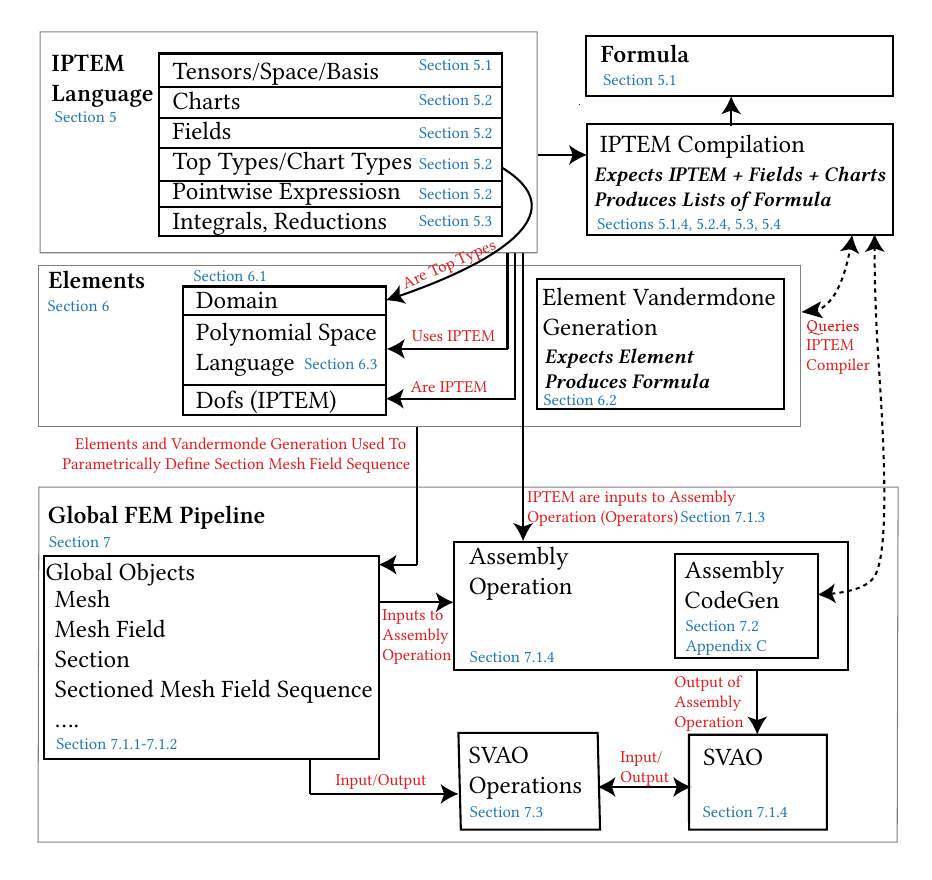}
    \caption{Our system can be decomposed into roughly three components mirroring the three methods/implementations sections. The first component consists of a core language, Integrated Pointwise Tensor Expressions on Manifolds (IPTEM), and an associated compiler. The second component features a software description of finite elements and infrastructure to generate \VA{} matrices, one of the core challenges of using general elements. The second component uses IPTEMs to describe element and uses IPTEM compiler to produce \VA{} matrices. The third component allows users to write global FEM code over general elements. In particular, we imagine that users will perform the following loop: (1) instantiate some initial global objects (meshes, input data), (2) combine these with IPTEM  (representing operators and other computations) to build global tensor objects called SVAO (partially via the IPTEM compiler), (3) use a few SVAO operations to produce new global objects, which can be used for more assemblies or passed as raw data to IO/solvers/dense array routines/sparse array routines/etc. Red text indicates how this outline maps on to our paper.}
    \label{fig:outline}
\end{figure}

To introduce our system from a user perspective and to discuss the fundamental needs of the system, we will (in \cref{sec:examlples}) discuss an example program (the cotangent Laplacian essentially) that we adapt to other PDEs and finite elements.
These examples motivate the three implementation sections: expressing local DOFs and operators leads to IPTEM (\cref{sec:swrdoIPTEM}); deriving bases for new elements leads to the transformation algorithm (\cref{sec:SRFE}); and assembling operators and handling boundary data leads to the global interface (\cref{sec:GFDC}).
After this, we provide mathematical background (in \cref{sec:background}) to properly define certain FEM concepts; we will define the quantities that our system automatically provides to further the automation of FEMs.
In the next three sections, we will describe our system; we will show how we actually implement previously non-automated aspects of FEMs.
We outline these sections in \cref{fig:outline}.
After we finish describing our system, we apply it to several different simulations of interest to computer graphics (in \cref{sec:eval}). 
Therein, we analyze how the automations impact our ability to explore these simulations and the utilities of the tradeoffs of different explored paths. 

\section{Examples}
\label{sec:examlples}

To illustrate the scope of our system and our way of thinking, we run through a few examples that use our highest-level interface.
\subsection{Laplace's Equation}

 Given a surface $\Omega$ and a boundary condition $f\colon \partial\Omega\to\mathbb{R}$, we want to find a function $u\colon \Omega\to\mathbb{R}$ such that
\begin{equation}\begin{split}
\label{eq:lap1}
    \Delta u(x) &= 0 \mbox{ for } x\in\Omega \\
    u(x) &= f(x) \mbox{ for } x \in\partial\Omega \;.
\end{split}\end{equation}

Our program, \cref{lst:simplestExampleListing}, uses linear Lagrange finite elements to produce the familiar cotangent Laplace matrix and solve Laplace's equation.
\Cref{lst:simplestExampleListing} shows how this works in our system: we read a mesh (\cref{line:loadMesh}), express a formulation of the Laplacian (\cref{line:laplace}), assemble the corresponding Laplacian matrix (\cref{line:laplaceMatrix}), assemble boundary DOFs (\cref{line:prop}) by propagating marked boundary edges to marked boundary DOFs, determine the boundary values (\cref{line:propFin}), eliminate the boundary conditions from the system (\cref{line:elim1}-\cref{line:elim2}), solve our system (\cref{line:solve}), and post-process (\cref{line:post}).

\begin{lstlisting}[float, language=PythonPlus, caption={Solving the Laplace equation with FEM in \SYS{}. },label={lst:simplestExampleListing},escapechar=|, captionpos=b,abovecaptionskip=5pt]
from EF import *

# Load a mesh.
mesh = TriangleVertexListReader(filename) |\label{line:loadMesh}|

@pointwiseIntegral
def laplace(u: field(R), v: field(R)): |\label{line:laplace}|
    return grad(u).dot(grad(v)) 

# Make a sparse matrix, the cotan Laplacian.
laplaceMatrix = assemble(mesh,[laplace],[LinearLagrange,LinearLagrange]) |\label{line:laplaceMatrix}|

# Create boundary DOFs
# on the boundary, which DOFs in the input spaces are relevant:
boundaryDofs = propagate(mesh.topologicalBoundary(), laplaceMatrix.inputSections[0]) |\label{line:prop}|

# setup boundary values
f = lambda x: ... 
boundaryValues = assemble(mesh, LinearLagrange, [f]) |\label{line:propFin}|

# Setup linear system via elimination method for bcs
Ainterior = laplaceMatrix.reduce(["inputDofs0", "inputDofs1"]).getSlice([~boundaryDofs, ~boundaryDofs]).asScipyMatrix() |\label{line:elim1}|
Abdry = laplaceMatrix.reduce(["inputDofs0", "inputDofs1"]).getSlice([~boundaryDofs, boundaryDofs]).asScipyMatrix()
b = boundaryValues.reduce(["outputDofs"]).getSlice(boundaryDofs).asDenseVector() |\label{line:elim2}|
# Solve
uvals = spsolve(Ainterior, -Abdry.dot(b)) |\label{line:solve}|
# wrap values back into the system:
from EF import GlobalDofField
u = GlobalDofField(mesh, LinearLagrange, uvals) |\label{line:post}|
# u can be visualized, used for further assemble calls, etc.
\end{lstlisting}

A few aspects of this program raise questions, especially when it comes to solving other problems: 
\begin{enumerate}
    \item Why does our Laplace function not directly implement $\Delta$ as seen in ~\cref{eq:lap1}?
    \item What really is \lstinline{propagate} doing to help enforce boundary conditions and how does this generalize to other conditions?
    \item Where does our element, \lstinline{LinearLagrange}, come from and how does it generalize?
\end{enumerate}

All these concerns are tied together via the concept of weak formulations.
To deeply understand the fundamental issues of matrices of PDEs, boundary conditions, and elements, we must understand the weak formulation.

\subsection{Weak Formulations of Laplace and the Biharmonic Equation}

Prior to writing \cref{lst:simplestExampleListing}, we transformed the initial problem. 
By choosing an appropriate space of functions, $V$, we can integrate by parts to go from ~\cref{eq:lap1} to the weak formulation \cite{themathematicaltheoryoffiniteelementmethods2008}:
find a $u\in V$ that satisfies the BCs in \cref{eq:lap1} such that for all $v\in \{v\in V\colon v|_{\partial\Omega} = 0 \}$:

\begin{equation}
    \label{eq:lapwf2}
    \int_{\Omega} \nabla v \cdot \nabla u = 0.
\end{equation}

Similar weak formulations and matrices for PDEs (stiffness, mass, etc) emerge via two steps: (1) using integration by parts to transform a strong form problem with order $2n$ derivatives to a weak one with order $n$ derivatives and then (2) picking a space of functions $V$ such that (at a minimum) the integration by parts is valid and the homogeneous version of the boundary condition can be identified~\cite{discretedifferentialgeometryoperatorsfortriangulated2manifolds2003}.
These two steps tie the weak formulation to the  mathematical structures of elements and boundary conditions.

\subsubsection{The Biharmonic Equation and Other Elements}

Applications such as elasticity,  Stokes flow, cage coordinates, and data interpolation can all be studied with the biharmonic equation.
Given a domain $\Omega$ and BCs $f,g\colon\partial\Omega\to\mathbb{R}$,  we seek a $u\colon\Omega\to\mathbb{R}$ such that 
\begin{equation}
\label{eq:biharm1}
    \Delta\Delta u = 0 \mbox{ on } \Omega
\end{equation}
and 
\begin{equation}
    \label{eq:biharm2}
    u = f  \mbox { and } \frac{\partial u}{\partial n} = g \mbox{ on } \partial\Omega.
\end{equation}
Together, these prescribe Dirichlet data for the fourth-order biharmonic problem: both the value and the normal derivative are fixed. The homogeneous case $f=g=0$ gives clamped boundary conditions.

We transform the problem into its weak form with a space of functions $V$ and $V_{0} = \{v\in V\colon v|_{\partial\Omega} =\frac{\partial v}{\partial n}|_{\partial\Omega} = 0\}$.
Our weak formulation reads: find a $u\in V$ subject to \eqref{eq:biharm2} such that for all $v\in V_{0}$:
\begin{align*}
    0 = \int_{\Omega} 0v &= -\int_{\Omega}(\Delta\Delta u)v
    =\int_{\Omega}(\nabla\Delta u)\cdot \nabla v - \int_{\partial\Omega} v \frac{\partial\Delta u}{\partial n}\\
    &=\int_{\Omega} \Delta u \Delta v - \int_{\partial\Omega} \Delta u \frac{\partial v}{\partial n} - \int_{\partial\Omega} v \frac{\partial\Delta u}{\partial n}
     =\int_{\Omega} \Delta u \Delta v\\
\end{align*}
where the last step uses $V_{0}$ and the rest use integration by parts.

Based on this analysis, a natural start to solving the biharmonic equation would be to replace the function \lstinline!laplace! in \cref{lst:simplestExampleListing}:\cref{line:laplace} with the biharmonic operator in \cref{lst:biharmonicExample} and just ignore boundary conditions for the moment.

\begin{lstlisting}[float, language=PythonPlus, caption={The Biharmonic Operator. This operator utilizes a \lstinline{chart} which allows access to the metric of the geomtry through the co-tangent space's inner produt. When we use a chart as an argument type, we are providing access to an arbitrary paramtermization of some unknown geometry.}, label={lst:biharmonicExample},abovecaptionskip=5pt]
@pointwiseIntegral
def biharmonic(c: chart, u: field(R), v: field(R)):
    # trace of Hessian with respect to metric
    # because the inner product of tensors is a trace
    deltau = (c.CT.inner()).dot(D(D(u)))
    deltav = (c.CT.inner()).dot(D(D(v)))
    return deltau * deltav
\end{lstlisting}

However, if we do this, then \cref{lst:simplestExampleListing}:\cref{line:laplaceMatrix} will produce a zero matrix.
To understand why, we consider the definition of \lstinline{LinearLagrange} in \cref{lst:llexample}.
\begin{lstlisting}[float, language=PythonPlus, caption = {[Linear Lagrange Dof]{How to write down linear Lagrange FEs.
%
We construct a pointwise evaluation on an abstract vertex by using the \lstinline!ChartType(0)!.
%
The \lstinline!ChartType(0)! indiciates the use of parameterization of an object of dimension zero, which is to say the provision of a point.
%
We interpret pointwise operators as being replicated across all objects in a domain (e.g. two evaluations per line, three per triangle, four per quad).
%
The final elements are defined by pairing a list of DOFs with a representation of a domain (an int here) and a polynomial space (Pure).
}},label={lst:llexample},abovecaptionskip=5pt]
cty = ChartType(0) # on objects of dimension 0
@pointwiseEval(cty, barycentricCoordinate=0.0)
def vertexEval(u: field(cty, R)):
    return u(0)  # u(0) is the field value i.e., the zeroth derivative.
# face type, degree/domain of polynomials, List[Dof]
LinearLagrangeEdge = FE(1, Pure(Simplex(1), 1), [vertexEval])
LinearLagrange = FE(2, Pure(Simplex(2), 1), [vertexEval])
\end{lstlisting}
Returning to \cref{lst:biharmonicExample}, we see the problem is at first just the polynomial degree: second derivatives of affine functions are zero.
A natural second attempt is to use quadratic functions, as in \cref{lst:ll2example}.

\begin{lstlisting}[float, language=PythonPlus, caption={[A Quadratic Lagrange DOF]{A Quadratic Lagrange DOF, specifying a pointwise evaluation on an edge.}}, label={lst:ll2example},abovecaptionskip=5pt]
cty1 = ChartType(1) # on objects of dimension 1
@pointwiseEval(cty1, barycentricCoordinate=(0.5, 0.5))
def edgeMidEval(u: field(cty1, R)):
    return u(0)
    
# face type, degree/domain, List[Dof]
quadEdge = FE(1, Pure(Simplex(1), 2), [vertexEval, edgeMidEval])
quadTri = FE(2, Pure(Simplex(2), 2), [vertexEval, edgeMidEval])
\end{lstlisting}
However, this will still result in a catastrophic error as the matrix will be singular.
How can we debug?
We can question the fundamental assumptions that got us to this point.
Since the original PDE problem has a unique solution, the two manipulations that we used to deduce the weak formulation were integration by parts and the definition of $V_{0}$.
We can test if these manipulations that we assumed actually work in practice with our system.
We start with integration by parts.
If we integrate by parts on a simple mesh, $\Omega = T_{1}\cup T_{2}$, we get
\begin{align*}
    0 &= \int_{\Omega}\Delta^{2}u v
    = \int_{T_{1}}\Delta^{2}u v + \int_{T_{2}}\Delta^{2}u v  \\
    &= \sum_{i=1,2}(\int_{T_{i}}\Delta u \Delta v - \int_{\partial T_{i}}\Delta u\frac{\partial v}{\partial n} - \int_{\partial T_{i}} v\frac{\partial\Delta u}{\partial n}) \\
    &= \int_{\Omega}\Delta u\Delta v - \int_{\partial \Omega}\Delta u\frac{\partial v}{\partial n} + \int_{\partial\Omega} v\frac{\partial\Delta u}{\partial n} + \int_{\partial T_{1}\cap \partial T_{2}} \sum_{i=1,2}( \Delta u \frac{\partial v}{\partial n} - v\frac{\partial\Delta u}{\partial n})\mid_{T_{i}}
\end{align*}

Thus, we need to test:
\begin{equation}
    \int_{\partial T_{1}\cap \partial T_{2}} ( \Delta u \frac{\partial v}{\partial n} - v\frac{\partial\Delta u}{\partial n})\mid_{T_{1}} - ( \Delta u \frac{\partial v}{\partial n} - v\frac{\partial\Delta u}{\partial n})\mid_{T_{2}} = 0
\end{equation}
for all $u,v$ in our quadratic Lagrange space.
We can generate this quantity in our system as a sparse matrix (though we simplify since third derivatives are zero).
To do this (\cref{lst:dgllisting}), we must integrate on all cell boundaries and take the difference of values computed for the same DOF on different boundaries.

\begin{lstlisting}[float, language=PythonPlus, caption={A patch test for the suitability of an element for the biharmonic equation.},label={lst:dgllisting},captionpos=b,abovecaptionskip=5pt]
@pointwiseIntegral(BoundaryChart, reduction=difference)
def interior(c: BoundaryChart, u: field(R), v: field(R)):
    n0 = c.normalVector(v.chartOf())
    deltau = u.chartOf().CT.inner().dot(u(2))
    return deltau * v(1).dot(n0) # deltau * dv/dn

interiorMatrix = assemble(mesh, [interior], [quadTri,quadTri])
# This should be zero except at boundary DOFs, but it won't be.
\end{lstlisting}

Since derivatives of quadratic Lagrange functions are discontinuous, this test will fail.
However, the structure of the test shows us how similar quadratic functions could succeed.
The Morley element, implemented in listing~\ref{lst:mordofs}, is the minimal element that satisfies \cref{lst:dgllisting}.
\begin{lstlisting}[float, language=PythonPlus, caption={The Morley DOFs demonstrate three key aspects of our system. First, we offer sophisticated tensor algebra such as the Hodge star to compute gradient normals.
%
Second, we are disciplined about how vectors and derivatives interact with meshes/physical space: \lstinline!euclideanDerivative=True! means that derivative values live in Euclidean space as opposed to the tangent space, ensuring the quantity is defined uniquely per edge.
%
In contrast, the Laplace operator uses a tangent space which is defined uniquely per cell.
%
Similarly, \cref{lst:dgllisting}, uses a normal vector because we want to measure a quantity that is computed differently per edge-cell relationship.
%
Third, our tensor algebra is disciplined in its interpretation of different spaces, preventing various mistakes: the tangent vector must be embedded into Euclidean space via \lstinline!c.E! before the derivative, defined in Euclidean space, uses it.
%
%
},label={lst:mordofs},abovecaptionskip=5pt]
@pointwiseEval(cty1, barycentricCoordinate=(0.5, 0.5))
def morleyDof(c: cty1, u: field(cty, R, euclideanDerivative=True)):
    return hodge(u(1))(c.E(c.tangentVector(0)))
Morley = FE(2, Pure(Simplex(2), 2), [vertexEval, morleyDof])
\end{lstlisting}

\subsubsection{The Biharmonic Equation and Boundary Conditions}
\label{sec:bebcerg}
In \cref{lst:simplestExampleListing}, we eliminated boundary conditions from the system around \cref{line:elim2}.
Despite the simplicity of this,  no system that we know of has consistent support for elimination methods on more complex FEs.
In particular, the complexity lies in determining if an elimination method is feasible (if $V_{0}$ in the weak formulation has a nice form).
(\Cref{lst:simplestExampleListing}:\cref{line:prop} only does the right thing when all DOFs defined on the boundary exactly correspond to the boundary condition).
Returning to the biharmonic equation, we can use the weak formulation to identify what the basis functions in $V_{0}$ must actually do.
For the biharmonic equation, if we want to use the naive weak formulation, we must ensure that
\begin{equation}
\label{eq:otest}
    (\int_{\partial \Omega}\Delta u\frac{\partial v}{\partial n} - \int_{\partial\Omega} v\frac{\partial\Delta u}{\partial n})  = 0
\end{equation}
for all $v\in V_{0}, u\in V$.
And to actually eliminate the DOFs, we need that if $u,v\in V$ then
\begin{equation}
\label{eq:test}
    \int_{\partial\Omega} (u + \frac{\partial u}{\partial n})(v + \frac{\partial v}{\partial n}) = 0.
\end{equation}
iff $u\in V_{0}$  or $v\in V_{0}$.
This is almost tautological, but the key point is that we can use our system to extract $V_{0}$ by building the above operators just as we could use it to analyze the smoothness of elements for a problem.
We could also use the above operators to implement Nitsche's method, an alternative for cases  where elimination methods fail~\cite{constructingnitschesmethodforvariationalproblems2024}.
Given the importance of boundary conditions though, we do offer a wrapper, really a simple elaboration on the above ideas, to apply boundary conditions as an alternative to using the system raw: given a mesh, an FE, a BC, and potentially a right hand side for a BC ($f_{i}$), we offer a variety of utilities to check if elimination methods can work or to utilize alternatives such as pointwise constraints.
These utilities along with our patch test and more complex elements allows us to build a more complete robust biharmonic solver from the framework of the basic Laplace solver.

\section{Background}
\label{sec:background}
Abstractly, our goal is to automate the labor of building discretized operators with FEMs.
Naively, our goal requires that we figure out how to write down bases of functions across arbitrary meshes.
However, FEMs already reduce this work through two steps: first, they stitch together many local bases defined on many mesh entities (\cref{sec:globfem}), and, second, they somehow describe each local basis in terms of a \textit{reference} basis (\cref{sec:mathfem}).
Unfortunately, the second step
is a tricky informal idea at present: we must still somehow specify the basis at each mesh element.
Thus, we will start (in \cref{sec:bptac}) by developing mathematical machinery to write down bases efficiently and exactly across many mesh elements.
In particular, we will recall linear algebra that allows for primal and dual vectors (contra vs covariant) and we will understand how these notions respond to changes of coordinates.
With this machinery, we can mathematically state (\cref{sec:probmain}) what it means to write " every single  basis in terms of a reference basis".
One of our chief contributions is to solve this problem (\cref{sec:SRFE}) for finite element bases specified via a language of coordinate independent linear algebra (\cref{sec:swrdoIPTEM}).
But to even state the problem, we must start with linear algebra and how we can write tensors efficiently and exactly.

\subsection{Writing Down Tensors}
\label{sec:bptac}
Given a point $x$ on the surface of the sphere $S^{2}$, consider the co-tangent space, $\mathcal{T}^{\star}_{x}S^{2}$. 
Roughly speaking, $\mathcal{T}^{\star}_{x}S^{2}$ is the space of vectors that could be $D_{x}f$ for any scalar function $f\colon S^{2}\to\mathbb{R}$.
A vector in the co-tangent space can be described as a list of numbers only after some contortion to pick a basis.
Alternatively, without a choice of basis or any information about the co-tangent space, we can refer to vectors in the co-tangent space by taking the derivative of scalar functions at a point $x$.
The definition of the co-tangent space lends itself to a natural and rather specific way of writing down vectors, but the modifier co hints at a germ of generality.

The cotangent space works because derivatives at a point $x$ are defined via: $f(x+t) = f(x) + Df_{x}(t)+ o(|t|)$ where $Df_{x}$ is linear in $t$.
This is to say that $Df_{x}\in \mathcal{T}^{\star}_{x}S^{2} :=  (\mathcal{T}_{x}S^{2})^{\star} := L(\mathcal{T}_{x}S^{2}, \mathbb{R})$ where $L(V,W)$ is the vector space of linear functions from a vector space $V$ to a vector space $W$ and where $\mathcal{T}_{x}S^{2}$ is the standard tangent space at $x\in S^{2}$.
We have just used the \textbf{dual space}, which is defined as follows: for a vector space $V$, the dual space is $V^{\star}:= L(V,\mathbb{R})$.

We can often more easily write down vectors in the dual space $V^{\star}$ than in the primal space $V$.
Consider the space of functions from a set $A$ to a vector space $V$: $\mathcal{F}(A,V)$.
In the case $\mathcal{F}(A,\mathbb{R})$, absent structure on $A$, a function/vector in  $\mathcal{F}(A,\mathbb{R})$ requires a table of information, $\{(a, f(a))\colon a\in A\}$. 
For each $a\in A$, the point-evaluation functional $F_{a}(f)=f(a)$ is a simple example of a vector in $\mathcal{F}(A,\mathbb{R})^{\star}$.
This is an example, not a characterization of all dual vectors: other linear functionals, such as integral moments on suitable function spaces, need not be evaluations at a single point.
We can generalize this pattern to a class of bases that allow us to compromise between the array prospective of Euclidean space and coordinate free vectors that are easy to write down:
\begin{definition}[Primal and Dual Basis]
        \label{def:pdbdag}
    Given a finite-dimensional vector space $V$ with a basis $\{v_{i}\}$, we say that $\{v^{\prime}_{j}\}$, a basis for $V^{\star}$ is a \textbf{dual basis} for $\{v_{i}\}$ if $v_{j}^{\prime}(v_{i}) = \delta_{ij}$.
    We call $\{v_{i}\}$ the \textbf{primal basis} to $\{v_{j}^{\prime}\}$.
    Together, $\{v_{i}\}$ and $\{v_{j}^{\prime}\}$ are a \textbf{primal-dual basis pair}. 
\end{definition}
The main point of this definition is that the dual basis is easy to write down and yields an array representation for the primal space (via the primal basis) that is easy to understand (via the dual basis).

Another clue that the derivative of a scalar function is a dual vector is the chain rule: $\nabla_{x} (f\circ T) = DT^{T}_{x} \cdot \nabla_{T(x)} f = \nabla_{T(x)} f\cdot DT_{x}$.
The co-vector, $\nabla_{T(x)} f$, is said to transform \textbf{covariantly} because the change of basis map, $DT_{x}$, is used instead of the inverse (\textbf{contravariance}).
A simpler definition is that dual vectors are covariant and primal vectors are contravariant.
This definition naturally extends when we consider how to write down tensors instead of just vectors.
When we make such an extension, we are now considering how we can write down linear maps such as $DT_{x}$ or $DT_{x}^{T}$ as arrays or other objects.

We are used to writing down linear maps as arrays of numbers and we understand that this is the result of a choice of basis.
However, it is not just any basis as given some $L\in L(V,W)$ some contortion is required to use bases for $V$ and $W$.
We introduce this contortion via the sharp map:
\begin{definition}[Sharp, Flat]
    \label{def:dag}
    Suppose, $\{v_{i}\}$ and $\{v_{j}^{\prime}\}$ are a primal-dual basis pair for a vector space $V$.
    Then they define a \textbf{sharp isomorphism} between $V^{\star}$ and $V$:
    \begin{equation}
        (v_{i}^{\prime})^{\sharp} =v_{i}.
    \end{equation}
    We denote the inverse of the sharp mapping as $(\cdot)^{\flat}$.
    The sharp map is determined by the choice of either dual or primal basis.
    If the space $V$ is an inner product space with inner product $(\cdot, \cdot)_{V}$ and $v_{i}$ is an orthogonal basis in the inner product, then that sharp map can also be written
    \begin{equation}
        (v^{\prime})^{\sharp}  = v \mbox { if } \forall w\in V, (w,v)_{V} = v^{\prime}(w).\label{eq:inner}
    \end{equation}
\end{definition}

Now, given a basis $\{w_{i}\}$ for $W$ and $\{v_{i}\}$ for $V$, the array for $L$ is $L_{ij}=w_{i}^{\flat}(L(v_{i}))$; the contortion lies in the flat map.
Really, we should have picked a dual basis for $W$; this is because the linear map is one part covariant and one part contravariant.
To capture this and to also model higher order tensors found in many applications (e.g., stress strain relationships in elasticity, differential forms, high order derivatives), we introduce tensor products.

In our discussion here, we first seek to convey that tensor products are a coordinate free formalism that captures a wide variety of mathematical objects with explicit co and contravariance.
See \cite{abstractalgebra3rdedition2004, introductiontosmoothmanifolds2012} for a more detailed discussion on tensor products.

\begin{example}[Tensor Product]
\label{ex:tenprodex}
We consider the space of linear maps $L(V, W)$, the space of bilinear maps $L(V,W;\mathbb{R})$, and the tensor product space $V\otimes W$.
We restrict ourselves to finite dimensional spaces.
We fix arbitrary bases for $V, V^{\star}, W,$ and $W^{\star}$: $v_{i}, v_{i}^{\prime}, w_{i}, w_{i}^{\prime}$. 
We emphasize that the following holds for arbitrary choices of base and that all definitions do not actually depend on bases:
\begin{enumerate}
    \item The space $V\otimes W$ has a basis $v_{i}\otimes w_{i}$. Further, $\otimes$ is a bilinear map from $V$ and $W$ to $V\otimes W$. The space $V\otimes W$ is contravariant in $V$ and $W$.
    \item The space $V^{\star}\otimes W$ is isomorphic to the space $L(V,W)$ via the invertible mapping:
    \begin{equation}
        \sum_{ij} L_{ij} v_{i}^{\prime}\otimes w_{j} \mapsto (v\mapsto \sum_{ij} L_{ij}v_{i}^{\prime}(v)w_{j}).
    \end{equation}
    The space is covariant in $V$ and contravariant in $W$.
    \item The space $V^{\star}\otimes W^{\star}$ is isomorphic to the space $L(V,W;\mathbb{R})$ via the invertible mapping:
     \begin{equation}
        \sum_{ij} L_{ij} v_{i}^{\prime}\otimes w_{j}^{\prime} \mapsto (v, w\mapsto \sum_{ij} L_{ij}v_{i}^{\prime}(v)w_{j}^{\prime}(w).
    \end{equation}
    The space is covariant in both \(V\) and \(W\).
\end{enumerate}
\end{example}
These ideas can naturally be extended to tensor products of more than two spaces.

Now we move to the next part of our discussion: tensor products provide a way to write down many coordinate independent objects and to represent them in sensible bases.
We start by noting a general rule for writing tensor products in bases: we use dual bases for the contravariant components and primal bases for the covariant components.
In the case of $V\otimes W$, this suggests that we should use a basis in the space $V^{\star}\otimes W^{\star}$, but on the other hand the standard dual basis for this space would be in $(V\otimes W)^{\star}$.
\textbf{Tensor products are defined ultimately so that}
\begin{equation}
    \label{eq:lift}
    (V\otimes W)^{\star}\cong V^{\star}\otimes W^{\star} \;,
\end{equation}
\textbf{where the isomorphism is unique (compared to the non-unique sharp map).}
The force of this is that we can write down coordinate independent dual vectors for a tensor product space via writing coordinate independent multi-linear maps (which is much easier to do). 
For example, the trace of a matrix is coordinate independent because it can be written as a coordinate independent vector in $(V^{\star}\otimes V)^{\star} \cong V\otimes V^{\star}$ via $(v,v^{\prime}) \mapsto v^{\prime}(v)$.
Similarly, symmetry is preserved under change of basis because we can write: $(v,w)\mapsto v\otimes w - w\otimes v$.
This is actually using a more general rule that any linear map on a tensor product space corresponds to a multi-linear map in a way that preserves coordinate independence of definitions.
This leads to our final definition, of critical importance to \cref{sec:mathfem}, which returns to the change of basis in the chain rule.

\begin{definition}[$V^{p,q}$ Tensor Space, Adjoint, Pullback]
\label{def:pull5}
    Suppose $V$ is a finite-dimensional vector space.
    Then $V^{p,q} = \left( \bigotimes_{i=1}^{p}V^{\star} \right) \otimes \left( \bigotimes_{j=1}^{q}V \right)$ is a \textbf{tensor space}.
    Suppose we have another finite dimensional space $W$ with an isomorphism $T\in L(V,W)$.
    The \textbf{adjoint} of $T$ is a map $T^{\star}\in L(W^{\star}, V^{\star})$ given by $T^{\star}(\phi) = \phi\circ T $.
    The \textbf{pullback} is a linear map in $L(W^{p,q}, V^{p,q})$, defined by the following rule and extended linearly:
     \begin{equation}
    \begin{split}
        T^{\star}(w_{0}^{\prime}\otimes\cdots\otimes w_{p-1}^{\prime}\otimes w_{0}\otimes\cdots\otimes w_{q-1}) &=\\ T^{\star}(w_{0}^{\prime})\otimes\cdots\otimes T^{\star}(w_{p-1}^{\prime})\otimes T^{-1}(w_{0})\otimes\cdots\otimes T^{-1}(w_{q-1}).
        \end{split} 
    \end{equation}
\end{definition}
One can check that this definition reproduces the contravariant change of basis for vectors, the covariant one for tensors, and the mixed rule for matrices/linear maps. 
The pullback can be extended to operate on tensor fields, functions taking values in tensor spaces, which is the definition we utilize to reason about elements:

\begin{definition}[Pullback and Pushforward]
    Let $N$ and $M$ be surfaces. Let $T\in C^{1}(N, M)$ have an inverse $T^{-1}\in C^{1}(M, N)$.
    Let $A = (\mathcal{F}(N, (\mathcal{T}N)^{p,q}))$ and let $B = (\mathcal{F}(M, (\mathcal{T}M)^{p,q}))$.
    The \textbf{pullback} along $T$ is a mapping in $L(B, A)$ given by 
    \begin{equation}
        (T^{\star}f)(x) = (DT_{x})^{\star}(f(T(x))) \mbox{ for all } f\in B.
    \end{equation}
    
    The \textbf{pushforward} along $T$ is a mapping in $L(A^{\star}, B^{\star})$ given by 
    \begin{equation}
        (T_{\star}\phi)(f) = \phi(T^{\star}f) \mbox{ for all }\phi\in A^{\star}\mbox{ and } f\in B.
    \end{equation}
    The superscript star denotes pullback on fields; its pointwise action uses the tensor pullback defined above, extending the adjoint on covectors. The subscript star denotes the induced pushforward on dual functionals.
    We abbreviate $(T^{-1})^{\star}$ by $T^{-\star}$ and $(T^{-1})_{\star}$ by $T_{-\star}$.

\end{definition}

\subsection{Reference Space and World Space}
\label{sec:rsws}
Like many paradigms for simulation, FEMs utilize reference space and world spaces.
The reference space is where calculations occur in a standard domain and the world space is where actual physical models live.
This distinction will be core to how we reason about finite elements.
For now, we narrow the notions of reference and world space to a simpler class of known objects.

\begin{definition}[Simplex, Entity, Standard Entity]
\label{def:pos1}
    In this paper, we restrict \textbf{mesh entities} to simplices. An abstract \textbf{simplex} is a finite nonempty set of vertices $V$ with faces $P(V)$, where $P(V)$ is the power set of $V$. Ordered by $\subseteq$, it is a partially ordered set, or \textbf{poset}: inclusion is reflexive, antisymmetric, and transitive.
    A simplex of topological dimension $d$ has $d+1$ vertices. Its \textbf{standard entity} labels these vertices by $V=\{0,\dots,d\}$; these are abstract labels, not coordinates on the real line. An extension to other cell types would require a corresponding standard face structure for each type.
\end{definition}

The utility of the above definition is that it is discrete and closer to general definitions where entities are subsets of posets.
Though for simplicity in this work we limit ourselves to affine simplices, large portions can be generalized to curved and more complex polyhedra.
We now need two notions of dimension for entities:
\begin{definition}[Topological Dimension, Geometric Dimension]
    The \textbf{topological dimension} of a simplex $S=(F, V)$ is $|V|-1$. 
    An entity is said to be \textbf{geometric} if the vertices are all members of $\mathbb{R}^{n}$ where $n$ is at least the topological dimension. The number $n$ here is the \textbf{geometric dimension} of an entity.
    We refer to these as $\operatorname{tdim}$ and $\operatorname{gdim}$.
\end{definition}
Our reference spaces will typically be standardized geometric entities and they will relate to world space through affine maps that we call parametrizations or charts (in honor of differential geometric terminology).
\begin{definition}[Geometric Standard Entities]
    For a simplex of topological dimension $d$, the vertices of a geometric standard entity are given by $\{0\}\cup \{e_{1},\dots, e_{d}\}\subset\mathbb{R}^{d}$ where $e_{i}$ are the standard basis vectors.
\end{definition}
We can now define our parameterizations:
\begin{definition}[Chart]
\label{def:chart1}
    A chart is a triple $(S_{0},S_{1}, F)$ where $S_{i}$ for $i=0,1$ are two geometric entities and $F$ is an injective affine map from $\mathbb{R}^{\operatorname{gdim}(S_{0})}$ to $\mathbb{R}^{\operatorname{gdim}(S_{1})}$ and $F$ is a bijection from the vertices of $S_{0}$ to the vertices of $S_{1}$.
\end{definition}
These give us our notion of a reference space and our notion of a world space. 
Much of this work could be extended to allow $F$ to be a polynomial mapping, allowing for general quads and hexes. 

\subsection{The Mathematical Definition of a Finite Element}
\label{sec:mathfem}

Before providing the definition of a FE and illuminating the resulting computational issues, especially those relating to transforming elements, we recast the linear Lagrange element on a triangle.
The two key properties of the linear Lagrange basis functions are:
\begin{enumerate}
    \item The basis functions $f_{0}, f_{1}, f_{2}$ are affine functions on the triangle,  i.e., $f_i\in P(K)$ the \emph{vector space} of affine functions defined on a triangle $K$.
    \item Each basis function corresponds to a triangle vertex, $v_{0}, v_{1}, v_{2}$, so that if $f= \alpha_{0}f_{0} + \alpha_{1}f_{1} + \alpha_{2}f_{2}$, then $f(v_{i}) = \alpha_{i}$.
\end{enumerate}

We can recast this definition by means of a dual basis. 
This may seem circuitous.  Why not just define a primal basis, the functions  $f_{0}, f_{1}, f_{2}$, directly?
Consider our triangle, $K$ and the linear Lagrange basis functions again.  Depending on the coordinates of our triangle, $f_0, f_1, f_2$ may have all different sorts of arbitrary coefficients.  However, consider the following \emph{dual} basis functions:
$\phi_{0}(f) = f(v_{0}),  \phi_{1}(f) = f(v_{1}),  \phi_{2}(f) = f(v_{2})$.
These these dual basis functions just say ``the value at each of the vertices,'' which is a simple, coordinate-independent concept. 
Finding the coefficients of the primal basis functions then reduces to solving $n$ linear systems:
\begin{equation}
\label{eq:solveFEM}
    \phi_{i}(f^{\prime}_{j}) f_{jk} = \delta_{ik}.
\end{equation}
where $f^{\prime}_{0}, f^{\prime}_{1}, f^{\prime}_{2}$ is some arbitrary basis for the function space (e.g., monomials or Legendre polynomials).
Note as a consequence of~\cref{eq:solveFEM}, that $f_{k} = \sum_{j} f^{\prime}_{j} f_{jk}$. 
Some readers may recognize the matrix $\phi_{i}(f^{\prime}_{j})$ as a generalized \textbf{\VA{}} matrix~\cite{approximationtheoryandapproximationpracticeextendededition2019}.

The key step in the above exercise is that we no longer write down the primal basis $f_{0}, f_{1}, f_{2}$ explicitly, but we instead write down the dual basis.
This has a few key advantages: the mathematical theory works mostly on the dual basis, the dual basis functions make the interaction between neighboring triangles easier to reason about, interpolation methods come for free via the dual basis, and the dual basis is easier to write down.
Thus the mathematical definition centers the dual basis:

\begin{definition}[Finite Element]
\label{def:finiteElement}
A \textbf{finite element} $(K, V, P, \Sigma)$ consists of
\begin{enumerate}
    \item $K\subseteq\mathbb{R}^n$, the domain of the element (e.g. a triangle, quad, hex, tetrahedron),
    \item $V$ a vector space of values (e.g., real numbers $\mathbb{R}$ or vectors $\mathbb{R}^{3}$)
    \item $P(K)$ is a finite dimensional vector space of functions from $K$ to $V$ (usually some class of polynomial functions)
    \item $\Sigma(K)$ is a dual basis for $P(K)$ (e.g., point evaluations)
\end{enumerate}
\end{definition}

\begin{figure}

  \centering
  \includegraphics[width=\linewidth]{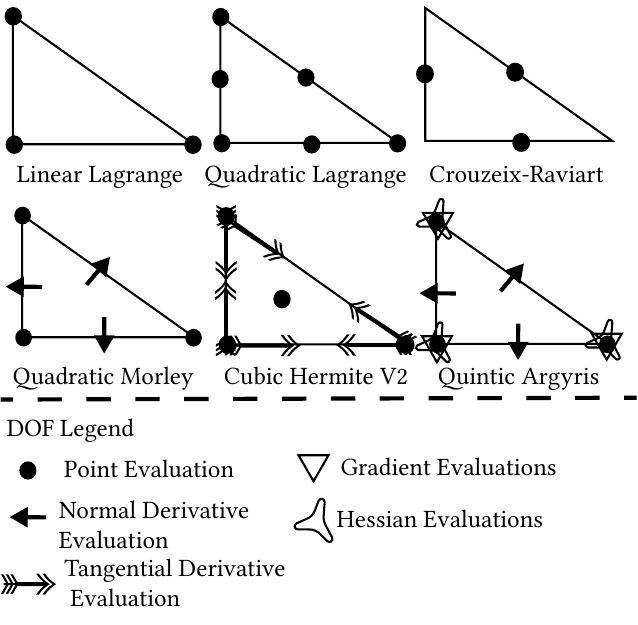}
  \caption[Examples of Finite Element DOFs]{Examples of finite elements depicted graphically via their dual basis functions. This figure was inspired by graphical depictions going back to the Ciarlet definition of the finite element, but especially those in~\cite{automatedsolutionofdifferentialequationsbythefiniteelementmethod2016}.}
  \label{figure:elemExamples}
\end{figure}

This definition dates back to Ciarlet~\cite{finiteelementmethodforellipticproblems2002} and variations of it have been reproduced in many FEM textbooks~\cite{themathematicaltheoryoffiniteelementmethods2008,finiteelementstheoryfastsolversandapplicationsinsolidmechanics2007,mathematicalfoundationsoffiniteelementsanditerativesolvers2022,automatedsolutionofdifferentialequationsbythefiniteelementmethod2016}.
Typically a diagram is used to depict a given FE, representing various dual basis functions via glyphs (such as dots and circles) as we do in~\cref{figure:elemExamples}.

These diagrams raise a critical issue in the definition: we have instantiated this FE on a particular set $K$, but our definition should be coordinate free and independent of $K$.
Moreover, as practitioners will have anticipated, we want to instantiate a FE on arbitrary $K$.
And if we simply use the recipe of using~\cref{eq:solveFEM} to find basis functions for many different $K$, we suffer from cubic, $O(|\Sigma|^{3})$, scaling in solutions to~\cref{eq:solveFEM}, not to mention the sub-optimal performance qualities of matrix inverses.
To avoid this issue, we observe that a single formula, dependent on $K$, might capture many different related FEs on different sets $K$.
To see how this might work, we return to the linear Lagrange element on a triangle.

Throughout the transformation discussion, $T$ maps the reference domain to the physical domain. Let $K_{0}$ be the unit reference triangle and $K_{1}$ be some arbitrary non-degenerate physical triangle, related to $K_{0}$ via an injective affine map $T\colon K_{0}\to K_{1}$.
Suppose we are given two finite elements, $(K_{i}, \mathbb{R}, P_{i}, \Sigma_{i})$ for $i=0,1$
where $P_{i}$ are the affine functions on $K_{i}$
 and $\Sigma_{i}$ consists of evaluations at corresponding vertices.
Let $p_{0}, p_{1}, p_{2}$ be the primal basis corresponding to $\Sigma(K_{0})$. 
With the formula for Lagrange interpolating polynomials on a triangle, we can verify that if $T\colon K_{0}\to K_{1}$ is an injective affine map and $q_{0}, q_{1}, q_{2}$ is the primal basis corresponding to $\Sigma(K_{1})$, then $q_{i} = p_{i}\circ T^{-1}$.
We use $\mathsf{P}_{ij}(T)$ for the coefficients that combine mapped reference basis functions into physical basis function $i$; its rows index physical basis functions and its columns index reference basis functions. In particular, if we consider the pullback map on scalar functions $T^{-\star}\colon P(K_{0})\to P(K_{1})$, given by $f\mapsto f\circ T^{-1}$, we can see that it is linear and its matrix representation in the bases $p_{0}, p_{1}, p_{2}$ and $q_{0}, q_{1}, q_{2}$ is the identity matrix, i.e.,
\begin{equation}
\label{eq:lrie}
    q_{i}=\sum_j\mathsf{P}_{ij}(T)\,T^{-\star}(p_j),
    \qquad \mathsf{P}_{ij}(T)=\delta_{ij}.
\end{equation}
The consequence of~\cref{eq:lrie} is that any computation on the primal basis corresponding to an arbitrary $K_{1}$ can be done in terms of $p_{i}$ and $T^{-1}$, avoiding the need to actually find $q_{i}$ for each cell.
This possibility is referred to as \textbf{affine equivalence}~\cite{finiteelementmethodforellipticproblems2002}:

\begin{definition}[Affine Equivalence for Scalar-Valued Elements]
\label{def:affine equivalence}
Two finite elements $(K_i,\mathbb{R},P_i,\Sigma_i)$ for $i=0,1$ are affine equivalent under an injective affine map $T\colon K_0\to K_1$ if $T^{-\star}\colon p\mapsto p\circ T^{-1}$ maps each reference nodal basis function to its corresponding physical nodal basis function. In our convention, $\mathsf{P}(T)=I$.
\end{definition}

However, many FEs are not affine equivalent, and so other notions of equivalence have been developed, including affine-interpolation equivalence for the Hermite element~\cite{finiteelementmethodforellipticproblems2002} and compatible nodal completion equivalent for the Morley and Argyris elements~\cite{ageneralapproachtotransformingfiniteelements2018}.
We do not deeply investigate these notions here, but the key idea is that the basis-transformation matrix $\mathsf{P}(T)$ in~\cref{eq:lrie} need not be the identity matrix but now takes some other forms, possibly non-linearly dependent on the map $T$ and/or $T^{-1}$. We retain this symbol and reference-to-physical direction throughout; the corresponding pullback of physical fields is $T^{\star}$, while $T^{-\star}$ maps reference fields to the physical domain.
In particular, for the Morley element, the basis-transformation matrix was computed by ~\cite{ageneralapproachtotransformingfiniteelements2018}; in our convention it is as follows:
\begin{example}[Morley Basis-Transformation Matrix]
\label{ex:mpm}
Let $K$ be the unit reference triangle, $K^{\prime}$ a non-degenerate physical triangle, and $T\colon K\to K^{\prime}$ affine with $A=DT$. Number corresponding vertices $1,2,3$ and let edge $i$ be opposite vertex $i$. Orient each unit tangent from the lower-numbered endpoint to the higher-numbered endpoint, and obtain its unit normal by a clockwise quarter turn. Write these vectors as $t_i,n_i$ on $K$ and $t_i^{\prime},n_i^{\prime}$ on $K^{\prime}$, and let $l_i$ be the physical edge length. Define
    \begin{equation*}
        G_i=\begin{bmatrix}n_i^T\\t_i^T\end{bmatrix},\quad
        G_i^{\prime}=\begin{bmatrix}(n_i^{\prime})^T\\(t_i^{\prime})^T\end{bmatrix},\quad
        B^i=G_i A^T(G_i^{\prime})^T.
    \end{equation*}
    Order the six DOFs as vertex values followed by normal derivatives at edge midpoints, on both triangles. Then
    \begin{equation}
    \label{eq:morc}
        \mathsf{P}(T) = \begin{bmatrix}
        1 & 0 & 0 & 0 & 0 & 0\\
         0 & 1 & 0 & 0 & 0 & 0\\
          0 & 0 & 1 & 0 & 0 & 0\\
           0 & -B_{12}^{1}/l_{1} & B_{12}^{1}/l_{1} & B_{11}^{1} & 0 & 0\\
            -B_{12}^{2}/l_{2} & 0 & B_{12}^{2}/l_{2} & 0 & B_{11}^{2} & 0\\
            -B_{12}^{3}/l_{3} & B_{12}^{3}/l_{3} & 0 & 0 & 0 & B_{11}^{3}
        \end{bmatrix}^{T}.
    \end{equation}
    Kirby uses the opposite geometric map $F=T^{-1}$, with $J=DF=A^{-1}$. Thus our $B^i$ is Kirby's $(G_i^{\prime}J^TG_i^T)^{-1}$, with normals before tangents. The matrix inside the brackets is his nodal matrix $V$; the transpose is essential because our $\mathsf{P}$ is his basis matrix $M=V^T$. With reference basis $p_j$ and physical basis $q_i$, we have $q_i=\sum_j\mathsf{P}_{ij}(T)(p_j\circ T^{-1})$.
    This matrix $\mathsf{P}(T)$ from~\cref{eq:morc} again lets us compute with the Morley element basis functions on an arbitrary triangle by only computing the matrix and the Morley element basis functions on a single reference triangle.
\end{example}

At present, these matrices are computed by hand, as in ~\cite{ageneralapproachtotransformingfiniteelements2018,codegenerationforgenerallymappedfiniteelements2019,transformationsforpiolamappedelements2023,aconstructionofedgebsplinefunctionsforac1polynomialsplineontwotrianglesanditsapplicationtoargyristypesplines2021,areissnermindlinplateformulationusingsymmetrichuzhangelementsviapolytopaltransformations2023}.
One of the core aspects of our system is that we will compute these automatically from a software specification of~\cref{def:finiteElement}.
\textbf{Naturally, some of these formulas are available in papers, but there are several other virtues to automation}.
\textbf{First}, for an unknown element, a formula is not already available.
\textbf{Second}, as we will describe in the next section, the formula is not enough. 
The formula must be integrated into an FE code with global function spaces, which is a finicky task that requires awareness of the various conventions and methods a given code employs.
\textbf{Third}, the first two reasons compound: exploring new elements requires changing other aspects of the system in subtle ways, and changing the system might require one to revisit and change existing formulas.
To facilitate rapid exploration of FEMs, we need to automate the discovery of the relevant formula in the context of an expressive system for other aspects of FEs.

\textbf{With this motivation and background in hand, we can now formally state the core problem that we will solve in ~\cref{sec:SRFE}}.
We are interested in finding formulas for the primal basis functions of a class of FEs $(K^{\prime}, V^{\prime}, P^{\prime}, \Sigma^{\prime})$ so that the primal basis functions depend (linearly) on a fixed FE $(K, V, P, \Sigma)$ and (non-linearly though ideally at cost at most $O(|\Sigma|^{2})$) on a transformation $T\colon K\to K^{\prime}$.
We can now formally define the object that we will compute in~\cref{sec:SRFE}.
\begin{definition}[Transformation of a Finite Element]
\label{sec:probmain}
    Fix a finite element $(K, V, P, \Sigma)$.
    Suppose we have a collection of sets $\{K^{\prime}\}$ that are diffeomorphic to $K$ via a $C^{1}$ mapping $T_{K^{\prime}}\colon K\to K^{\prime}$.
    Suppose further that we have a collection of finite elements $\{(K^{\prime}, V^{\prime}, P^{\prime}, \Sigma^{\prime})\}$ indexed by $K^{\prime}$.
    Let $\{f_{i}\}$ be the primal basis for $(K, V, P, \Sigma)$.
    Let $\{f_{i}^{{K^{\prime}}}\}$ be the primal basis for $(K^{\prime}, V^{\prime}, P^{\prime}, \Sigma^{\prime})$.
    For each $T=T_{K^{\prime}}$, we seek a matrix $\mathsf{P}(T)\in\mathbb{R}^{\dim P^{\prime}\times\dim P}$, with physical basis functions indexing rows and reference basis functions indexing columns, such that
    \begin{equation}
    \label{eq:formelempre}
        T^{\star}_{K^{\prime}} f_{i}^{K^{\prime}} = \sum_{j}\mathsf{P}_{ij}(T)f_{j}.
    \end{equation}
    This is the same basis-transformation convention as~\cref{eq:lrie}. In the usual convention of multiplying column vectors of coefficients, the matrix of $T^{\star}\colon P^{\prime}\to P$ is $\mathsf{P}(T)^T$. If the pullback is an isomorphism, the coefficient matrix of $T^{-\star}\colon P\to P^{\prime}$ is $\mathsf{P}(T)^{-T}$.
    Further, this means the primal basis for the target element can be written as:
    \begin{equation}
    \label{eq:elemuse}
        f_{i}^{K^{\prime}}(T(x)) = \sum_{j}\mathsf{P}_{ij}(T)\bigl((DT_x)^{-1}\bigr)^{\star}f_{j}(x).
    \end{equation}
    Here $(DT_x)^{-1}=D(T^{-1})_{T(x)}$ acts on tensor values; for scalar fields this factor is the identity.
\end{definition}

Ultimately our system will compute this from a software implementation of the mathematical definition of a FE. Our system uses \cref{eq:elemuse} as the formula for primal basis functions on elements.
\textbf{Based on our discussion of pullbacks, we can now anticipate some core challenges.}
\textbf{First}, to even state the problem in software, we must specify elements simultaneously across all possible domains.
This is a key language design problem.
\textbf{Second}, the above formulation assumes that  $(T^{\star}_{K^{\prime}}) f_{i}^{K^{\prime}}\in P$, which in general does not need to be the case.
Thus, mapping $\mathsf{P}$ does not obviously exist in many cases.
\textbf{Third}, even if it does exist, $\mathsf{P}$ might be a complex symbolic object, especially because pullbacks depend polynomially on derivatives and inverse matrices turn polynomials into large complex rational functions.
We return to this in~\cref{sec:SRFE} and for now move on to the other big challenge of using FEs: the global FE space and operations on it.

\subsection{The Mathematical Definition of Global Finite Element Spaces and Associated Operations}
\label{sec:globfem}

\begin{wrapfigure}[7]{r}{0.6in}
      \vspace{-0.17in}
      \hspace{-0.24in}
      \includegraphics[width=0.8in]{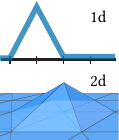}
\end{wrapfigure}
An FE defines a local space of functions on individual mesh elements.
Now we seek to construct a global FE space, a space of functions defined on the whole (global) mesh. Like the individual FEs, this construction will consist of a domain, a space of functions, and a dual basis. 
Returning to the linear Lagrange example from~\cref{sec:mathfem}, the primal bases of the corresponding global space are the well-known linear hat functions, one for each vertex in the mesh.

\begin{wrapfigure}[8]{r}{0.6in}
      \vspace{-0.17in}
      \hspace{-0.24in}
      \includegraphics[width=0.8in]{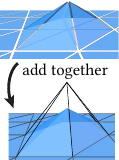}
\end{wrapfigure}
A hat function associated to the vertex $v$ matches a primal basis function in each local linear Lagrange FE space on a triangle incident to the vertex.
However, the hat function is not simply the sum of these local functions, as this would result in over-counting values on the shared edges between triangles.
The problem of how to handle boundaries between elements also arises when we seek to describe spaces of discontinuous functions, whose values at such boundaries are ill-defined. For instance, this occurs when modeling fractures, creases, or interfaces between different materials.

We need a few auxiliary notations from topology in order to talk coherently about the boundary vs. interior of our elements.  We write $\overline{S}$ to mean the closure of the set $S$, and $S^\circ$ to mean the interior of the set $S$. We say that a set $S$ is \textbf{regular closed} if $\overline{S^{\circ}} = S$.

\begin{definition}[Restriction of Functions]
\label{def:rof}
Suppose $K \subseteq \Omega$ is a regular closed set and $V$ is a vector space. 
Given $f\in \mathcal{F}(\Omega, V)$ such that $f$ is continuous and bounded on $K^{\circ}$, then $\rho_{K}(f)$, the \textbf{restriction} of $f$ to $K$, is the unique continuous extension of $f|_{K^{\circ}}$ to $K$.
\end{definition}

\begin{wrapfigure}[11]{r}{1in}
      \vspace{-0.17in}
      \hspace{-0.24in}
      \includegraphics[width=1.2in]{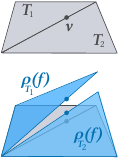}
\end{wrapfigure}
A key application of this concept is the construction of discontinuous functions: consider two triangles $T_{0}, T_{1} \subseteq \Omega$ sharing a vertex $v$, as well as a function $f$ defined on $\Omega$.  If $f$ is discontinuous at $v$, then we can evaluate $\rho_{T_{1}}(f)$ and $\rho_{T_{2}}(f)$ at $v$ to get two distinct values.  This allows us to have two different DOFs at $v$ and thus to construct \emph{global} FE spaces that include discontinuous functions. 
We can transpose the preceding construction to extend local dual functions back to operating on the global space.

\begin{definition}[Extension of Dual Functions]
Suppose $K \subseteq \Omega$ is a regular closed set and $V$ is a vector space. 
Given a $\phi\in \mathcal{C}^{\star}(K,V)$, the \textbf{extension} to $\Omega$ is a dual function $E_{K}(\phi)\in \{f\in \mathcal{C}(K^{\circ}, V)\colon f\in\mathcal{F}(\Omega, V)
\}^{\star}$ defined via $E_{K}(\phi)(f) = \phi(\rho_{K}(f))$ for all $f$. More concisely, $E_{K}(\phi) = \phi\circ\rho_{K}$. Hence $E_{K}$ is dual (a.k.a. adjoint) to $\rho_{K}$.
\end{definition}

Given some point evaluation of a vertex on a mesh, $\phi_{v}(f) = f(v)$, then for a triangle with $v\in K$, $E_{K}(\phi_{v})$ is now not just a point evaluation, but a more complex object: it is the limit of all point evaluations as they approach the vertex. 
Thus, as emphasized above the extension of a dual function by different elements yields different dual functions on the global domain.  These are the dual functions to which DOFs are most naively associated.
Every graphics programmer is familiar with this phenomenon as it applies to texture coordinates, or mesh normals that vary according to triangle, not merely according to vertex.

This machinery is sufficient to define what we call a broken global FE space, which models a fully discontinuous space.
\begin{definition}[Broken Global Finite Element Space]
\label{def:bgfes}
Let $\Omega\subset\mathbb{R}^{n}$ be a domain meshed by a finite collection of elements $\{T_i\}$, each a regular closed subset of $\Omega$ and forming a partition of $\Omega$ in the sense that $\Omega = \cup T_{i}$ and $T_{i}^{\circ}\cap T_{j}^{\circ} = \emptyset{}$ if $i\neq j$.  Furthermore, let $(T_{i}, V, P_{i}, \Sigma_{i})$ be the FEs associated to each $T_i$.
Then define
\begin{enumerate}
    \item  a global space of piecewise functions 
    \begin{equation*}
        P = \left\{p\colon \Omega\to V\mid \forall i.\rho_{T_{i}}(p)\in P_{i} \right\}
    \end{equation*}
    \item a global set of dual function $\Sigma = \cup_{i} \{E_{T_{i}}(\phi) \colon \phi\in\Sigma_{i}\}$ so that $\Sigma\subset P^{\star}$.
\end{enumerate}
We call $(\Omega, V, P, \Sigma)$  the \textbf{broken global finite element space} associated to a collection of finite elements.
\end{definition}

Note that the union of local dual basis functions, (2), in the above construction is disjoint.  We have exactly one global dual function for each local dual function on each element.  Further note that the construction of $P$ given above is intuitively accurate, but requires tricky functional analysis to specify precisely\footnote{Let us consider an example of the issue that occurs.  Consider a mesh of the interval $[0,2]$ consisting of two sub-intervals $[0,1]$ and $[1,2]$.  Even if  the local function spaces consist of affine functions, $P$ as constructed above does not specify the value of $f(1)$ for $f\in P$; it can be arbitrarily modified.  This problem occurs in much of functional analysis and is resolved by drawing functions from Lebesgue spaces, such as $L_1$.  These spaces technically consist of equivalence classes of functions, rather than individual functions.  This is almost entirely a distraction from any of our development here, so we choose to not needlessly complicate our presentation further.}.

The broken space does not yet allow us to produce our hat functions. Intuitively, we still need to ensure that all the DOFs corresponding to point evaluation at the same vertex evaluate to the same value for all functions in $P$.
To do this, we must group or equate DOFs to construct the global FE space.  We accomplish this via an equivalence relationship on the global broken degrees of freedom.

\begin{definition}[Global Finite Element Space]
\label{def:gfes}
Suppose we have a collection of finite elements $(T_{i}, V, P_{i}, \Sigma_{i})$ that form a broken global finite element space $(\Omega, V, P, \Sigma)$.
Further suppose $\equiv$ is an equivalence relation on $\Sigma$ such that basis functions from the same element are not made equivalent: in other words, $\phi_{1}, \phi_{2}\in \Sigma_{i}$ for some $i$ and $\phi_{1}\neq\phi_{2}$, then $E_{\Omega}\phi_{1}\not\equiv E_{\Omega}\phi_{2}$.
Then we can define $P_{\equiv}\subset P$  via
\begin{equation*}
     P_{\equiv} = \{p\in P\colon \forall \phi,\phi^{\prime}\in\Sigma\colon \phi\equiv \phi^{\prime} \implies \phi(p) =\phi^{\prime}(p)  \},
\end{equation*}
and we can define 
\begin{equation*}
    \Sigma_{\equiv} = \Sigma/\equiv.
\end{equation*}
The \textbf{global finite element space} is then $(\Omega, V, P_{\equiv}, \Sigma_{\equiv})$.%

\end{definition}

We make two observations.
First, a good choice of equivalence relationship can sometimes be generated via the space of continuous functions, $C(\Omega)$:
\begin{equation*}
    \phi_{1} \equiv \phi_{2} \leftrightarrow \forall f\in \mathcal{C}(\Omega, V)\colon \phi_{1}(f) = \phi_{2}(f).
\end{equation*}
This requires that all dual basis functions are well defined on functions in $\mathcal{C}(\Omega, V)$.
In the FE literature, equivalence relations are used implicitly by way of a choice of function space.
Second, although $\Sigma_{\equiv}$ is technically a set of equivalence classes, the construction of $P_{\equiv}$ as a quotient allows us to use any set of representatives of the classes $\Sigma_{\equiv}$ as a dual basis for $P_{\equiv}$.

As an initial application of the above machinery of global FE spaces, we can define a notion of interpolation.
Traditionally, interpolation uses a set of points $\{x_{i}\}$ and a corresponding set of measurements $\{y_{i}\}$.
Interpolation seeks to find a function that satisfies $f(x_{i}) = y_{i}$.
For FEs, we can replace the pairs $(x_{i}, y_{i})$ with dual basis functions, $\phi_{i}$, and values for these dual basis functions, $\alpha_{i}$.
However, we are not often given a set of values $\alpha_{i}$ as actual numbers. 
Instead, we often are given some other function $h$, not necessarily in the FE space, that we use to compute the $\alpha_{i}$, which then defines a function $f$ that approximates $h$.
\begin{definition}[Well Defined, Finite Element Interpolation]
\label{def:interp}
    Let $E = (\Omega, V, P_{\equiv}, \Sigma_{\equiv})$ be a global finite element space.
    Let $h\in \mathcal{F}(\Omega, V)$ be such that  for all $\phi\in \Sigma_{\equiv}$, $\phi(h)$ is well defined, which is to say that for all $\psi,\psi^{\prime}\in \phi$, $\psi(h) = \psi^{\prime}(h)$.
    Then the interpolation of $h$ into the global finite element space is a function $I_{E}h\in P_{\equiv}$ given by
    \begin{equation}
        I_{E}h = \sum_{\phi_{i}\in \Sigma_{\equiv}} \phi_{i}(h)f_{i}.
    \end{equation}
\end{definition}
FE interpolation lets one approximate a wide variety of functions with FE functions.
Practically, interpolation allows one to get data into the system in a consistent way.
In particular, if some FEM converges at an order $k$ with some given FE, then interpolating incoming data will often not disrupt this order of convergence.\footnote{In some systems, e.g., Firedrake/FEniCS, inappropriate representations of DOFs have led to interpolation changing the order of convergence. Here is a recent example of this problem: \url{https://github.com/gridap/Gridap.jl/pull/1048}.} 

FE interpolation also reveals another \textbf{fundamental challenge} of using complex FE global spaces (beyond the correct software implementation of the above definitions and their usage to build global operators/sparse matrices):
\textbf{FE interpolation requires that the equivalence classes of DOFs are well defined but sometimes they are actually multi-valued.}
As a simple example, interpolating from a discontinuous linear Lagrange space to a continuous one does not make sense in the above definition. 
In DEC, this problem is often encountered when converting between edge, vertex, and face fields as one can't interpolate between these and must use global Hodge star like objects.
\textbf{For our system to support more general FEMs, we must systematically handle how quantities derived from finite element spaces are sometimes multi-valued and how this renders some computations ill-defined.}
As it turns out, our system will systematically provide warnings when such problems emerge and this mechanism will be at the heart of how our system supports all the different ways one might want to use functions defined in global finite element spaces (e.g., operator assembly, residuals, interpolation).

\section{Related Work}
\label{sec:related}
\subsection{Finite Elements in Graphics}
\label{sec:feing}

FEMs are a standard part of the toolkit of computer graphics, especially piece-wise linear FEs.
The piecewise linear discretization of the Laplacian operator appears ubiquitously in computer graphics in the form of the cotangent Laplacian~\cite{Pinkall1993ComputingDM,discretedifferentialgeometryoperatorsfortriangulated2manifolds2003}.
Moreover, piece-wise linear Lagrange FEs are used to simulate fluids, simulate solids, simulate fluid structure interaction, mesh objects, render objects, smooth meshes, perform 4D variations on these themes, and more~\cite{afiniteelementmethodforanimatinglargeviscoplasticflow2007,volumeconservingfiniteelementsimulationsofdeformablemodels2007,simulatingliquidsandsolidliquidinteractionswithlagrangianmeshes2013,stableneohookeanfleshsimulation2018,anisotropicpolygonalremeshing2003,aradiativetransferframeworkforrenderingmaterialswithanisotropicstructure2010,computationalpwillmoreflowwithconformalpenalty2020,screenedpoissonsurfacereconstruction2013, shi2025hyper, sin2013vega}.

In several subfields of computer graphics, alternatives to piece-wise linear Lagrange FEs are under consideration due to various problem specific needs.
The primary alternatives under consideration are higher-order Lagrange elements or similar constructions for squares and hexahedra (such as B-splines)~\cite{alargescalecomparisonoftetrahedralandhexahedralelementsforsolvingellipticpdeswiththefiniteelementmethod2022,polysplinefiniteelementmethod2019,higherorderfiniteelementsforembeddedsimulation2020, mezger2008interactive}.
Most notably, higher-order Lagrange elements (typically not beyond cubic) have been employed to deal with locking phenomena in elasticity problems~\cite{higherorderfiniteelementsforembeddedsimulation2020,bargteil2014animation, fernandez2025symx, weber2015deformation}.
In elasticity in particular, more exotic elements have been used specifically to deal with many phenomena~\cite{lockingprooftetrahedra2021,embeddedthinshellsforwrinklesimulation2013, bargteil2014animation}.
In fluids in particular, B-Spline constructions have been employed as FEs, mainly to allow for smoother bases~\cite{anefficientbsplinelagrangianeulerianmethodforcompressibleflowshockwavesandfracturingsolids2022} though \cite{coadjointorbit} also employs a scheme similar to some FEEC elements.
Moreover, spline  bases have a long history in fluids/continuum simulation as a discretization in FEM-like components of particle/grid type simulations~\cite{Fu2017,jiang2016material, hu2018moving}.
A few works have directly or potentially indirectly\footnote{Indirect usage is possible for discontinuous elements because they often show up through tricks to break connections between elements or to model discontinuous geometries (e.g., periodic geometries involve a discontinuous paramterization). However, it is difficult to tease out the exact relationship.} utilized something like lower order discontinuous Lagrange elements~\cite{kaufmann2009flexible,chen2018numerical, animatingdevelopablesurfacesusingnonconformingelements2008, edwards2014detailed}.
Finally, the lower-order non-linear Lagrange elements have been employed in a variety of other problems, such as meshing and surface deformation~\cite{spectralquadrangulationwithfeaturecurvealignmentandelementsizecontrol2015,secondorderfiniteelementsfordeformablesurfaces2023}.
Beyond higher order Lagrange, a few have used the FEEC elements (mainly the Whitney elements via subdivision representations) and their relatives (Crouzeix-Raviart), especially when dealing with the structure of harmonic functions~\cite{edgesubdivisionschemesandtheconstructionofsmoothvectorfields2006,discretedifferentialformsforcomputationalmodeling2006,subdivisiondirectionalfields2020,mixedfiniteelementsforvariationalsurfacemodeling2010,asmoothnessenergywithoutboundarydistortionforcurvedsurfaces2020,animatingdevelopablesurfacesusingnonconformingelements2008,boundaryalignedsmooth3dcrossframefield2011,modelingnsymmetryvectorfieldsusinghigherorderenergies2018}.
In the latter case, the elements are always of the lowest orders.
Much more rarely, authors have used elements that we cannot find implemented elsewhere, such as the Fraeijs de Veubeke's element
~\cite{avectorialsolverforfreeformvectorgradients2012}.
The same work also uses Morley's element but does not provide any related details on its useage.
Lastly, we should also note that a many works have considered basis functions built from fundamental solutions to PDEs, complex geometrical features, complex higher order sub-cell/multi-scale constructions, similar to various extended or enriched finite element schemes~\cite{martin2008polyhedral, efemex,koschier2017robust, chen2019material,hafner2019x,kharevych2009numerical}; we consider these and many other approaches as potential future work as they are often built on top of FEM pipelines.

\subsection{Automation for the Finite Element}
\label{sec:affe}
In this section, we survey the related work with respect to the ability of other FE software to use~\cref{def:finiteElement} of an FE automatically and parametrically.
Outside of providing basis function values, several other aspects of using FEs have been partially automated in a few libraries.
The systems that support non-affine equivalent elements support using their DOFs to perform interpolation or to enforce BCs to varying degrees, though in all cases the support is partial~\cite{codegenerationforgenerallymappedfiniteelements2019,fullyandsemiautomatedshapedifferentiationinngsolve2020,c1finiteelementsfordune,scikitfemapythonpackageforfiniteelementassembly2020,getfemautomatedfemodelingofmultiphysicsproblemsbasedonagenericweakformlanguage2020}.
Similarly, some of these systems make assumptions on the mesh to make certain types of FEs work~\cite{aparalleledgeorientationalgorithmforquadrilateralmeshes2016} while others, mainly FEniCS via the Basix library for basis functions, allow users to specify how basis functions are impacted by generators of symmetry groups~\cite{constructionofarbitraryorderfiniteelementdegreeoffreedommapsonpolygonalandpolyhedralcellmeshes2022}.
Lastly, \citet{marsden2023redefining} proposed extending Ciarlet's definition with information about continuity across shared mesh entities and structured DOF representations.
FUSE~\cite{marsden2026fuse} provides a unified, serialisable element representation, with DOFs expressed through kernels, integral pairings over entities, and group actions.
Its generation of DOF families through group actions serves several roles, including one closely related to our topological enumeration (\cref{sec:TEP}): both expand compact DOF specifications across entities.
Our enumeration does this by matching pointwise expressions with topologically compatible charts and fields, whereas FUSE uses group actions to express cell symmetries and their action on DOFs.
Among the systems discussed here, FUSE is the closest related work to our element-specification language.

For an element whose physical-cell transformation is not already supplied by a library, implementing the element has traditionally required the developer to derive and encode that transformation.
For elements with known transformation formulas, this may mean implementing an existing result; for others, it requires a new derivation establishing some version of~\cref{def:affine equivalence} or~\cref{eq:lrie}.

For example, Symfem constructs finite element bases symbolically~\cite{symfemasymbolicfiniteelementdefinitionlibrary2021}. \citet{scroggs2026doftransformations}, part of the FEniCSx effort\cite{baratta2023dolfinx} and related to Symfem,  automates DOF transformations when the element can be written on a general domain and when the transformation can be decomposed block-wise based on the topological sub-entities of the reference element.
Their algorithm, like ours, directly attacks \cref{eq:lrie}, but under an assumption of a block-wise sparsity pattern found in elements that are affine equivalent under appropriate pushforward, mainly the FEEC families.
Their transformations may produce a nicer formula for some of these while we believe ours would probably work on a larger class of elements.
Further, at the level of the specification language, Symfem currently doesn't build all simplicial elements on arbitrary affine geometries, so the infrastructure for the problem is not yet fully present there.

After promulgation of aspects of this work at PDESoft 2024~\cite{collin2024pdesoft} and documentation of its mathematical and algorithmic foundations in our 2025 thesis~\cite{collin2025automated}, we presented further aspects of this work at PDESoft 2026~\cite{collin2026pdesoft}. To our knowledge, related generic transformation work in FIAT/FInAT began in July 2026~\cite{brubeck2026zanyinitial}.
Section~6.2 of the thesis develops the two-stage \VA{} construction and recursive block inversion; its Appendix~A develops the dual-basis-cover construction and support analysis. \Cref{sec:AAFET} of this paper gives the cover factorization and its algorithmic consequences.
The FIAT/FInAT implementation~\cite{fiat2026generictransformations} uses constructions closely related to those developed in our thesis and presented here: weighted derivative representations of DOFs (\cref{thm:main,sec:AAFET}); generalized \VA{} matrices (\cref{sec:SFIVFG,eq:transformVander}); support-based block algorithms (\cref{sec:SFSI,sec:SFVM,sec:AAFET}); and simplification using pullback structure (\cref{sec:TEBA,sec:TEFG,sec:SCP}; see also our related Grassmann work~\cite{collin2026grassmann}). Its public development notebook also discusses the construction and support-based block elimination~\cite{fiat2026zanynotebook}.
FIAT/FInAT integrates these methods into its established element infrastructure for affine simplices and supported physical-node conventions~\cite{fiat2026generictransformations}.

Before generic transformation automation, libraries supporting non-affine equivalent elements (\cref{def:affine equivalence}) allowed developers to supply geometry-dependent matrix representations of the pullback.
Examples include Firedrake, FEniCS, NGSolve, Dune, scikit-fem, and GetFEM~\cite{codegenerationforgenerallymappedfiniteelements2019,fullyandsemiautomatedshapedifferentiationinngsolve2020,c1finiteelementsfordune,scikitfemapythonpackageforfiniteelementassembly2020,getfemautomatedfemodelingofmultiphysicsproblemsbasedonagenericweakformlanguage2020,schoberl2014cpp}.
These libraries offer a wide variety of FEs, including Firedrake's support for generally mapped elements and macroelements~\cite{fiatenablingclassicalandmodernmacroelements2025,fiatimprovingperformanceandaccuracyforhighorderfiniteelements2024,firedrakeusermanual2023,multigridsolversforthederhamcomplexwithoptimalcomplexityinpolynomialdegree2024,codegenerationforgenerallymappedfiniteelements2019}.
FIAT, a foundation for this line of work and a library used in Firedrake and originally FEniCS, automatically generates reference basis functions for a wide variety of DOFs, polynomial spaces, and domains. Earlier FInAT work supplies element-specific transformations and other symbolic data needed for generally mapped elements~\cite{firedrakeautomatingthefiniteelementmethodbycomposingabstractions2016,algorithm839fiatanewparadigmforcomputingfiniteelementbasisfunctions2004,finatfinatasmarterlibraryoffiniteelements,ageneralapproachtotransformingfiniteelements2018,codegenerationforgenerallymappedfiniteelements2019}.

Other FEM systems provide different balances of built-in element families, weak-form interfaces, and specialized operators~\cite{baratta2023dolfinx,  mfemamodularfiniteelementmethodslibrary2021,thedealiifiniteelementlibrarydesignfeaturesandinsights2021,dunefemageneralpurposediscretizationtoolboxforparallelandadaptivescientificcomputing2012,anobjectorientedfiniteelementframeworkformultiphysicsphasefieldsimulations2012,openfoamaclibraryforcomplexphysicssimulations2007,febiofiniteelementsforbiomechanics2012,mpsalsaafiniteelementcomputerprogramforreactingflowproblemspart1theoreticaldevelopment1996,newdevelopmentinfreefem2012,nektarenhancingthecapabilityandapplicationofhighfidelityspectralelementmethods2020,nekrsagpuacceleratedspectralelementnavierstokessolver2022}.
Supporting a predefined element family or parametrizing a variational form does not by itself make transformation generation, interpolation, and boundary-condition routines parametric over a new mathematical element specification.

Our focus is to express DOFs in IPTEM and use the same compiler for element transformations and variational forms, including the generation and structured inversion of geometry-dependent \VA{} matrices, as well as the parametric use of elements in interpolation, boundary-condition enforcement, and assembly. These routines operate on the resulting element representation, so they adapt to the element specification.

\subsection{Automation for Other Aspects of the Finite Element Method}
\label{sec:afofem}
Outside of the FE itself, many other aspects of the FEM have been somewhat automated, often via DSLs.
Discussions of meshing software are not germane to our purposes, but the specification of weak formulations is heavily related to our enterprise. 

Essentially three approaches have emerged to allow users of FEM software to specify their problems. 
First, systems such as PolyFEM, Vega, or nekRS offer parameterized PDEs that users can configure with expressions for fixed tensors or BCs, meshes, or other parameters~\cite{polysplinefiniteelementmethod2019, sin2013vega, nekrsagpuacceleratedspectralelementnavierstokessolver2022}.
Second, systems such as MFEM, Trixi.jl, and Nektar$++$~ offer libraries of operators that can be combined to specify the variational formulation of a PDE~\cite{mfemamodularfiniteelementmethodslibrary2021,Ranocha2023,nektarenhancingthecapabilityandapplicationofhighfidelityspectralelementmethods2020}.
These systems sometimes also expose internal constructs for building such operators, such as element loops and tools to represent specific components of an operator; packages such as deal.ii and MFEM exemplify this approach~\cite{thedealiifiniteelementlibrarydesignfeaturesandinsights2021, kronbichler2012generic, andrej2024high}. 
Third, several systems such as Firedrake, FEniCS, and Dune rely on the Unified Form Language, a DSL for specifying weak forms or energies as expressions~\cite{unifiedformlanguageadomainspecificlanguageforweakformulationsofpartialdifferentialequations2014,automatedsolutionofdifferentialequationsbythefiniteelementmethod2016,firedrakeautomatingthefiniteelementmethodbycomposingabstractions2016,dunefemageneralpurposediscretizationtoolboxforparallelandadaptivescientificcomputing2012}. 
Within computer graphics, SymX is closet cousin of these, as it has a DSL for energies that it can minimize~\cite{fernandez2025symx}.
Some packages such as Gridap.jl~\cite{Badia2020,verdugo2022software} or LibCEED~\cite{brown2021libceed,libceedusermanual2020} operate potentially between the second and third approaches, via novel ways of expressing and decomposing finite element assembly loops that utilize compiler technologies to achieve varying advantages.

Form languages are the line of work closest to our own as the specification form is similar to the specification of elements, but they are not used this way.
Form language compilers produce quadrature loops, which are heavily optimized for performance, as opposed to symbolic outputs that can be used to compute pullback matrices~\cite{exposingandexploitingstructureoptimalcodegenerationforhighorderfiniteelementmethods2017,ontheefficiencyofsymboliccomputationscombinedwithcodegenerationforfiniteelementmethods2010,acompilerforvariationalforms2006,automatedsolutionofdifferentialequationsbythefiniteelementmethod2016}.
As a minor example of this, when users wanted to use FEEC elements in form compilers, form compilers had to be modified to deal with pullbacks associated with FEEC elements as opposed to symbolically deducing such pullbacks~\cite{efficientassemblyofhdivandhcurlconformingfiniteelements2010}.
However, we note that form compilers have other symbolic capabilities such as various forms of automatic differentiation~\cite{unifiedformlanguageadomainspecificlanguageforweakformulationsofpartialdifferentialequations2014,automatedderivationoftheadjointofhighleveltransientfiniteelementprograms2013,fireshapeashapeoptimizationtoolboxforfiredrake2021}, the ability to perform various automated optimizations~\cite{exposingandexploitingstructureoptimalcodegenerationforhighorderfiniteelementmethods2017,russell2013optimized,automaticcodegenerationforhighperformancediscontinuousgalerkinmethodsonmodernarchitectures2020,Bergbauer2025} (and this is not to mention a lot of compiler work aimed at the performance of FEM systems in general), and the ability to express many highly specialized time-steppers and preconditions~\cite{slateextendingfiredrakesdomainspecificabstractiontohybridizedsolversforgeoscienceandbeyond2020,automatedcodegenerationfordiscontinuousgalerkinmethods2009,solvercompositionacrossthepdelinearalgebrabarrier2018,irksomeautomatingrungekuttatimesteppingforfiniteelementmethods2021}.
A particularly valuable contribution is automatic sum factorization for high-order elements, enabled by FInAT's explicit representation of tensor-product structure and TSFC's exploitation of that structure in Firedrake~\cite{exposingandexploitingstructureoptimalcodegenerationforhighorderfiniteelementmethods2017}.
This reduces the computational complexity of assembly, making it an important consideration in performance comparisons with \SYS{}, which does not currently implement sum factorization.

At this time, we do not pursue this higher-level work, but we do produce a form language (the same language we use to specify elements).
Our form language inherits some advantages and disadvantages from a non-specialized form language: for example, our language can be more concise for many forms that operate on high-level tensor expressions, especially those involving exterior calculus, but our language can also be more verbose as we do not build in specialized constructs for form differentiation, mixed methods, or for discontinuous Galerkin methods.

\subsection{Less Well-Known Finite Elements}
\label{sec:lwkfe}

We highlight the many FEs that appear in the literature, briefly commenting on their various utilities. 
For example, there are many families of FEs that offer various forms of continuity~\cite{nonconformingfiniteelementspacesfor2mthorderpartialdifferentialequationsontextbackslashmathbbvphantomrvphantomnsimplicialgridswhenmn12018,minimalfiniteelementspacesfor2mthorderpartialdifferentialequationsinrn2012,forthereducedhsiehcloughtochertriangle,thetubafamilyofplateelementsforthematrixdisplacementmethod1968,weakgalerkinmethodsbasedmorleyelementsongeneralpolytopalpartitions2022,theminimalconforminghkfiniteelementspacesonrnrectangulargrids2014,arefinedtriangularplatebendingfiniteelement1969,aconstructionofcrconformingfiniteelementspacesinanydimension2021}.
Sometimes FEs are defined because they allow elegant and correct solutions to certain problems~\cite{mixedfiniteelementsforelasticity2002,thehellanherrmannjohnsonmethodwithcurvedelements2020,finiteelementdifferentialformsoncubicalmeshes2013,adiscretederhamcomplexwithenhancedsmoothness2006,infsupstablefiniteelementsonbarycentricrefinementsproducingdivergencefreeapproximationsinarbitrarydimensions2018,asystematicconstructionoffiniteelementcommutingexactsequences2016}. 
Some FEs offer superior convergence speeds or conditioning properties over easier to use options~\cite{afamilyofcrouzeixraviartfiniteelementsin3d2017,spectralhpelementmethodsforcfd1999,ahermitelikebasisforfastermatrixfreeevaluationofinteriorpenaltydiscontinuousgalerkinoperators2019,asystematicapproachtoconstructingpreconditionersforthehpversionmassmatrixonunstructuredandhybridfiniteelementmeshes2022,highorderfiniteelementsonpyramidsapproximationspacesunisolvencyandexactness2010,afamilyofnonconformingelementsandtheanalysisofnitschesmethodforasingularlyperturbedfourthorderproblem,computationalserendipityandtensorproductfiniteelementdifferentialforms2019} or sometimes they offer highly efficient code for the same convergence~\cite{fastsimplicialfiniteelementalgorithmsusingbernsteinpolynomials2011,ashortnoteonabernsteinbezierbasisforthepyramid2016,analysisofsomefiniteelementsforthestokesproblem,orthogonalbasesforvertexmappedpyramids2016,adualfiniteelementcomplexonthebarycentricrefinement2005}. 
Some FEs are uniquely suited for certain applications using some portion of the above properties~\cite{generalrelativitywithoutcoordinates1961,structurepreservingdiscretizationofthehessiancomplexbasedonsplinespaces2021,mixedfiniteelementsforelasticity2002,mixedfiniteelementmethodsfornonlinearcontinuummechanicsandshells,commutingdiagramsforthetntelementsoncubes2014,asystematicconstructionoffiniteelementcommutingexactsequences2016,thehellanherrmannjohnsonmethodwithcurvedelements2020,nonconformingtetrahedralmixedfiniteelementsforelasticity2014,infsupstablenonconformingfiniteelementsofarbitraryorderontriangles2005,arobustfiniteelementmethodfordarcystokesflow2002}.
We believe that \SYS{} can implement most of these elements with the exception of the small minority of elements relying on rational function spaces and NURBS/Spline type constructions. 
We do require an extension for piecewise polynomial spaces in the element space, but these are known to be simple.
Further, projects such as tIGAr~\cite{Kamensky2023,Kamensky2021} has shown how to extend code generation approaches such as FEniCS to isogeometric analysis, which we believe we could adapt to our system with non-trivial effort.

\section{Software Representation of DOFs and Operators: Integrated Pointwise Tensor Expressions on Manifolds}
\label{sec:swrdoIPTEM}
We capture both DOFs and operators via a new language for Integrated Pointwise Tensor Expressions on Manifolds (IPTEM).
The IPTEM language balances the expression of DOFs/operators against both succinctness and automation.
Our IPTEM languages facilitate a succinct yet expressive representation that can be manipulated automatically to produce simple formulas for use in FEM code. 

We introduce our IPTEM language and compiler in stages, each of which introduces a novel language and novel compilation algorithm.
The languages are tensor expressions (\cref{sec:TE}), pointwise tensor expressions on manifolds (\cref{sec:PTE}), abbreviated PTEM, and, finally, integrated pointwise tensor expressions on manifolds, resulting in IPTEM (\cref{sec:IRIPTEM}).
We organize \cref{sec:swrdoIPTEM} in this order. 
Each language addition has an associated compilation algorithm and together they amount  to a single compiler. The composition of these compilers is depicted in \cref{lst:IptemCompilerArchitecture}. Notably, the organization of the compiler runs contrary to that of the languages, going from PTEM to tensors to integrals.
The rest of our implementation sections utilize the overall compiler depicted in \cref{lst:IptemCompilerArchitecture} as a black-box to generate code for \VA{} matrices (\cref{sec:SFVA}) and for the local parts of global FEM code (\cref{sec:IAGF} and \cref{sec:AGC}).

\begin{lstlisting}[float, language=PythonPlus, caption={
Architecture of the IPTEM compiler. 
%
At the bottom language level, we introduce a compiler from tensor expressions to formula, simple array programs described in \cref{sec:TEF}, that represent the expression in an input basis (\cref{sec:TEFG}).
%
%
We can treat this level as a computer algebra system compiler hybrid that other levels query.
%
This operates in the middle of our overall IPTEM compiler (in \cref{lst:IptemCompilerArchitecture}: \cref{line:tensorEnd}).
%
To get to the point where we can compile tensor expressions to formula, we must manage PTEMs, the middle language level, and translate them to situations that our tensor compiler can handle (\cref{sec:SCPTE} and  \cref{line:testart} - \cref{line:signEnd}).
%
We do this in two phases.
%
We call the first phase topological enumeration  (\cref{sec:TEP} and \cref{line:teend}) and the second phase single compilation (\cref{sec:SCP}, \cref{line:signEnd}).
%
Topological enumeration enumerates the different ways a PTEM can be matched to input data (fields and charts - see \cref{sec:cftv}) and the second phase converts a PTEM to a tensor expression given a matching of PTEM to input data.
%
At the top language level, we can add integration and reduction of PTEMs, but in our overall compilation algorithm we handle integration and reduction last (\cref{line:redStart}-\cref{line:endnend1}) via two post-processing phases.
%
First, our compiler generates code for integrating formulas produced from PTEMs matched to input data (\cref{sec:DSII} and \cref{line:integs}). Second, our compiler generates code to combine formulas for various IPTEM together based on reductions across shared chart and field argument matches (\cref{sec:Red} and \cref{line:redEnd}).
%
Particular attention to the relationship between topological enumeration and reductions is given in \cref{fig:megaTE} as it is critical to how we solve problems emerging from multi-valued computation in FEM code (see \cref{sec:globfem}).
%
},label={lst:IptemCompilerArchitecture},escapechar=|, captionpos=b,abovecaptionskip=5pt]
matchTy =  dict[PTEMArgument, Union[Chart,Field]]
|{\color{red}(See \cref{sec:PTEG} and \cref{sec:vii})}|
iptemTensorTy = Tuple[Tensor, Basis, dict[Tuple[str, Basis], Formula]]
|{\color{red}(See \cref{sec:TESEM}, \cref{sec:TEBA}, and \cref{sec:TEF})}|
def iptemCompiler(iptems: List[IPTEM],|{\color{red}( See \cref{sec:IRIPTEM})}| |\label{line:iptemInput}|
                  charts: List[Chart], |{\color{red}( See \cref{sec:vii})}|  |\label{line:chartInput}|
                  fieldArgs: dict[Argument, List[Field]], |{\color{red}( See \cref{sec:vii})}|  |\label{line:argField}|
                  evaluateIntegral: bool |{\color{red}(See \cref{sec:DSII}})|
                  ) -> DataFrame: |{\color{red}(See \cref{fig:megaTE})}||\label{line:argOut}|
    # PTEM Phase: |{\color{red}(See \cref{sec:PTE})}||\label{line:testart}|
    iptemMatches : dict[IPTEM, List[matchTy]] = {}
    for iptem in iptems:
        iptemMatches[iptem] = topologicalEnumerate(iptem.ptem, charts, fieldArgs) |{\color{red}(See \cref{sec:TEP} and \cref{fig:megaTE})}||\label{line:teend}|
     |\label{line:singStart}|
    iptemTensors: dict[Tuple[IPTEM, matchTy], iptemTensorTy]  = {}|\label{line:multiComp}|
    for (iptem, matches) in iptemMatches:
        for m in matches:
            iptemTensors[(iptem, m)] = singleCompilation(m) |{\color{red}(See \cref{sec:SCP})}||\label{line:signEnd}|
    # Tensor Compiler Phase: |{\color{red}(See \cref{sec:TESEM})}|
    ptemFormulas: dict[Tuple[IPTEM, matchTy], Formula]  = {} |\label{line:tensorStart}|
    for (key, (tensor, basis, env)) in iptemTensors:
        ptemFormulas[key] = representTensor(tensor, basis, env) |{\color{red}(See \cref{sec:TEFG})}||\label{line:tensorEnd}|
    # Integrals/Reductions phase: |{\color{red}(See \cref{sec:IRIPTEM})}|
    iptemFormulas: dict[Tuple[IPTEM, matchTy], Formula]  = {}  |\label{line:redStart}|
    for ((iptem, m), Formula) in ptemFormulas:
        iptemFormulas[iptem, m] = applyIntegration(iptem.integral, m, formula, evaluateIntegral) |{\color{red}(See \cref{sec:DSII})}| |\label{line:integs}|
    table = applyReductions(iptemFormulas) |{\color{red}(See \cref{sec:Red})}| |\label{line:redEnd}|
    # Table has colums for IPTEM, IPTEM arguments, and Formula |\label{line:tensorcomp}|
    return table|\label{line:endnend1}|
\end{lstlisting}

\subsection{Tensor Expressions: Tensors, Tensor Spaces, Basis}
\label{sec:TE}
Our system balances succinctly expressing tensors against our need for the automatic generation of simple formulas.
Along the lines of \cref{sec:bptac}, our system achieves this balance by utilizing the idea of coordinate independence.
We introduce a grammar for tensor expressions (\cref{grammar:tensorExpr1}) that only allows coordinate independent concepts, leaving the choice of basis (\cref{grammar:tensorBasis1}) to automation and optimization via compiler algorithms. 
We do not fully elucidate the formal aspects of this language as these are the subject of another work, which describes a similar language in fully formal terms.~\cite{collin2026grassmann}.

\newcommand{\gmr}[1]{\mathtt{#1}}

\subsubsection{Grammar and Semantics for Tensor Expressions}
\label{sec:TESEM}
We clarify the mathematical meaning of the terms in \cref{grammar:tensorExpr1}.
In what follows, we follow the grammar: we examine each space, $\s$, and some associated tensors, $\e$, in turn, defining inner products recursively as a running class of examples.
Through these examples, we show that terms are coordinate independent yet expressive.
We also introduce syntactic sugar, which is present in our implementation as a Python EDSL.
Here, \lstinline!V! is a DSL object representing a vector space, with methods such as \lstinline!V.inner!; an annotation \lstinline!v: V! declares a symbolic vector in that space. These annotations are interpreted by the DSL rather than as ordinary Python class hints. The \lstinline!@linear! decorator constructs a symbolic multilinear map from the function, so the result supports tensor operations such as \lstinline!reshape!. For space objects, \lstinline!V | W! denotes the direct sum $V\oplus W$, while \lstinline!V * W! denotes the tensor product $V\otimes W$; the former is not Python's union-type notation.

\begin{enumerate}
    \item The first space in \cref{grammar:tensorExpr1} is $\Rspace$, which is associated to the operations/vectors  available in every vector space: $\Zero$, $\boldplus$,  scalar multiplication via $\boldtimes$, and specific scalars (noted by $\cvar$ in the grammar). Scalar multiplication provides an inner product for this space.
    \item Given a space $\gV$, \cref{grammar:tensorExpr1} allows for the dual space $\vstar{V}$ (see \cref{sec:bptac}). The dual space is closely associated to the sharp map $\boldsharparg{v}$ (\cref{def:dag}), 
    $\Abstract$, and $\Eval$.
    For two tensor expressions $\attemptmath{{e, e^{\prime}}}$ (corresponding to tensors $v\in V^{\star}$ and $v^{\prime}\in V$), we use $\Eval$ to produce a scalar via $\attemptmath{Eval(e^{\prime}, e)}$, representing $v^{\prime}(v)$. We can abbreviate this as $\attemptmath{e^{\prime}(e)}$.
    Similarly, we can write $\attemptmath{e^{\boldsharp}}$ for $v^{\sharp}\in V$.
    Since spaces in our grammar are inner product spaces, this sharp, just like evaluation, does not reference specific coordinates (see \eqref{eq:inner}).
    To express the inner product, we introduce multi-linear evaluation, $\attemptmath{\Eval(e, e_{0},\dots, e_{n})}$ for evaluating the first $n$ linear arguments of some $\attemptmath{e}$.
    We also support partial evaluation: $\attemptmath{\Eval(e, (i_{0}, \dots, i_{n}), e_{0},\dots, e_{n})}$
    where $\attemptmath{i_{j}}$ indicates which argument of $\attemptmath{e}$ will be supplied $\attemptmath{e_{j}}$.
    Finally, we can produce expressions representing vectors in $V^{\star}$ using abstract: given an $\attemptmath{e}$ representing a vector in the space $V$ and an inner product on $(\cdot, \cdot)_{V}$ on $V$, represented with $\Inner(\gV)$, we can write
    \begin{equation*}
        \attemptmath{\Abstract(\tx_{0}, \gV, \Rspace, \Inner(\gV)(e,\tx_{0}))}
    \end{equation*}
    which should be read as mapping a variable $\tx_{0}$ representing a value in the space denoted via $\gV$ to a value in the space denoted via $\Rspace$ via an expression.
    The expression evaluates the inner product on $V$ on an input variable and $\attemptmath{e}$.
    Since evaluating the inner product is linear, this represents a value in $V^{\star}$.
    We can use abstraction and multi-linear evaluation to define an inner product on $V^{\star}$:
    \begin{equation*}
        \attemptmath{\Abstract([\tx_{0}, \tx_{1}], [Dual(V), Dual(V)], \mathbb{R}, \Inner(\gV)(\tx_{0}^{\boldsharp}, \tx_{1}^{\boldsharp}))}
    \end{equation*}
    We could translate this into our Python syntax:
        \begin{lstlisting}[language=PythonPlus]
        @linear # Abstract
        def f(tx0: Dual(V), tx1: Dual(V)) -> R:
            return V.inner(tx0.sharp(), tx1.sharp()) 
    \end{lstlisting}

    \item To characterize linear and multi-linear maps, we utilize tensor products (\cref{ex:tenprodex}) as a coordinate independent representation. 
    Tensor product tensor expressions can be constructed with $\boldotimes$, but also via $\Abstract$ and any node that provides a multi-linear map. For example, the inner product of $\gV$ represents a bilinear map, denoted with $\gV^{\boldstar}\boldotimes \gV^{\boldstar}$.
    The reshape map can be used to convert this to a map on $(\gV\boldotimes \gV)^{\boldstar}$ via $\attemptmath{\Reshape(\Inner(\gV), [0, 1], 0)}$, as in \eqref{eq:lift}.
    The integer arguments in $\Reshape$ and $\Eval$ handle the case of a more general tensor product space.
    Suppose we have a tensor product of $k$ dual spaces and $m$ other spaces.
    The integers for  $\Eval$ and $\Reshape$  allow one to access the $k$ dual spaces.
    The last integer in $\Reshape$ places a reshaped space in the $k+m$ spaces.
    Finally, we note that given a $\gV$ and $\gW$ with associated inner products, we can define an inner product on $\gV\boldotimes \gW$.
    To hint at our Python EDSL, we show this one in Python syntax:
    \begin{lstlisting}[language=PythonPlus]
        @linear # Abstract
        def f(v: V, w: W, vp: V, wp: W) -> R:
            return V.inner(v, vp) * W.inner(w, wp)
        inner = f.reshape([0,1], 0).reshape([1,2], 1)
    \end{lstlisting}

    \item Given two vector spaces expressions $\gV$ and $\gW$, we have a direct sum $\gV\boldoplus \gW$. This allows the representation of $\mathbb{R}^{2}$ ($\Rspace\boldoplus\Rspace$) and a vector in $\mathbb{R}^{2}$ might be written as $\attemptmath{1\boldoplus 2}$.
    The operations $\Project$ and $\Embed$ allow us to extract an entry from a direct sum or to create a direct sum with a specific entry.
    By conflating $\Project$ with indexing, we have a Pythonic way to write the inner product:
    \begin{lstlisting}[language=PythonPlus]
        @linear # Abstract
        def inner(x: V | W, y: V | W) -> R:
            return V.inner(x[0], y[0]) + W.inner(x[1], y[1])
    \end{lstlisting}
    With this addition, our language starts to resemble a NumPy like language that respects co- and contra-variance and utilizes coordinate free quantities.

    \item The final four spaces of the grammar all utilize the same idea. 
    Given a linear map $L\in V^{\star}\otimes W$, we can factor $V$ into the kernel and co(mplement)-kernel as well as $W$ into image and co(mplement)-image.
    Our final four spaces provide exactly these with some additional metadata.
    We can indicate if $L$ is injective and/or surjective via properties ($\gprop$), as these sometimes simplify these spaces.
    These spaces allow our tensor expression language to express a surprisingly wide variety of mathematical concepts.
    For example, symmetric tensors found in elasticity\footnote{sometimes the symmetric tensors in elasticity are considered to be symmetric tensors of anti-symmetric tensors, which we can also model with this} can be expressed using the kernel of the anti-symmetrization map and the exterior algebra can be expressed using the co-image of the symmetrization map. 
    Also, the tangent space can be expressed as the image of the derivative of a paramterization/chart at a point.\footnote{The former definition is not the coordinate free definition of the tangent space, but it is defined independently of basis once the derivative at a point is supplied. In other words, the linear algebra is still coordinate independent.}
    The operator $\Embed$ allows us to embed vectors from these subspaces into their ambient spaces, leading to simple definitions of inner products via those on $V$ or $W$.
    \begin{lstlisting}[language=PythonPlus]
        @linear 
        def inner(x: Kernel(L, V), y: Kernel(L, V)) -> R:
            return V.inner(x.embed(V), y.embed(V))
    \end{lstlisting}
    The operator $\Project$ allows us to enter these spaces and we do not support indexing for it.
    Lastly, we encode the first isomorphism theorem in our language: users can access the induced mapping from the co-kernel to the image with the $\Restrict$ mapping and the inverse\footnote{the inverse can be used on other mappings but might fall back on the first isomorphism and projection} via $\Inverse$.
\end{enumerate}

As a final note on expressiveness, we can recover arbitrary Einstein index notation from this language, but we still need other operations on tensors in various mathematical models, necessitating an extensible escape hatch in the $\NonLinear$ term. 
The escape hatch is intended for non-linear scalar functions such as $\cos$ or $\sqrt{}$ as these often appear in weak formulations.

\begin{figure}
\footnotesize
\begin{tabular}{rcll}
\toprule
$\tx$  &       &                                      & tensor variable \\
$\nvar$   &       &                                      & constant integer \\
$\bvar$   &       &                                      & constant boolean \\
$\fvar$   &       &                                      & Non-Linear Function \\
\midrule
$\s$  & $::=$ & $\Rspace$                                  & Real Numbers \\
    &  $|$  & $\sstar$                                  & Dual Space  \\
     &  $|$  & $\szero \boldotimes\ \sone$                                  & Tensor Product \\
      &  $|$  & $\szero \boldoplus\ \sone$                                  & Direct Sum \\
      &  $|$ & $\Kernel(\e, \szero ?, \sone ?, \gprop)$ & Kernel\\ 
       &  $|$ & $\Image(\e, \szero ?, \sone ?, \gprop)$ & Image\\
        &  $|$ & $\CoImage(\e, \szero ?, \sone ?,\gprop)$ & Co-Image\\
    &  $|$ & $\CoKernel(\e, \szero ?, \sone ?,\gprop)$ & Co-Kernel\\
\midrule
$\gprop$  & $::=$ & $\Injective \mid \Surjective \mid \Bijective$  & Linear Map Properties\\
\midrule
$\e$  & $::=$ & $\tx$                                  & Tensor variable name \\
    &  $|$  & $\cvar$                                  & Constant real number \\
    &  $|$  & $\Zero(\s)$                                  & Zero of a space\\
    &  $|$  & $\ezero \boldplus\ \eone$                                  & Addition of Tensors\\    
    &  $|$  & $\ezero \boldtimes\ \eone$                                  & Scaling of Tensors\\        
    &  $|$  & $\esharp$                                  & Sharp operation\\          
    &  $|$  & $\ezero \boldotimes\ \eone$                                  & Tensor Product\\      
    &  $|$  & $\Eval(\ezero, (\nzero,\dots)?, \eone, \dots)$                                  & Evaluation\\            
    & $|$ & $\Abstract((\txzero, \dots), (\szero,\dots)?, \e, \s ?)$                                  & Multi-Linear Function\\    
    &  $|$  & $\Reshape(\e, (\nzero,\dots, \nk), \nvar)$                                  & Universal Property of Tensors\\            
    &  $|$  & $\Inner(\s)$                                  & Inner Product of  $\s$\\            
    &  $|$  & $\ezero \boldoplus\ \eone$                                  & Direct Sum \\            
    &  $|$  & $\Project(\szero, \sone, \nvar ?)$                                  & Projection\\            
    &  $|$  & $\Embed(\szero, \sone, \nvar ?)$                                   & Embedding\\            
    &  $|$  & $\Inverse(\e)$                                & Inverse\\            
    &  $|$  &  $\Restrict(\e)$                                   & Restriction\\            
    &  $|$  & $\NonLinear(\fvar, \e)$                                & Black-box evaluation\\    
\bottomrule
\end{tabular}
\caption[Tensor Expression Grammar]{A Grammar to Describe Tensor Expressions and Tensor Spaces.}
\label{grammar:tensorExpr1}
\end{figure}

\subsubsection{Formulas}
\label{sec:TEF}

In this paper, a \textbf{formula} is a straight-line array program: a sequence of intermediate assignments followed by a returned array. Formulas are the target language of our tensor expression compiler, with syntax given in~\cref{grammar:RF}.
However, formulas as a representation of programs constrain $\Array$ entries to be rational functions ($\rf$) (in the intermediate variables of the program).
Critically, formulas constrain arrays to enable symbolic manipulation via manipulation of rational functions.
Consequently, we have two notions of simplicity.
First, we have the degree of the rational functions, a natural number for the polynomial degree of the numerator  and a natural number for the polynomial degree of the denominator,  in terms of the inputs to the formula (which is potentially infinite due to non-linear functions).
This roughly captures the simplicity of the \VA{} matrix of the Morley element (\cref{ex:mpm}).
Moreover, this captures the literal simplicity of evaluating a formula.
The degree and the number of variables allows one to bound the evaluation cost.
Second, we have the degree of the rational function outputs in terms of all intermediates, which is necessarily finite. 
This second notion captures our ability to manipulate formulas with polynomial manipulation routines (e.g., from Sympy) to structurally simplify formulas.
Our compiler for tensor expressions prioritizes the first notion unless a formula becomes too big to manipulate in some context (e.g., symbolic inversion of a large dense symbolic matrix).

\begin{figure}
\footnotesize
\begin{tabular}{rcll}
\toprule
$\sh$  &       &                                      & constant nat tuple \\
$\av$  &       &                                      & array variable \\
$\fvar$  &       &                                      & External Function \\
$\en$  &       &                                      & Einsum string \\
\midrule
$\rf$  & $::=$ & \textbf{c}                              & Constant Real Number \\
     &  $|$  & $\av[\sh]$                                  & Array Variable Access \\
     &  $|$  & $\rfzero \boldplus\ \rfone$                                  & Sum \\
     &  $|$  & $\rfzero \boldtimes\ \rfone$                                  & Product \\
     &  $|$  & $\rfzero \bolddiv\ \rfone$                                  & Division \\
\midrule

$\ar$  & $::=$ & $\Array((\rfzero,\dots,), \sh)$   & Rational Function Array \\
\midrule

$\stmt$  & $::=$ & $\av_{0} = \Einsum(\en, \av_{1},\dots)$                                  & Einsum \\
    &  $|$  & $\av_{0} = \Inv(\av_{1})$  & Inverse \\
    &  $|$  & $(\av_{0}, \av_{1}) = \QR(\av_{2})$  & QR \\
    &  $|$  & $(\av_{0}, \av_{1}, \av_{2}) = \SVD(\av_{3})$  & SVD \\
    &  $|$  & $(\av_{0},\dots, \av_{N}) = \fvar(\av_{N+1}\,\dots)$  & Black-Box Evaluation \\
    &  $|$  & $\av_{0} = \ar$  & Array Definition \\
    &  $|$  & $\Return(\ar)$  & Array Return \\

\midrule

$\formula$  & $::=$ & $(\av_{0},\dots, \av_{n})\mapsto (\stmt_{0},\dots, \stmt_{N})$   & Formula \\
\bottomrule
\end{tabular}

\caption[Formula Grammar]{A grammar to describe formulas, code that computes arrays that represent a tensor expression in a given basis.}
\label{grammar:RF}
\end{figure}

\subsubsection{Bases}
\label{sec:TEBA}
Given a well formed\footnote{We don't go into this detail here, but obviously we need to exclude various expressions that are without meaning (e.g., non-match arguments, sharp of a non-dual space, etc).} tensor expression in~\cref{grammar:tensorExpr1}, the expression will have a clear mathematical meaning. However, we cannot assign the expression to a formula in~\cref{grammar:RF} unless we select bases.
We introduce our basis language in \cref{grammar:tensorBasis1}.
A basis simply specifies a list of tensor expressions and the grammar mostly mirrors the structure of the tensor space grammar.
For example, we note that every space has a corresponding standard basis, which is orthonormal in the inner product described in \cref{sec:TESEM}.
Here, we only discuss the last two basis constructions, which are crucial for our compiler and constitute a useful but confusing contrast.

For any of the four subspaces that depend on a linear function (e.g., the kernel), we can use a $\ProjectedBasis$ to project a basis to the subspace.
Practically, these can be used in formulas via $\SVD$ or $\QR$ though such bases are not unique in all cases\footnote{For injective maps, both are unique up to ordering. In general though, we do not care that these are not unique as we simply need a basis to be consistently chosen. Since we need to make a consistent choice, we do need a unique basis to realize the linear map in, which is why require two basis arguments for $\ProjectedBasis$.}.
Furthermore, a critical alternative exists.
Mainly, if a linear map $F\colon V\to W$ is injective\footnote{Note that charts give us injective linear maps}, then the image and co-kernel have special basis constructions via $\ImageBasis$.
Suppose we have another basis $\bzero$  (representing a basis $\{b_{i}^0\}$ for $W$).
We have $\ImageBasis(\bzero, \Image(\gF, \Injective))$ (representing the basis for $V$ given by $\{F(b_{i})\}$).
Via use of $\Inverse$ and $\Restrict$, we have a similar basis for $\CoKernel$.
The distinction between  $\ProjectedBasis$ and $\ImageBasis$ can model the difference between an immersed inner product for a tangent space (i.e., a surface in Euclidean space uses the Euclidean inner product) vs. an inner product induced by the parameterization of a surface. 
The former choice is the correct one for mathematical modeling, but the latter is easier to compute with.
Our compiler can simplify away transformations by recognizing when an $\ImageBasis$ does not need to be converted to a  $\ProjectedBasis$ to represent a tensor (or vice-versa).

\begin{figure}
\footnotesize
\begin{tabular}{rcll}
\toprule
$\bvar$  & $::=$ & $\{1\}$                                  & Scalar Basis \\
 &  $|$  & $\RawBasis(\ezero,\dots,)$                                  & Basis of tensors\\
 &  $|$  & $\DualBasis(\bzero)$                                  & Dual Basis\\
 &  $|$  & $\PrimalBasis(\bzero)$                                  & Primal Basis\\
 &  $|$  & $\bzero \boldotimes\ \bone$                                  & Tensor Product \\
 &  $|$  & $\bzero \boldoplus\ \bone$                                  & Direct Sum \\
 &  $|$ & $\ProjectedBasis(\bzero, \bone, \s)$ & Projection Basis\\ 
 &  $|$ & $\ImageBasis(\bzero, \s)$ & Image Basis\\
\bottomrule
\end{tabular}

\caption[Tensor Basis Grammar]{A grammar to describe bases for tensor spaces.}
\label{grammar:tensorBasis1}
\end{figure}

\subsubsection{Formula Generation}
\label{sec:TEFG}
We now sketch our compiler, which produces a formula representing a given tensor in a given basis.
And we study some examples that show how the separation between tensor and basis enables automatic derivation of simple formulas for FEM.

We implement our compiler via two mutually recursive functions, depicted in \cref{fig:lst:a}, and a helper function to access the standard orthonormal basis of a space.
These functions carry around an environment to handle variables bound in functions and to handle externally defined representations of variables.
Our compiler, \lstinline[language=PythonPlus]!representTensor!, recursively de-structures a tensor and a basis to produce a formula via various builtin rules.
We provide a sample rule in \cref{fig:lst:b}. 
As we see in  \cref{fig:lst:b}, when \lstinline[language=PythonPlus]!representTensor! encounters a tensor that can't be represented in a basis due to a mismatch between the basis structure and tensor structure,  then \lstinline[language=PythonPlus]!representTensor! uses the standard basis and calls \lstinline[language=PythonPlus]!changeOfBasis! to put the result in the desired basis.
This ensures termination\footnote{up to user defined bases with no conversion to the standard basis} as the standard basis always allows for the rules in \lstinline[language=PythonPlus]!representTensor! to work without a change of basis.

We implement both functions via lists of rules for various tensor and basis combinations.
We discuss a few rules and an example to understand how coordinate independence buys us useful automation.
In ~\cref{fig:lst}, we see rules for inner product, evaluation, and tensor product.
Change of basis rules are similar so we defer examples to \cref{sec:scofb}.
By contrasting the rules for inner product and evaluation, we can learn a core lesson: since our system has some freedom of basis choice in the evaluation rule, our system can use these to find simpler representations.
In particular, the inner product rule must convert to a particular basis whereas the evaluation rule just needs a primal and dual relationship between them.
In \cref{fig:lst:e}, the first two scenarios show the consequence of this: we see that in the first scenario a change of basis is required, leading to a symbolic output, while in the second it can be avoided, leading to a simple numerical output.
Though mathematically we don't actually get to pick between these two scenarios as a final output, we might get to pick as some intermediate path. Of course, our choice of intermediate path must be offset later, but our third scenario shows how the path still practically impacts formulas, especially given rules such as the tensor product rule.
The third scenario is mathematically equivalent to the second one, but we have erased the basis structure and placed it into the resulting formula.
Unless one can reliably detect $(FF^{-1})[a,b] = \delta_{ab}$ on symbolic formulas, we lose the much simpler numerical representation.
The key point is that though our compiler can't change the mathematical answer with an intermediate basis, our compiler can make symbolic computation more reasonable by picking and tracking bases. 
Though we could feasibly detect $(FF^{-1})[a,b] = \delta_{ab}$ with Sympy for small matrices, tensor products could produce rather larger matrices (e.g., $6$ by $6$, which generically has a huge determinant expression) where in contrast the freedom in the tensor product rule could allow us to use simpler representations.
Since the tensor product is coordinate independent, we can sometimes use whatever basis is best for the sub-terms, avoiding symbolic coordinate operations until we need to and retaining the basis structure to try to ease these\footnote{
If the symbolically simpler basis is not an option, our change of basis operations use basis structures similarly. 
Our only innovation with change of basis rules (compared to representation rules) is we compute both directions at the same time to better exploit the basis structure (e.g., a change of basis between two tensor products of bases can be inverted by inverting the constituent change bases, evading a larger matrix inverse). See \cref{sec:scofb}.}

The core point again is that our compiler can avoid a symbolic explosion through picking bases that result in simpler expressions and by tracking bases to avoid having to do hard and expensive symbolic math. 

\begin{figure}
  \centering

  \begin{subfigure}{0.5\textwidth}
    \centering
    \begin{lstlisting}[numbers=none,basicstyle=\fontsize{4pt}{4.125pt}\selectfont\ttfamily]
Env = dict[Tuple[str, Basis], Formula]
def representTensor(tensor: TensorExpr, basis: Optional[Basis], env: Optional[Env]) -> Tuple[Formula, Basis]:
def changeOfBasis(basisSrc: Basis, basisDst: Basis, env: Optional[Env]) -> Tuple[Formula, Formula]:
def standardBasis(space: Space, env: Optional[Env]) -> Basis:
    \end{lstlisting}
    
    \caption{Core Compiler Functions} \label{fig:lst:a}
  \end{subfigure}\hfill
  \begin{subfigure}{0.5\textwidth}
    \centering
        \begin{lstlisting}[numbers=none,basicstyle=\fontsize{4pt}{4.125pt}\selectfont\ttfamily]
def innerRule(t1: TensorExpr, t2: TensorExpr, basis: Optional[Basis], env: Env) -> Tuple[Formula, Basis]:
    (f1, _) = representTensor(t1, standardBasis(spaceOf(t1), env)) # Result in std basis.
    (f2, _) = representTensor(t1, standardBasis(spaceOf(t1), env))
    if basis is None or isStandardBasis(basis):
        # make formula that dots formulas, symbolically.
        return (f1.dot(f2), standardBasis(R))
    else:
        (fpre, preBasis) = innerRule(t1, t2, standardBasis(R), env)
        (fromStandard, _) = changeOfBasis(basis, preBasis)
        return (fromStandard @ fpre, basis)
    \end{lstlisting}
    \caption{Inner Product Rule.} \label{fig:lst:b}
  \end{subfigure}\hfill
  \begin{subfigure}{0.5\textwidth}
    \centering
                \begin{lstlisting}[numbers=none,basicstyle=\fontsize{4pt}{4.125pt}\selectfont\ttfamily]
def simpleEvalRule(t1: TensorExpr, t2: TensorExpr, basis: Optional[Basis], env: Env) -> Tuple[Formula, Basis]:
    if basis is None:
        (f1, b1) = repesentTensor(t1, None, env) # pick simple formula
        (f2, b2) = representTensor(t2, None, env)
        if isPrimalDual(b1, b2):
            return (f1.dot(f2), standardBasis(R))
        else:
            # there are many options here
            # We show one
            (f1p, _) = representTensor(t1, DualBasis(b2), env)
            return (f1p.dot(f2), standardBasis(R))
    else:
        match basis:
            case Dual(b1) * b2:
                (f1, _) = repesentTensor(t1, b1, env)
                (f2, _) = representTensor(t2, b2, env)
                return (f1.dot(f2), standardBasis(R))
            case _:
                recurse = simpleEvalRule(t1, t2, None)
                (fromBasis, _) = changeOfBasis(basis, recurse[1])
                return (fromBasis @ recurse[0], basis)
    \end{lstlisting}
   
    \caption{Evaluation Rule} \label{fig:lst:d}
  \end{subfigure}\hfill
    \begin{subfigure}{0.5\textwidth}
    \centering
            \begin{lstlisting}[numbers=none,basicstyle=\fontsize{4pt}{4.125pt}\selectfont\ttfamily]
def tensorProdRule(t1: TensorExpr, t2: TensorExpr, basis: Optional[Basis], env: Env) -> Tuple[Formula, Basis]:
    if basis is None:
        (f1, b1) = repesentTensor(t1, None, env) # pick simple formula
        (f2, b2) = representTensor(t2, None, env)
        return (f1.oprod(f2), b1 * b2)
    else:
        match basis:
            case b1 * b2: # tensor product
                (f1, _) = representTensor(t1, b1, env)    
                (f2, _) = representTensor(t2, b2, env)            
                return (f1.oprod(f2), b1 * b2)
            case _:
                (f1, b1) = representTensor(t1, None, env)    
                (f2, b2) = representTensor(t2, None, env)  
                (fromProd, _) = changeOfBasis(basis, b1 * b2)
                return (fromProd @ (f1.oprod(f2)), basis)
    \end{lstlisting}
    \caption{Tensor Product Rule} \label{fig:lst:c}
  \end{subfigure}\hfill
  \begin{subfigure}{0.5\textwidth}
    \centering
    \begin{lstlisting}[numbers=none,basicstyle=\fontsize{4pt}{4.125pt}\selectfont\ttfamily]
import sympy as sp
env = EmptyEnv()
env[("F", DualBasis(standardBasis(R^3)) * standardBasis(R^3))] = sp.Matrix("F")
img = Image(Var("F"), R^3, R^3, Bijection)
# Scenario 1:
env[("a", standardBasis(Dual(img))] = np.array([1, 0])
env[("b", imageBasis(standardbasis(R^3), img)] = np.array([1, 0])
ret = representTensor(Var("a")(Var("b")), None, env)
assert ret[0] == F[0,0]
# Scenario 2:
env[("ap", DualBasis(imageBasis(standardbasis(R^3), img))] = np.array([1, 0])
env[("bp", imageBasis(standardbasis(R^3), img)] = np.array([1, 0])
ret = representTensor(Var("ap")(Var("bp")), None, env)
assert ret[0] == 1.0
# Scenario 3:
env[("ap", DualBasis(ProjectedBasis(standardBasis(R^3), standardBasis(R^3), img)] = F.inverse() @ np.array([1, 0])
ret = representTensor(Var("ap")(Var("bp")), None, env)
assert ret[0] == (F.inverse() @ F)[0,0]

    \end{lstlisting}
    \caption{Sketch of possible scenarios, showing how simpler symbolic intermediates could result from greater freedom of choice in bases.} \label{fig:lst:e}
  \end{subfigure}

  \caption{Example compilation rule with an example illustrating the different structures of the rules.}
  \label{fig:lst}
\end{figure}

\afterpage{
\clearpage
\begin{figure*}
    \centering
    \includegraphics[width=.99\textwidth]{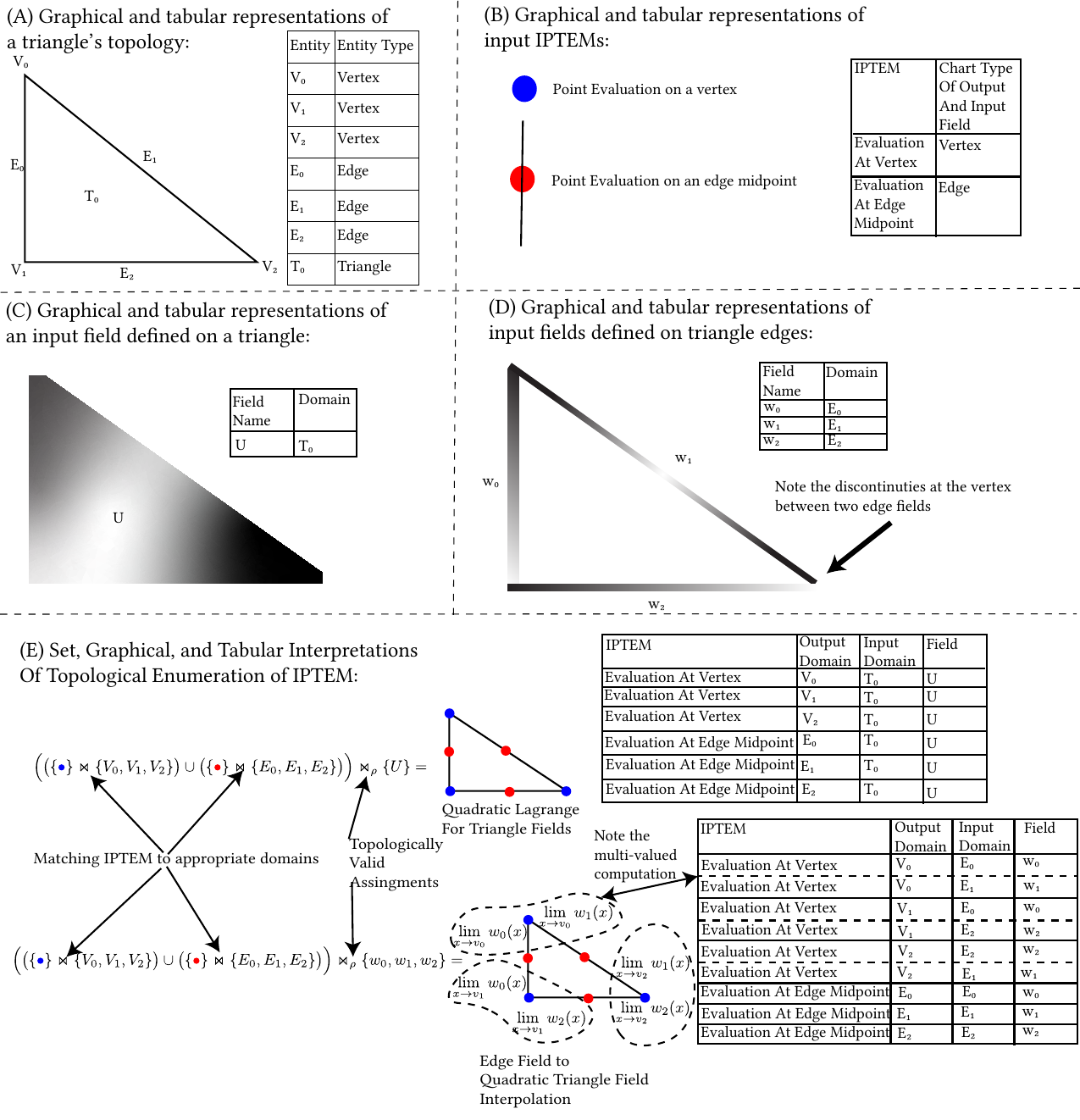}
    \caption{On a triangle topology (A), we have two groups of field inputs, (C) and (D), and we have one group of IPTEMs (B). By changing the field inputs, we have two different topological enumeration calculations, depicted graphically with unions and joins in (E). The first reproduces quadratic Lagrange elements while the second produces a rule to transfer edge fields to a quadratic Lagrange element, up to a resolution of discontinuities at vertices. The inputs and outputs are represented tabularly too.}
    \label{fig:megaTE}
\end{figure*}
}

\subsection{Pointwise Tensor Expressions on Manifolds (PTEM)}
\label{sec:PTE}

Pointwise expressions are of the form $F(x, u(x), Du(x),\dots, t(x),\dots)$ where $F$ only has tensor operations, $x$ is a point, $u$ is some field,  $Du$ is the derivative, and $t(x)$ is geometric information such as a tangent vector.
(The expression is pointwise because it depends on $u(x)$ not $u(x+y)$ or $u(v(x))$.)
Due to the geometric nature of these expressions, we call them pointwise tensor expressions on manifolds (PTEM).
For the purposes of automating FEM, we seek a succinct representation of pointwise expressions that allows us to translate $F(x, u(x), Du(x),\dots, t(x),\dots)$  into a pure tensor expression so we can utilize our tensor compiler.
Our tensor compiler can exploit the exact numerical representation of an entry in a basis, but specifying the exact numerical representation of geometric information and fields poses an incredible burden on users.

To keep PTEM succinct for users while preserving automation with our tensor compiler, we embrace more automation.
In particular, PTEM uses chart and field types (\cref{sec:cftv}) as placeholders for concrete charts and fields. Upstream automation (\cref{sec:IAGF}, \cref{sec:SFVA}, and \cref{sec:AGC}) supplies matching charts and fields (\cref{sec:vii}) through the interfaces in \cref{lst:IptemCompilerArchitecture}.
Users will not ever write the fields and geometries mentioned in this section; they will write corresponding types  (\cref{sec:texi}) to write pointwise expressions (\cref{sec:PTEG}) that our compiler will automatically match (\cref{sec:TEP}) with actual charts and fields (\cref{sec:vii}) to produce tensor expressions (\cref{sec:SCP}).
We call our compiler's process of matching PTEM with charts and fields topological enumeration (\cref{sec:TEP}) and we call the final production of a tensor expression from a match, single compilation (\cref{sec:SCP}).
To use the previous section as a metaphor, pointwise expressions with field/geometry types are tensor expressions, fields/geometries are bases, and the tensor expressions derived from their combinations are formulas.

Through this design, we also solve another problem raised in \cref{sec:globfem}: fields are multi-valued so some computation is multi-valued, which must be carefully handled in FEM codes.
However, our process of topological enumeration actually finds all combinations of inputs where the output is a single well defined value.
Thus, as will be shown in \cref{fig:megaTE}, our automation to evade cumbersome specification of geometries and fields will actually solve higher level problems and simplify the specification of high level FEM computations.

In this section, we work from the ground up: we start with charts and fields, move to PTEM, and then discuss single compilation, and the finally topological enumeration.

\subsubsection{Charts and Fields: Types and Values}
\label{sec:cftv}

We detail both charts and fields as the external interface (types) and the internal interface (values). 

\paragraph{Values: Internal Interface}
\label{sec:vii}
Our compiler and system internally think of a chart as just an affine map with some metadata (\cref{def:chart1}), but the matrix and vector are replaced with tensor expressions, which are allowed to potentially depend on inputs such as vertices.
Though charts can just use the general formula for an affine map induced by sets of vertices, we generally want our tensor compiler to use more specific information where possible.
The internal interface for fields, depicted in ~\cref{lst:fieldInterfaceLocal},  is similarly straightforward : given a position with a reference position and world position with respect to some chart, we can sample the field and its derivatives in some basis.
We note that just like the vertices in the chart, the field can be sampled at a formula for a point.
Both of these interfaces make patching general or specific fields/charts into the tensor compiler easy, but users should not have to construct such objects.
Our PTEM compiler just needs these interfaces to (1) provide the necessary data to our tensor expression compiler and (2) to track charts and fields internally so we can automatically reason about how to combine them with pointwise expressions.

\begin{lstlisting}[language=PythonPlus, caption={Field interface for pointwise expressions.}, label={{lst:fieldInterfaceLocal}},abovecaptionskip=5pt]
class Chart:
    inputs: List[Tuple[str, Basis]] # e.g vertex symbols
    A: TensorExpr
    b: TensorExpr
    inputTopType: TopType
    # omitting topology/geometry of domain/image

class Field(Protocol):
    type: FieldType
    chart: Chart # where is the field defined
    def sample(refPoint: Tuple[Formula, Basis], 
               worldPoint: Tuple[Formula, Basis], 
               chart: Chart,
               derivatives: int,
               basis: Optional[Basis] = None) -> Tuple[Formula, basis]: ...
\end{lstlisting}

\paragraph{Types: External Interface}
\label{sec:texi}
Users utilize types in our pointwise expressions, depicted in ~\cref{grammar:pointwiseTensorType}.
For topological types, the entities that we consider (\cref{def:pos1}) leads to a simple grammar for our types.
Using this grammar, users can pick out subsets of the set of affine maps that correspond to paramterization of entities of a given topological type:
\begin{definition}[$\operatorname{Aff}, \operatorname{Conv}$]
    Let $\operatorname{Aff}$ be the set of all charts.
    The set $\operatorname{Aff}(t)$ for a topological type $t$ from our grammar is the set of charts whose domain corresponds to the given topological type $t$. These are the charts given by $(S_{0}, S_{1}, F)$ (see \cref{def:chart1}) where $S_{0}$ has topological type $t$.
    Given a chart $c=(S_{0}, S_{1}, F)\in \operatorname{Aff}(t)$, we let $\operatorname{Conv}(c)$ be the convex hull of the images under $F$ of the vertices of $S_{0}$; this is the domain that the chart parametrizes.
\end{definition}
Our system includes two context-dependent placeholders in chart types: the largest charts available on a mesh and charts of the types found on the boundary of a mesh.
These are eliminated internally. 

For field types, users specify a domain (defaults to $\TopChart$), a range (defaults to $\Rspace$), and a choice of Euclidean or tangential derivatives (defaults to $\Tangential$).
The range is a tensor space that depends only on the chart type, allowing for the use of spaces that depend on the given type (e.g., $\TangentSpace$).
Finally, a property indicates if a derivative is taken in the tangent space or a Euclidean space.\footnote{Practically, this is the difference between a field defined on $N\subset\mathbb{R}^{k}$ taking values in a tensor space $V$ having a first derivative in in $\mathcal{T}_{x}N^{\star}\otimes V$ vs $(\mathbb{R}^{k})^{\star}\otimes V$.}
This models derivatives defined multiply vs. singly on shared mesh entities/internal boundaries (Recall \cref{lst:mordofs} and \cref{sec:globfem}).

\subsubsection{Ordering of types}
\label{sec:OT}
We partially order topological types via $\Simplex(d)\leq \Simplex(d^{\prime})$ iff $d\leq d^{\prime}$. 
With just simplices, we technically have a total order, but we have an intuitive purpose for this order: one type is less than another iff an entity of the former type can always be embedded via a chart into a sub-entity of another.
If we added more general entities, such as rectangles or hexes, our intuition renders these incomparable to a triangle.
Hence, we call this a partial order.
\subsubsection{Grammar for Pointwise Tensor Expressions on Manifolds (PTEM):}
\label{sec:PTEG}
We supply a grammar for pointwise tensor expressions on manifolds in~\cref{grammar:pointwiseTensor}.
The grammar is minimal, focusing on the core geometric interface and the separation between types and arguments.
Our actual implementation for example contains a wide variety of primitives from exterior algebra, but these are easy extensions because our language is implemented via tracing\cite{frostig2019compiling} in a Python EDSL. See \cref{sec:additional} for details.

The grammar has two components that go into the pointwise tensor expression: arguments and expressions.
Arguments bind names to types and expressions extend tensor expressions with concepts that provide data from the arguments (e.g., we can get a field value from a field argument).
Our system also supports default arguments (see \cref{sec:defaultArgs}).
The expressions should be understood to modify the standard tensor expression grammar with special values, including tangent vectors, the current position $x$, derivatives of expressions with respect to $x$, and field access. 
Further, since we give access to the spaces of arguments via an extension to tensor spaces, we also give access to inner products of the the tangent space (the metric).
The pointwise tensor expressions themselves combine linear and non-linear arguments (the first and second groups), an optional output field type, and a tensor expression value that depends only on the arguments to the expression.

\begin{figure}
\footnotesize
\begin{tabular}{rcll}
\toprule
$\cv$ & & & Chart Variable \\
$\fv$ & & & Field Variable \\
$\nvar$ & & & Constant Natural \\
\midrule

$\s$  & $\cup =$ & $\TangentSpace(\cv)$     & Tangent Space \\
      &  $|$     & $\EuclideanSpace(\cv)$  & Euclidean Space\\
      &  $|$     & $\spaceOf(\e)$          & An Expression's Space\\
\midrule

$\e$  & $\cup =$ & $\TangentVector (\cv, \nvar ?)$     
                                 & Tangent Vector \\
      &  $|$     & $\CoTangentVector(\cv, \nvar ?)$  
                                 & Co-Tangent Vector\\
      &  $|$     & $\NormalVector(\cv, \cv ?, \nvar ?)$  
                                 & Normal Vector (in)\\
      &  $|$     & $\CoNormalVector(\cv, \cv ?, \nvar ?)$  
                                 & Co-Normal Vector (in)\\[2pt]

      &  $|$     & $\Xop()$  
                                 & Point $x$\\[2pt]

      &  $|$     & $\FieldAccess(\fv, \nvar ?)$  
                                 & Field at $x$\\
      &  $|$     & $\Dop(\e)$  
                                 & Derivative w.r.t. $x$\\
\midrule

$\argVar$  & $::=$ & $\cv : \chartTy \mid \fv : \fieldTy$ & \\

$\args$    & $::=$ & $(\argVar_{0}, \dots, \argVar_{N})$ & \\
\midrule

$\PTEM$  & $::=$ & 
$\Pointwise(\args_{0},\ \args_{1},\ \fieldTy ?,\ \e)$
& Pointwise Expression\\

\bottomrule
\end{tabular}

\caption[Pointwise Tensor Expression Grammar]{A grammar to describe pointwise tensor expressions on manifolds, operating on tensor fields.}
\label{grammar:pointwiseTensor}
\end{figure}

\begin{figure}
\footnotesize
\begin{tabular}{rcll}
\toprule
$\tdim$ & & Topological Dimension \\
\midrule

$\dervType$ & $::=$ & $\Euclidean$ & $\mathbb{R}^{n}$\\
            &  $|$  & $\Tangential$ & $\mathcal{T}_{x}$\\
\midrule

$\topType$  & $::=$ & $\Simplex(\tdim)$                        & Simplex \\
\midrule

$\chartTy$  & $::=$ & $\TopChart$                              & Largest Chart(s) \\
            &  $|$  & $\BoundaryChart$                         & Boundary Charts\\
            &  $|$  & $\Chart(\topType_{0})$                   & Fixed Top Type\\
\midrule

$\s$        & $\cup =$ & $\TangentSpace(\chartTy)$             & Tangent Space \\
            &  $|$     & $\EuclideanSpace(\chartTy)$           & Euclidean Space\\
\midrule

$\fieldTy$  & $::=$ & $\field(\chartTy?, \s?, \dervType?)$     & Field Type\\

\bottomrule
\end{tabular}

\caption[Pointwise Tensor Expression Types]{A grammar to express types in pointwise expressions.}
\label{grammar:pointwiseTensorType}
\end{figure}

\subsubsection{Semantics, Single Compilation, and Topological Enumeration of PTEM}
\label{sec:SCPTE}

\paragraph{Semantics of PTEM}
Semantically, PTEM define an expression parametrically over charts.
To see this and its complexities, we step through PTEM as defining a function of the incoming charts and fields, the values of \cref{sec:vii}, leading to  \eqref{eq:ptei}.
Let us consider a PTEM with $M$ chart arguments with topological types $ty_{i}$ and $N$ field arguments, each associated to another chart argument with topological type $fty_{i}$. 
There may also be an output field with type $fty_{0}$.
We see immediately that we must provide at least $M$ charts of the associated types to the PTEM.
We can't immediately supply fields because these depend on spaces that depend on more charts.
Each chart argument of a field argument parametrizes the range of the field via a tensor space of the field arguments\footnote{plus some number of derivatives via $\oplus$}, roughly as $R_{i}\colon \operatorname{Aff}(fty_{i})\to \operatorname{Vect}$\footnote{the collection of all vector spaces.} for $i=1,\dots, N$. We may also have a tensor space for the output parametric on some chart: $R_{0}\colon \operatorname{Aff}(fty_{0})\to \operatorname{Vect}$.
These define the field output spaces or the input spaces to the PTEM.
Thus, we must first supply charts for all the input fields and then we can supply fields that map from the given charts to the derived vector spaces.
Finally, we also need the point in $\mathbb{R}^{d}$.\footnote{When charts are supplied, they must all have a common geometric dimension $d$ for this to make sense.}
With all these ingredients we can provide the signature for the ways that we can interpret an IPTEM:
\begin{align}
    (c_{1},\dots, c_{M})\colon \times_{i=1}^{M}\operatorname{Aff}(ty_{i})&\mapsto (fc_{0}, fc_{1},\dots, fc_{N})\colon \times_{i=0}^{N}\operatorname{Aff}(fty_{i}) \nonumber\\ 
    &\mapsto (f_{1},\dots, f_{N})\colon\times_{i=1}^{N}C^{\infty}(\operatorname{Conv(fc_{i})}, R_{i}(fc_{i})) \nonumber\\
    &\mapsto (x)\colon \mathbb{R}^{d}\to  R_{0}(fc_{0}). \label{eq:ptei}
\end{align}
Our system can compile PTEM to a program with the above signature.
Though pragmatically,  \eqref{eq:ptei} suggests a large  breadth of what a single PTEM expresses: many calculations over many combinations of geometries, fields, and points.
Really, our compilation system compiles a PTEM to a collection of tensor expressions that represent \eqref{eq:ptei} given different combinations of a given set of chart and field values (as in \cref{lst:IptemCompilerArchitecture}).
In other words, the PTEM compiler will be given a PTEM and some charts/fields to produce tensor expressions that correspond to $(x)\colon \mathbb{R}^{d}\mapsto  R_{0}(fc_{0})$ for every reasonable way to plug the charts and fields into \cref{eq:ptei}.
Though, of course, by supplying very general chart and field (which \cref{lst:fieldInterfaceLocal} allows for), we can completely recover \eqref{eq:ptei}.

Our PTEM compiler will operate in stages based on \eqref{eq:ptei}.
For the first two groups of arguments, the charts drawn from $\operatorname{Aff}$, our compiler must resolve all the valid/useful combinations of arguments, which we do via topological enumeration (\cref{sec:TEP}).
For the next groups of arguments, our compiler has determined the geometric settings and our compiler now tries to produce nice tensor expressions, which we do via single compilation (\cref{sec:SCP}).
\paragraph{Single Compilation}
\label{sec:SCP} (\cref{lst:IptemCompilerArchitecture}:\cref{line:signEnd})
 refers to the second half of \cref{eq:ptei}: given a suitable pairing charts and fields to PTEM arguments, our system compiles a given PTEM to a tensor expression in the output space that depends on a position in space (parametrized with a reference and world space position) as well as any other variables used in chart, space, and field definitions. 
Single compilation is mostly mechanical, consisting mostly of substitutions, symbolic differentiation, change of coordinates, and other bookkeeping for the tensor compiler.
First, given that the charts supply spaces, our compiler replaces place holder tensor spaces and expressions such as tangent space with a definition depending on the chart. 
In cases where a supplied chart is larger than the type associated to a field or when a supplied field used a Euclidean (Tangent) space where a type used a Tangent (Euclidean) space, our compiler may also modify expressions with pullbacks\footnote{Technically, our definition of the pullback does not exactly achieve this purpose. Using our tensor expression language, we can recursively define a generalize pullback given base cases for certain primal and dual spaces. If our base cases use the project and embedding maps to/from (co)tangent spaces, we yield a generalize pullback suitable for correcting between tangent and Euclidean space inputs.} (\cref{def:pull5}) to correct the mismatch.
Second, similar to many expression based languages, our language can be symbolically differentiated, which our compiler performs to remove all derivatives except as directly applied to field inputs.
Third, our compiler rewrites all field expressions in terms of a query to reference and world space coordinates and then fields are queried via the field interface to replace field accesses with tensor expressions. 
At this point, our pointwise tensor expression is just a tensor expression with dependencies on a point input and other variables that defined charts and fields.
Fourth, our tensor compiler needs a basis for input spaces, output spaces, and intermediate ones; our compiler can build these with field/chart queries/metadata except for the basis of the output space. For a real valued DOF this is not a hard choice, but for more general spaces our compiler defaults to a standard basis that is orthonormal in the inner product of the space.
This can be considered part of more general bookkeeping done throughout single compilation: as our compiler lowers geometry and fields down to tensor expressions, our compiler introduces values coming from charts and fields, meaning our compiler must track the introduction of new terms and their corresponding representations in formulas and bases for the tensor compiler.

\paragraph{Topological Enumeration}
\label{sec:TEP}
Via \cref{eq:ptei}, we raise three potential problems with using the PTEM grammar.
\textbf{First}, an expression might not make sense if the domains of some functions don't overlap (e.g., $u(x)*v(x)$ requires $u$ and $v$ to share some domain.)
\textbf{Second}, as we have already noted, users should not have to explicitly pair fields and charts to PTEM, which is potentially a lot of complex information in \cref{eq:ptei}.
\textbf{Third}, we note that the pairing of fields/geometries with PTEMs in different ways is often at the core of multi-valued functions: if multiple fields or charts are possible options for a PTEM in context (a vertex sampling edge fields for example), then there may be multiple distinct values for some computation.
We address these three issues with topological enumeration (\cref{lst:IptemCompilerArchitecture}:\cref{line:teend}), which proves itself to be a useful bit of automation.

Topological enumeration finds all the "topologically valid" ways to match a PTEM's arguments to a concrete collection of charts and fields.
By valid, we mean an approximation to the rule that pointwise calculations mean that all fields should share a common domain.
We formally defined validity:

\begin{definition}[Topological Validity of PTEM Arguments]
\label{def:TVPTEA}
    Suppose we have a PTEM with $N$ field arguments and $M$ chart arguments.
    Thus we have $N$ topological types $fty_{i}$ for the fields and $M$ topological types $ty_{i}$ for the charts.
    We further suppose a topological type associated to the output $fty_{0}$.
    Suppose we assign to each $ty_{i}$ an entity of the corresponding type $S_{i}$.
    For each $fty_{i}$, we assign an entity $FS_{i}$  of type greater than or equal to $fty_{i}$ except for $i=0$ where equality must hold.
    (These entities will come from the ranges of charts.)
    Then such an assignment is valid if
    \begin{enumerate}
        \item There is an entity $FS_{I}$ such that  $\cap_{i=1}^{N} FS_{i} \supseteq FS_{I} \supseteq FS_{0}$.
        \item If $ty_{i} \leq  ty_{i^{\prime}}$, $S_{i}\subseteq S_{i^{\prime}}$.
        \item If any $ty_{i} = fty_{0}$, $S_{i} = FS_{0}$.
    \end{enumerate}
    Recall that we defined a partial order on topological types in \cref{sec:OT}.
\end{definition}
The content of this definition is that the pointwise combination of fields and geometric information is uniquely defined on a unique output domain, $FS_{0}$.
Our compiler implements topological enumeration by enumerating all combinations and keeping only those that satisfy \cref{def:TVPTEA}.
This definition resolves our first conceptual issue by default, but to see that this also resolves the latter two we consider some examples.

To make topological enumeration less abstract, we work through two examples in \cref{fig:megaTE} using both graphical and table based vocabularies.
In particular, what is remarkable is that topological enumeration reproduces key patterns in FEM codes, revealing them to be part of fundamental patterns of pointwise manipulations.
Thus by computing \cref{def:TVPTEA} to implement topological enumeration, our compiler rediscovers these patterns automatically. 

In one scenario of  \cref{fig:megaTE}, the combination of different entities of a triangle and different IPTEMs with a single triangle field reproduces quadratic Lagrange elements.
Our definition of validity allows us to reproduce all elements in \cref{figure:elemExamples} by just changing the IPTEM inputs to \cref{fig:megaTE}.
Thus, users do not need to explicitly manage a geometry and its consequences when writing down elements.
And the automation is quite simple: our compiler simply pairs the chart types with charts of the same or greater types and then filters for when the criteria of topological validity are met.

In the other scenario of  \cref{fig:megaTE}, we see that the multi-valued character of moving from edge fields to a triangle field is highlighted by the results of topological enumeration.
Our tabular representation of the output of topological enumeration for this example indicates this issue by grouping entries that have the same IPTEM and domain.
This shows the utility of this bit of automation: the compiler can detect when multi-valued quantities emerge by looking for such groupings and then the user needs to supply a resolution.
If a user uses quadratic Lagrange DOFs on an edge field, then the resulting calculation can't become the pattern in the first table of (E) in  \cref{fig:megaTE}, a proper set of quadratic Lagrange DOFs,  unless the user specified how to combine the grouped rows.
We describe how to supply a resolution in \cref{sec:IRIPTEM}.

\subsection{IPTEM: Integration and Reductions}
\label{sec:IRIPTEM}
In \cref{grammar:reduction}, we augment PTEM with integration  and reductions (to handle multiple values induced by topological enumeration though we also gain more expressive capacity in this freedom).
Both of these additions are succinct and affect a small almost post-processing change to the overall pipeline (\cref{lst:IptemCompilerArchitecture}).

\subsubsection{Integration}
\label{sec:DSII}
In \cref{grammar:reduction}, we provide integration.
Integrations are simply a chart type to integrate over.
Point evaluation is also provided under integration as a barycentric coordinates on a specific topological type.

Integration modifies the PTEM  compilation algorithm only as a post-processing step because our compiler implements all integrals via quadrature that can integrate $d$ degree polynomials exactly\footnote{For product simplices, it is a product of degrees}.
All other pre-steps remain unchanged (e.g., topological enumeration's output type is the integral type and the optional new storage chart type for the IPTEM can participate as another topological type.)

Our compiler integrates via quadrature rules.
For a given chart $c$ with a domain that we assume is numeric, a quadrature rule gives us points $x_{i}$ and weights $w_{i}$. With these, we can evaluate the integral of a formula $F$ via something like
\begin{equation*}
    \sum w_{i}|\det Dc(x_{i})| F(x_{i}, c(x_{i}),\dots).
\end{equation*}
In some cases (\cref{sec:SEFEMOS}), our compiler symbolically computes this sum by substituting quadrature points into formulas, summing them together, and simplifying with Sympy (in particular if we are computing \VA{} matrices or if all the formula's return values happen to be numeric), but for other purposes (\cref{sec:GFDC}) our compiler places this sum into generated code.
Consumers of our compiler must specify this choice (\cref{lst:IptemCompilerArchitecture}).
However, all of this requires someone to choose a quadrature degree $d$, which is potentially error prone and not ergonomic.
Ideally, users only think about the quadrature degree if there is not an obvious choice. 

\paragraph{Quadrature Degree Estimation:}
\label{sec:DSQDE}
Given a  formula acting on inputs that all have some polynomial dependence on the point $x$, our compiler can estimate the polynomial dependence of the output of the formula on the point.
To do so is fairly trivial for any formulas that are polynomial\footnote{This does require some plumbing, which leads to polynomial degrees being attached to field inputs so this information can be attached to inputs to formulas.}.
For non-polynomial functions, our compiler relies on one important simplification or just fails: for $0$ degree dependence, non-linear functions still have $0$ degree dependence.
This simplification means that formulas for all operators and DOFs can determine quadrature degrees automatically on affine geometries.
When quadrature estimation fails, users at a higher level on the interface must specify the polynomial dependence of some inputs or simply specify the quadrature degree.
We note that this analysis can be interpreted via a simple lattice \cite{cousot1977abstract}: $\mathbb{N}\cup\{\infty\}$.

\subsubsection{Reductions}
\label{sec:Red}
In \cref{grammar:reduction}, we also provide reductions, which our compiler applies to the topological enumeration of IPTEMs.
In particular, reductions should be considered as reductions on the tabular representations of the topological enumeration of IPTEMs (see \cref{fig:megaTE}). 
Later, they will also apply to the global FEM problem (see \cref{sec:oogo})

Reductions are specified via two mechanisms.
First, we need a reduction operation, such as arbitrary choice, average, or sum.
Second, we need to pick columns to reduce over, which we can do so by selecting arguments to a pointwise expression and then sub-selecting charts/fields and local/global.
For local, these sub-selections occur on the output of topological enumeration.
By sub-selecting to reduce over charts (fields), we reduce together rows (interpretations) that are the same up to the chart (field) used in a given augment.
These sub-selections allow for surprisingly flexibility, such as a DOF defined on the entire boundary of a cell via reducing over boundary integration charts.
Our compiler implements these at the local level through simple group-by operations on the output of topological enumeration (\cref{lst:IptemCompilerArchitecture}:\cref{line:redStart}).

The integration mechanism chooses the default reduction.
If a point evaluation is used, we reduce over fields via a \textbf{choiceAllEq}, which will warn us if a point evaluation is ill-defined due to multiple and non-equal choices at runtime.
Thus, topological enumeration and reduction provide a flexible solution to the problem of partially defined finite element interpolations (see \cref{def:interp}).
If an integration is used, we use sum reductions over fields.
These create the desired results for most DOFs (typically point evaluations) and operators (typically integrals).
In this work, we only need to overriding system defaults when we specify integral FEEC DOFs (see \cref{sec:feecres})

\begin{figure}
\footnotesize
\begin{tabular}{rcll}
\toprule
$\bc$ & & & Barycentric Coord.\\
\midrule

$\integral$  
    & $=$ 
    & $\PointEval(\topType,\bc)$     
    & Point Evaluation\\
    & $|$  
    & $\IntegralOp(\chartTy)$  
    & Integral\\
\midrule

$\op$  
    & $ =$ 
    & $\MinOp \mid \MaxOp \mid \SumOp$     
    &  \\
    & $\mid$  
    & $\Difference \mid \Avg $ & \\
    & $\mid $ & $\choice \mid \choiceAllEq$  
    &  \\
\midrule

$\overVar$  
    & $ =$ 
    & $\LocalChart \mid \LocalField$     
    &  \\
    & $|$  
    & $\GlobalChart \mid \GlobalField$ 
    & \\
    & $|$ 
    & $\ChartOp \mid \Basis \mid \ALL$ 
    & \\
\midrule

$\reduction$ 
    & $::=$ 
    & $\Argument(\nvar,\ \op,\ \overVar)$ 
    & \\
    
$\reductions$ 
    & $::=$ 
    & $(\reduction_{0},\dots)$ 
    & \\
    
$\Storage$ 
    & $::=$ 
    & $\Default \mid \chartTy$ 
    & \\
\midrule

$\IPTEM$  
    & $::=$ 
    & $\IPTX(\PTEM,\ \integral,\ \reductions,\ \Storage)$
    & \\
\bottomrule
\end{tabular}

\caption[Integrated Pointwise Tensor Expression Grammar]{A grammar to describe integrated pointwise tensor expressions with reductions.}
\label{grammar:reduction}
\end{figure}

\subsection{Miscellaneous Aspects}
For clarity of discussion, we have left out three important aspects of our system.
However, since these are actually quite critical for the implementation, we have included them in appendices.
These are permutations of mesh entities (\cref{sec:permutations}), default arguments to IPTEM (\cref{sec:defaultArgs}), and additional syntax/library content for IPTEMs (\cref{sec:additional}).
Though these issues first emerge in this section, these aspects actually impact content in \cref{sec:SRFE} up to \cref{sec:eval}.
We organize them in terms of how they impact other sections and suggest consulting them if further understanding is required.

\section{Software Representation of Finite Elements}
\label{sec:SRFE}

We can now provide a software specification of finite elements via the machinery of the previous section.
Further, we solve the problem posed by \cref{sec:probmain}, the core obstacle to automatically using finite elements.
We provide our specification in \cref{sec:SEFEMOS} and solve \cref{sec:probmain} in \cref{sec:SFIVFG}.
Finally, in \cref{sec:SFPS}, we detail our specification of the function spaces in finite elements, as these do not emerge as easily as other components from our pre-existing machinery.

\subsection{Overall Specification} 
\label{sec:SEFEMOS}
Our software definition (\cref{lst:dofinput}) of an FE naturally mirrors \cref{def:finiteElement}, but our definition is carefully constructed to balance expressing many elements against the automatic generation of formulas, specifically via the problem specified in~\cref{sec:probmain}.

\begin{lstlisting}[language=PythonPlus, caption={Interface for constructing finite elements. The topological type represents the possible domain sets (e.g., triangles). The polynomial space represents polynomial spaces paramterized by the domain. The list of IPTEM represents the degrees of freedom, also paramterized by the domain. Note that the value space is in the polynomial space.}, label={{lst:dofinput}},abovecaptionskip=5pt]
class FiniteElement:
    V: TopType # Domain
    P: PSpace # Polynomial Space
    Sigma: Sequence[IPTEM] # Dofs
    
\end{lstlisting}

Our construct contains two familiar objects that represent classes of objects parametrized by a domain: a topological type for the domain and a list of IPTEMs for the dual basis functions.
While our infrastructure for IPTEM handles many of the problems of representing many dual basis functions and generating formulas for them on symbolic domains\footnote{In particular, we can supply charts with symbolic variables in their definitions when we compile IPTEMs}, we must resolve similar problems for spaces of functions.
In particular, two function spaces defined similarly on two domains are unlikely to be isomorphic via pullback, which makes computing the \VA{} matrix impossible or symbolically messy. 
To understand our representation of classes of polynomial function spaces for FEs (\lstinline[language=PythonPlus]!PSpace!), we first detail a software interface and algorithms for using the finite element, solving~\cref{sec:probmain}.
Then we can discuss the requirements on \lstinline[language=PythonPlus]!PSpace! in  \cref{sec:SFPS}.
To just solve \cref{sec:probmain}, we leave the polynomial space to an interface for now.

\subsubsection{PSpace Interface}
\label{sec:SFEPS}
In~\cref{lst:spaceInterface}, we provide our interface to polynomial spaces parametrized by domains.
At the core, the interface is simple and similar to that for fields (\cref{lst:fieldInterfaceLocal}): on a given domain, a given set of points, and a given basis, our compiler can sample some unknown basis of functions and its derivatives.
The main purpose of this interface is to provide fields as arguments to our compiler (\cref{lst:IptemCompilerArchitecture}: \cref{line:argField}), which is why we also include a maximum polynomial degree.

\begin{lstlisting}[language=PythonPlus, caption={Interface that local spaces, PSpaces, must provide.},label={{lst:spaceInterface}},abovecaptionskip=5pt]
class PSpace:
    def tabulateChart(self, chart: Chart, derivatives: int, refPoints: NDArray, basis: Optional[Basis] = None) -> Tuple[Formula, Basis]: ...
    def valueSpace(self, chart: Chart) -> Space: ...
    def dim(self) -> int: ...
    def maxDegree(self) -> int: ...
    def asFields(self, chart: Chart) -> List[Fields]: ...
    def validChartTypes(self) -> List[ChartType]:...
\end{lstlisting}

\subsection{Inverse \VA{} Formula Generation}
\label{sec:SFIVFG}

We now solve \cref{sec:probmain} up to the details of the space representation.
We depict the overall algorithm in~\cref{lst:vanderOverall}.
The algorithm consumes an FE, a reference chart, and a symbolic world space domain given by a chart.
 The algorithm conducts some validity checks (\cref{sec:SFVAL}), checking that the symbolic chart maps from the range of the reference chart and that the FE type checks.
 Then the algorithm constructs a \VA{} matrix (\cref{sec:SFVA}) and uses this to construct basis functions that can be used on the symbolic chart, exploiting that \lstinline!VRef! is numerical.
 Then using the \VA{} algorithm again, the algorithm can construct a symbolic \VA{} matrix from the symbolic chart and the reference function basis functions pulled back to the symbolic world space.
 Then, a symbolic inversion routine (\cref{sec:SFSI}) exploits the expected structure of this matrix to produce a formula representing the inverse.
 The symbolic output matrix provides the symbolic Vandermonde matrix problem (\cref{sec:probmain}), expressing the FE basis functions on the symbolic domain in terms of the reference domain.
 We prove the correctness of this approach up to an assumption on the space (\cref{sec:SFCOR}). \Cref{sec:SFVM} summarizes the matrix structure; \cref{sec:AAFET} gives the dual-basis-cover factorization that separates reference sampling, coordinate changes, and physical DOF weights. This suggests an alternative algorithm that factorizes the \VA{} matrix using DOF blocks corresponding to mesh entities, ordered according to the dependencies between those blocks. The factorization explains a guaranteed part of the matrix's sparsity, while symbolic construction and simplification can reveal further zeros.

\begin{lstlisting}[language=PythonPlus, escapechar=|, caption={Overall algorithm for Vandermonde inverse calculation.}, label={{lst:vanderOverall}},abovecaptionskip=5pt]
def vandermondeInverse(dofs: FiniteElement, refChart: Chart, symbolicChart: Chart) -> Formula:
    checkValid(refChart, symbolicChart, dofs)
    refVander = vandermondeAlgo(dofs.Sigma, refChart, dofs.space.asFields(refChart)) |\label{lst:line:va1}|
    VRef = refVander.asNumericArray()
    assert VRef.dtype == np.float64
    vanderInv = np.linalg.inv(VRef)
    assert vanderInv.shape == (dofs.P.dim(), dofs.P.dim())
    refFieldsOnSymbolic = vanderInv.T.dot(dofs.space.asFields(symbolicChart)) |\label{lst:line:va1f}|
    symbolicVander = vandermondeAlgo(dofs.Sigma, symbolicChart, refFieldsOnSymbolic) |\label{lst:line:va2}|
    symbolicVanderInv = symbolicInverse(symbolicVander) |\label{lst:line:sic}|
    return symbolicVanderInv
    
\end{lstlisting}

\subsubsection{Validity}
\label{sec:SFVAL}
For a finite element, \lstinline[language=PythonPlus]!E = FiniteElement(V, P, Sigma)! to be valid, the DOFs must consume elements of \lstinline[language=PythonPlus]!E.P.valueSpace()! and must not use any chart types larger than the chart type \lstinline[language=PythonPlus]!E.V!.
Similarly, the chart type  \lstinline[language=PythonPlus]!E.V! must be in \lstinline[language=PythonPlus]!E.P.validChartTypes()!.
And naturally the DOFs must all be linear in a single non-default argument.
The function \lstinline[language=PythonPlus]!checkValid! checks all of this and further checks that the two chart inputs are compatible: they must both map between domains of type \lstinline[language=PythonPlus]!E.V! and the range of the reference space chart should be the domain of the symbolic chart.

\subsubsection{\VA{} Algorithm}
\label{sec:SFVA}
The \VA{} Algorithm is the compilation procedure for a list of IPTEM.
The function \lstinline[language=PythonPlus]!vandermondeAlgo(dofs, chart, fields)! runs our IPTEM compiler (\cref{lst:IptemCompilerArchitecture}) with the following arguments: the IPTEMs are the DOFs, the charts are \lstinline!chart! plus charts for every sub-entity of the chart's world space, the fields are the fields of the polynomial space, and all quadrature sums are symbolically simplified.
We note that our construction of DOFs on a given polynomial space emerges naturally out of our compilation procedure (see \cref{fig:megaTE}).
The result will be a matrix that is the number of topologically enumerated DOFs\footnote{Note that the actual DOFs are defined via topological enumeration.} by the dimension of the polynomial space.

\subsubsection{Symbolic Inversion}
\label{sec:SFSI}
Our second \VA{} matrix is symbolic, meaning the formula returns an array of rational functions, potentially depending on other statements.
However, we expect that that this matrix will be roughly block triangular.
We will recursively exploit the Schur complement to invert such a matrix:
\begin{equation}
    \begin{pmatrix}
    A & 0 \\
    C & D
    \end{pmatrix}^{-1} = \begin{pmatrix}
        A^{-1} & 0\\
        -D^{-1}CA^{-1} & D^{-1}
    \end{pmatrix}.
\end{equation}

A sketch of the algorithm is depicted in~\cref{lst:symVan}.
The algorithm directly inverts small matrices.
The algorithm tries to find a choice for $A$ and $C$ that maximizes the size of the zero block.
Then the algorithm recurses on the blocks and recombines them with a triple matrix multiply.
We isolate this as a separate function as we want to control symbolic complexity here.
Our compiler introduces intermediates for non-zeros in the triple matrix multiplication if a rational function entry has a denominator with degree larger than one.
Additionally, when performing the triple matrix multiply or the base case inversion, we use Sympy to attempt to cancel fractions and detect structurally zero values as much as possible.
The actual implementation features two other tricks: (1) we check for diagonal matrices and (2) we have an environment variable to determine when to time out and use a direct inverse (we do not need this for any of our results).
We include this latter trick for the case where an element does not yield a reasonable symbolic \VA{} matrix so we can fall back to a horribly inefficient option - numerically inverting the \VA{} $\mathsf{C}$ matrix for each cell, with a warning to tell the users that this is going to happen. 
Inverting such \VA{}-like matrices per cell is an established practice for some methods where this is the case - such as in polygonal FEMs or virtual element methods, though it comes with performance and numerical problems that typically lie outside the realm of standard FEM~\cite{abriefreviewonpolygonalpolyhedralfiniteelementmethods2018,thevirtualelementmethodin50linesofmatlab2016}.

\begin{lstlisting}[float, language=PythonPlus, caption={Symbolic Inverse Algorithm },label={{lst:symVan}},abovecaptionskip=5pt]
def symbolicInverse(V: Formula) -> Formula:
    (mat, stmts) = V
    N = mat.shape[0]
    if N <= 6: # largest reasonable symbolic inverse
        (matInv, stmtsp) = directSymbolicInverse(mat, stmts)
        # Depending on the complexity of rational functions, we might introduce intermediate variables
        return Formula(matInv, stmtsp)
    else:
        (blockP, blockQ) = choosePQ(mat)
        A = mat[0:blockP, :][:, 0:blockQ]
        B = mat[0:blockP, :][:, blockQ:]
        
        D = mat[blockP:, :][:, blockQ:]
        C = mat[blockP:, :][:, 0:blockQ]
        assert allSymbolicZero(B)
        (Ainv, AinvStmts) = symbolicInverse(Formula(A))
        (Dinv, DinvStmts) = symbolicInverse(Formula(D))
        (schur, schurStmts) = tripleMatMul(Dinv ,C, Ainv)
        matInv = bmat([[Ainv, np.zeros(blockP, blockQ)], [-schur, Dinv]])
        return Formula(matInv, stmts + AinvStmts + DinvStmts + schurStmts)
    
\end{lstlisting}
\subsubsection{Correctness}
\label{sec:SFCOR}
We now establish what the inverse Vandermonde algorithm in \cref{lst:vanderOverall} computes and the conditions on the function spaces under which it is correct.
We need some intermediate results relating pullbacks (\cref{def:pull5}) and primal dual basis pairs (\cref{def:pdbdag}).

\begin{proposition}
\label[proposition]{prop:pre}
Let $E := (K, V, P, \Sigma)$ and $E^{\prime}= (K^{\prime}, V^{\prime}, P^{\prime}, \Sigma^{\prime})$  be finite elements where $V$ and $V^{\prime}$ are
subspaces of $\mathcal{T}(K)^{(p,q)}$ and $\mathcal{T}(K^{\prime})^{(p,q)}$ respectively. 
Let $T\in C^{1}(K, K^{\prime})$ have an inverse in $C^{1}(K^{\prime}, K)$.
The pullback and pushforward maps satisfy the inverse identities:
\begin{enumerate}
    
    \item $T_{\star}\circ T_{-\star} = Id$ on dual functionals over $K^{\prime}$,
    \item $T^{\star}\circ T^{-\star} = Id$ on fields over $K$.
    
\end{enumerate}
Further, if $(T^{-\star}){|_{P}}$ is an isomorphism onto $P^{\prime}$, then
\begin{enumerate}
    \item $T_{\star}{\mid_{ P^{\star}}}$ is an isomorphism onto $(P^{\prime})^{\star}$.
    \item For any $\phi\in P^{\star}$ and any $f\in P$, 
    \begin{equation}
    \label{eq:dr}
        T_{\star}(\phi)((T^{-\star})(f)) = \phi(f).
    \end{equation}
    This says the pushforward is analogous to an inverse transpose of a pullback.
    \item Given any primal-dual basis pair $(\{f_{a}\}, \{\phi_{b}\})$ for $P$ (i.e., $\phi_{b}(f_{a}) = \delta_{ab}$), we have that $(\{T^{-\star}f_{a}\}, \{T_{\star}\phi_{b}\})$ is a primal-dual basis pair for $P^{\prime}$.
    \item Similarly, for any primal-dual basis pair $(\{f^{\prime}_{a}\}, \{\phi^{\prime}_{b}\})$ for $P^{\prime}$ (i.e., $\phi^{\prime}_{b}(f^{\prime}_{a}) = \delta_{ab}$), we have that $(\{T^{\star}f^{\prime}_{a}\}, \{T_{-\star}\phi^{\prime}_{b}\})$ is a primal-dual basis pair for $P$.
\end{enumerate}
\end{proposition}
\begin{proof}
The field pullbacks along $T$ and $T^{-1}$ are inverse. By definition of pushforward,
\begin{equation*}
    (T_{\star}\phi)(T^{-\star}f)
      =\phi(T^{\star}T^{-\star}f)=\phi(f).
\end{equation*}
The same calculation with $T$ and $T^{-1}$ exchanged gives the inverse identity on dual functionals. Under the stated isomorphism of polynomial spaces, these maps restrict to inverse maps on their dual spaces. Applying the displayed pairing identity to each primal-dual basis pair preserves the primal-dual basis relations, proving the remaining assertions.
\end{proof}

Our core result, \cref{thm:pb1}, is our ability to solve \cref{sec:probmain}.
This result closely mirrors our algorithm and provides the main condition on the spaces.
As we prove it, we will indicate how steps of the proof refer back to \cref{lst:vanderOverall}.

\begin{theorem}
\label{thm:main}
    Assume the conditions of \cref{prop:pre}.
    Let $\Sigma = \{\phi_{b}\}$ and $\Sigma^{\prime} = \{\phi_{b}^{\prime}\}$.
    Let $\{f_{b}\}$ and $\{f_{b}^{\prime}\}$ be the associated primal bases. 
    Further, suppose 
 \begin{enumerate}
     \item  $T^{-\star}_{|P}$ is an isomorphism onto $P^{\prime}$.
     \item There exist points $x_{b,q}\in K$ and tensor valued rational functions $w_{b,q}$ such that for any $f^{\prime}\in P^{\prime}$,
     \begin{equation*}
         \phi^{\prime}_{b}(f^{\prime}) = \sum_{q} w_{b,q}(x_{b,q}, T(x_{b,q}), DT(x_{b,q}))((T^{\star}f^{\prime})(x_{b,q}), (T^{\star}Df^{\prime})(x_{b,q}), \dots)).
     \end{equation*}
     This usually boils down to the existence of a quadrature rule for integrating polynomial functions on $K$. Further, we wish to stress that this condition just aims to model rational dependency on geometry in the sort of DOFs that we see and could be varied quite a bit.
 \end{enumerate}

 Then there exists a matrix $\mathsf{P}$ of shape $|\Sigma^{\prime}|\times |\Sigma|$ (physical basis rows, reference basis columns) such that
 \begin{enumerate}
     \item $\mathsf{P}_{b^{\prime}b}$ is a rational function in the values $\{DT(x_{a,q}), x_{a,q}, T(x_{a,q})\}$ over all DOF indices $a$ and quadrature indices $q$,
     \item $\mathsf{P}$ transforms the mapped reference basis into the physical nodal basis, or equivalently
     \begin{equation}
     \label{eq:good}
          (T^{\star})(f^{\prime}_{b^{\prime}}) = \sum_{b} \mathsf{P}_{b^{\prime}b}f_{b}.
     \end{equation}
     Recall that this is \cref{eq:formelempre} in \cref{sec:probmain}.
 \end{enumerate}
    \label{thm:pb1}
\end{theorem}
\begin{proof}
By the assumed isomorphism, a computable basis for $P^{\prime}$ is the image of the reference primal basis under $T^{-\star}$.
We call this basis $f_{b}^{T} := (T^{-\star})(f_{b})$.
(We construct $f_{b}$ via Line \ref{lst:line:va1} and $f_{b}^{T}$ via Line \ref{lst:line:va1f}).
Because of \cref{prop:pre}, we also have a corresponding dual basis $\phi_{b}^{T} := T_{\star}(\phi_{b})$.
Evaluate physical DOFs on this mapped reference basis to form a generalized \VA{} matrix, with physical DOFs indexing rows and reference basis functions indexing columns:
\begin{equation}
\label{eq:transformVander}
    \mathsf{C}_{b^{\prime}b}
      := (T_{-\star}\phi_{b^{\prime}}^{\prime})(f_b)
       = \phi_{b^{\prime}}^{\prime}(f_b^T).
\end{equation}
Equivalently, $\phi_{b^{\prime}}^{\prime}=\sum_b\mathsf{C}_{b^{\prime}b}\phi_b^T$ on $P^{\prime}$.

\Cref{lst:vanderOverall}, \cref{lst:line:va2}, computes this matrix.

Writing $f_i^{\prime}=\sum_j\mathsf{P}_{ij}f_j^T$ and applying $\phi_k^{\prime}$ gives
\begin{equation*}
    \delta_{ki}=\sum_j\mathsf{C}_{kj}\mathsf{P}_{ij}
       =(\mathsf{C}\mathsf{P}^{T})_{ki}.
\end{equation*}
Both sets are bases of the same space, so $\mathsf{C}$ is square and invertible. Therefore
\begin{equation}
\label{eq:erg}
    \mathsf{P} = \mathsf{C}^{-T}.
\end{equation}
Thus, the inverse transpose combines mapped reference basis functions into physical basis functions.
(\Cref{lst:vanderOverall}:\cref{lst:line:sic} returns $\mathsf{C}^{-1}$; its columns contain the coefficients of individual physical basis functions. Transposing it gives the row convention used for $\mathsf{P}$.)
Applying $T^{\star}$ to the basis identity and using $T^{\star}f_b^T=f_b$ proves~\cref{eq:good}. In coefficient-vector convention, $T^{\star}$ is represented by $\mathsf{P}^T=\mathsf{C}^{-1}$ and $T^{-\star}$ by $\mathsf{C}$.

By the definition of pullback, we observe that for $y=T(x)$,  $f_{b^{\prime}}^{T}(y)$ is a polynomial in the coefficients $DT(x)$, $y$, and $DT^{-1}(y)$.
By the conditions of the theorem, $\phi_{b}^{\prime}$ will evaluate $f_{b^{\prime}}^{T}$ at points $\{T(x_{bq})\}$ and will weight each value by a rational function in $(x_{b,q})$, $T(x_{b,q})$, and $DT(x_{b,q})$.
Thus, each entry of $\mathsf{C}$ is a rational function $\{x_{b,q}, DT(x_{b,q}),  T(x_{b,q})\}$. Since the inverse of a matrix of rational functions is a matrix of rational functions, this completes the proof.
\end{proof}
The core extra step is that $T^{-\star}|_{P}$ is an isomorphism onto $P^{\prime}$.
We can weaken this if we are willing to weaken the results to just \cref{eq:good}: we just need an isomorphism from $T^{-\star}(P)$  to $P^{\prime}$, but the relationship of this isomorphism to $T$ might cost us rational function like-formulas and our precise relationship to the pullback.
Ideally, $T^{-\star}(P) = P^{\prime}$, but if not we only need some correction as a  function of $T$ that might make our formulas more complex and our code more expensive and ugly.
Thus, our representation of function spaces must provide a way to work with these two possibilities.
We discuss this in \cref{sec:SFPS} after two additional matters.

\paragraph{Relation to prior work}
We note that a key difference between our methods here and those employed by hand in~\cite{ageneralapproachtotransformingfiniteelements2018} is that our methods operate on $P(K^{\prime}, V^{\prime})^{\star}$ rather than $C^{k}(K^{\prime}, V^{\prime})^{\star}$; in~\cite{ageneralapproachtotransformingfiniteelements2018}, certain definitions\footnote{Mainly compatible nodal completion, which is a stronger and harder to construct facility than our dual basis cover (\cref{def:dualBasisCover} in~\cref{sec:AAFET}).} to facilitate manipulations of DOFs by hand arise out of concerns about DOFs' behaviors as objects in  $C^{k}(K^{\prime}, V^{\prime})^{\star}$ whereas we avoid this at the cost of having to use $f_{b}$ explicitly to manipulate objects in $P(K^{\prime}, V^{\prime})^{\star}$.
In particular, we note that the Morley DOFs on a physical element cannot be expressed as linear combinations of push-forwarded DOFs if we consider them as objects in $C^{k}(K^{\prime}, V^{\prime})^{\star}$, but this is the case for the Morley DOFs as objects in $P(K^{\prime}, V^{\prime})^{\star}$.
The approach in \cite{ageneralapproachtotransformingfiniteelements2018} is easier for hand calculation, but harder for automation. 
\paragraph{Structure of \VA{} Matrices}
\label{sec:SFVM}
\Cref{sec:AAFET} proves $\mathsf{C}=HFG$ (\cref{thm:pb2}) and explains the contribution of each factor: $H$ combines physical derivative samples into DOFs, $F$ applies local coordinate changes, and $G$ tabulates reference basis derivatives. For corresponding affine covers, $F$ is block diagonal by point and derivative order, while $G$ can be precomputed on the fixed reference element.
The factors expose which reference DOFs can contribute to each physical DOF. Their dependency pattern gives a useful lower bound on the sparsity of $\mathsf{C}$ by certifying zeros and explaining their origin. Running the construction with symbolic simplification can reveal a more precise pattern, including cancellations between $H$ and $F$. This pattern can suggest an ordering for the recursive inversion routine.
The alternative algorithm in~\cref{sec:AAFET} constructs $\mathsf{C}$ from these factors and then solves for $\mathsf{P}^{T}$. It separates reusable reference data from symbolic geometry calculations. Its limitation is that constructing $F$ independently of $H$ does not exploit how the DOFs' contractions can simplify the pullbacks. Our implemented construction compiles the complete DOF expressions using the coordinate-independent tensor and basis machinery in~\cref{sec:TEBA,sec:TEFG}, also developed in our Grassmann work~\cite{collin2026grassmann}. This machinery can avoid unnecessary basis conversions and simplify pullbacks in the context of the contractions that use them; in the factorization, these simplifications exploit the relationship between $H$ and $F$.

\subsection{Specification of Polynomial Spaces}
\label{sec:SFPS}
Our system must allow users to specify polynomial spaces such that the following two requirements are easy to satisfy automatically:
\begin{enumerate}
    \item provide a tabulate function as in the interface depicted in \cref{lst:spaceInterface}, ideally providing simple output formula,
    \item ensure that two polynomial spaces $P$ and $P^{\prime}$ instantiated at two different isomorphic domains  $K$ and $K^{\prime}$ are isomorphic under pullback or that there is some known mapping from $T^{-\star}(P)$ to $P^{\prime}$ dependent on $T$, allowing \cref{thm:main}.
\end{enumerate}
However, as system designers, we have a third requirement: our polynomial grammar must be expressive enough to capture a wide variety of polynomial spaces.
We had to balance the first and third against the second.

\begin{figure}
\footnotesize
\begin{tabular}{rcll}
\toprule
$\dvar$ & & & Polynomial degree\\
\midrule

$\IPTEMS$ 
    & $::=$ 
    & $(\IPTEM_{0},\dots)$ 
    & IPTEM List\\
\midrule

$\degVar$ 
    & $::=$ 
    & $\Deg(\dvar)$ 
    & Pure degree\\
    & $::=$ 
    & $\RangeOp(\degVar_{0},\degVar_{1})$ 
    & Range of degrees\\
\midrule

$\Psp$  
    & $::=$ 
    & $\Pure(\topType,\ \degVar,\ \s)$     
    & Tensor Space Polynomials\\

    & $|$ 
    & $\SumOp(\Psp_{0},\dots)$ 
    & Direct Sum \\

    & $|$ 
    & $\KernelOp(\Psp_{0},\IPTEMS)$ 
    & Kernel\\

    & $|$ 
    & $\IKernel(\IPTEMS,\ \Psp_{0})$ 
    & Invariant Kernel\\

    & $|$ 
    & $\BKernel(\IPTEMS_{0},\ \IPTEMS_{1},\ \Psp_{0})$ 
    & Kernel in a given basis\\

\bottomrule
\end{tabular}

\caption[Polynomial Space Grammar]{A grammar to describe collections of polynomial spaces, indexed by charts.}
\label{grammar:pspace}
\end{figure}

We provide a grammar for polynomial spaces in ~\cref{grammar:pspace}.
Polynomial degrees come in two\footnote{If we added hexes or quads, we would add a third type: products of degrees.} kinds: up to and including a given degree and a range of degrees.
Degrees, along with spaces that depend on a given $\topType$ (like in $\fieldTy$), parametrize tensor valued polynomial spaces, denoted with $\Pure$.
Naturally, we can define the direct sum of polynomial spaces, but we gain a great deal more expressiveness by allowing users to specify polynomial spaces as kernels of DOFs.
We provide three options to vary the pain of computing an isomorphism from $T^{-\star}(P)$ to $P^{\prime}$ in the case of kernels.
\begin{enumerate}
    \item a kernel of a collection of DOFs, requiring the kernel to be computed via an SVD or QR in the code,
    \item a kernel of a collection of DOFs that we promise is invariant, as a hack,
    \item an effective intermediate (due to the second use of compatible nodal completions in ~\cite{ageneralapproachtotransformingfiniteelements2018}) wherein a second super space $P_{0}$ is supplied along with a corresponding dual basis $\operatorname{IPTEMS}_{1}\supset \operatorname{IPTEMS}_{0}$; the idea is that our compiler can transform the space with the element defined with $P_{0}$ and $\operatorname{IPTEMS}_{0}$ and then extract the space $\cap_{\phi\in(\operatorname{IPTEMS}_{1}-\operatorname{IPTEMS}_{0})}\ker\phi$.

\end{enumerate}

Lastly, we note that the range degree construct is actually not affine invariant and is transformed via the middle case, as we mainly intend this as a useful hack.
Our compiler can implement tabulate by recursion on the grammar and we can inductively argue that our second property is satisfied through this interface.
For both issues, we handle the base case, $\Pure$ polynomial spaces of a given degree, via orthogonal polynomials, as in~\cite{algorithm839fiatanewparadigmforcomputingfiniteelementbasisfunctions2004}.

\section{Global FEM Data and Computation}
\label{sec:GFDC}

Our global interface provides a wide variety of FEM computations parametrically over elements.
However, as we saw in \cref{sec:examlples}, not all combinations of PDEs/elements result in reasonable solutions so we can't preserve a user's intention to solve a given PDE if they change an element.
At a more basic level, a truly parametric interface must deal with intentions, such as interpolation (\cref{def:interp}), where changing the element may lead to undefined quantities.
Moreover, many much more basic computations make no sense when users change elements: a manipulation of an interpolation matrix produced by a quadratic Lagrange element might make no sense for a linear Lagrange element.
To achieve an expressive parametric interface, we seek to ensure a more fundamental property: if a users write a computation using just components of the system, then the computation is always discretely meaningful even if they might not be meaningful in a continuous sense.
We believe this insurance will at least promote truly parametric FEM code.
In contrast, the state of the art simply either completely fails on or silently allows potentially undefined quantities.\footnote{e.g., Firedrake as of writing does not allow interpolation into spaces such as Hermite or Argyris even when it might be well defined.}

To achieve an expressive global parametric interface, we limit our interface to a small collection of consistently meaningful discrete objects and operations with a highly structured automated connection to elements.
In particular, we replace the standard easy to abuse outputs of FEM computations, sparse and dense tensors, with our novel structured output, a \textbf{single valued assembled object (SVAO)}.
SVAOs summarize many types of computation (e.g., boundary integrals, interior integrals, integrals, interpolation, boundary identification), shrinking the number of frontend objects.
In particular, users create SVAOs by interpreting a collection of IPTEM on a mesh given inputs such as FEM bases:
\lstinline!assembly(mesh: Mesh, iptems: List[IPTEM], linearArgs: List[MeshField | MeshFields], nonLinearArgs : dict[str, MeshField | MeshFields]) -> SVAO:!
Users only use elements to either construct SVAOs through assembly or to extract components from SVAOs, limiting the interface between elements and global code to code that our system automatically generates for each element on demand.
Thus, we build an interface that is parametric in the sense that users don't have to write new code if an element is changed for the old code to run and be meaningful in a discrete sense, though it might not achieve the desired PDE solution.
Further, our interface will use few operations and objects, and, our interface will resolve problems around multi-valued computation, such as in \cref{def:interp}.

In the following, we will develop the machinery to define SVAOs and assembly (\cref{sec:SVAOOGO}).
To define assembly in \cref{sec:definesvao}, we will heavily leverage our semantics for IPTEMs (\cref{sec:SCPTE}), especially topological enumeration (\cref{sec:TEP}) as assembly can be considered as a sort of mesh wide topological enumeration. 
We limit an implementation discussion to ensuring a parametric interface (\cref{sec:IAGF}), leaving many details to an appendix   (\cref{sec:iasa}).
Finally, in \cref{sec:oogo}, we will discuss other well defined discrete operations that we combine with SVAO and \lstinline!assembly! to express FEM computations in a parametric consistently discretely meaningful manner, leading to an overall workflow fully described in \cref{sec:OUW}. 

\subsection{SVAOs and other global objects}
\label{sec:SVAOOGO}
To define SVAOs and their assembly, we need to build the other objects in our system first.
Precise definitions of these objects (meshes/sections in \cref{sec:meshmath} global field(s) in \cref{sec:fieldbasis}, and IPTEMs in \cref{sec:globiptem}) help us ensure that SVAOs are always discretely meaningful.
We describe everything with succinct mathematical expressions that are not too far from the implementations.

\subsubsection{Mesh and Data On Meshes}
\label{sec:meshmath}
Mathematically, our global meshes are posets with a regularity condition that makes them meshes of entities.
\begin{definition}[Mesh]
\label{def:meshtop}
    Let $V\subset P(\mathbb{R}^{n})$ be the vertices of a mesh topology.
    Let $F\subset P(V)$ be a poset under $\subseteq$, called the set of faces. Here, \emph{face} includes entities of every dimension, such as vertices, edges, triangles, and tetrahedra; it is not restricted to two-dimensional entities.
    Then $(V, F)$ is a \textbf{mesh} if 
    \begin{enumerate}
        \item For every $v\in V$, $\{v\}\in F$,
        \item and if for every $f\in F$, $F_{f}:=\{f^{\prime}\in F\colon f^{\prime}\subseteq f\}$, a sub-poset of $P(V)$, is isomorphic as a poset to a standard entity (\cref{def:pos1}),
        \item The sets\footnote{we use convex hull to refer to the geometric entities; some additional machinery could remove this, but it is simple enough here.} $\{\operatorname{Conv}(f)\colon f\in F, \not\exists f^{\prime} f\subset f^{\prime}\}$ have disjoint interiors. 
    \end{enumerate}
\end{definition}
A triangle mesh is a running example: the definition makes its vertices and edges explicit alongside its triangles. The same definition applies to simplicial meshes in other dimensions.
We leave discussion of permutations/orientations of mesh entities to \cref{sec:permutations}.

Data on meshes must be laid out or associated with mesh entities.
Sections, like meshes, have a regularity condition, which we model by requiring a schema for storing data on meshes.
Users do not need to write these as they can be inferred from IPTEM (see \cref{sec:globiptem}).
Note, in this definition,  we rely on topological types and their order as in \cref{sec:OT}:

\begin{definition}[$\mathcal{T}$, Schema for data on a  mesh]
\label{def:storageSchema}
Let $L$ be an abstract set of labels (e.g. IPTEMs).
Let $\mathcal{T}$ be the collection of all topology types (\cref{grammar:pointwiseTensorType}).
A function $f\colon L\to\mathcal{T}^{2}$ is a \textbf{schema} for a mesh if for all $l\in L$ then $(t_{1}, t_{2}) = f(l)$ satisfies $t_{1}\leq t_{2}$.
\end{definition}
We can interpret schema as placing data on the edges of the mesh $\subset$ relationship.
We consider two examples of schema.
For linear Lagrange,  our schema might map the single point evaluation to vertices: $L_{0}=\{\operatorname{vertexeval}\}$, $ f_{0}(\operatorname{vertexeval}) = (\operatorname{Simplex}(0),\operatorname{Simplex}(0))$.
The two entries describe the entity carrying the datum and the containing entity in whose context it is defined. For vertex evaluation these are the same vertex, so both entries are zero-dimensional.
We might store the normal to an edge within a triangle with $L_{1}=\{\operatorname{edgenormal}\}$, $f_{1}(\operatorname{edgenormal}) = (\operatorname{Simplex}(1),\operatorname{Simplex}(2))$ because edge normals might not agree between two triangles that share an edge.
Sections simply describe data on a mesh that satisfies a given schema.
\begin{definition}[$\mathrm{Ty}$, Section]
\label{def:storageSection}
Let $\mathcal{T}$ be the set of topology types and let $L$ be an abstract set of labels.
Further, let $f\colon L\to\mathcal{T}^{2}$ be a schema.
If $(V, F)$ is a mesh, let $\mathrm{Ty}\colon F\to\mathcal{T}$ be a mapping from faces to the topology types corresponding to the standard entities a given face is isomorphic to.
A \textbf{section} $S_{V,F, f}$ is the set $S_{V,F, f} := \{(l, h, g)\colon l\in L, h\in F, g\in F, h\subset g, (t_{1}, t_{2})= (f(l)),  \mathrm{Ty}(h) = t_{1}, \mathrm{Ty}(g) =t_{2} \}$.
\end{definition}
Returning to our previous examples, $S_{V, F, f_{0}}$ is isomorphic to the set of linear Lagrange DOFs with $|S_{V, F, f_{0}}| = |V|$.
Similarly,  $|S_{V, F, f_{1}}| = 3 * |T|$ where $T$ is the set of triangles. 
A bijection from a section $S_{V,F,f}$ to $\mathbb{N}$ would be a numbering for DOFs.

\subsubsection{Global Fields}
\label{sec:fieldbasis}
Here, a \emph{field} is a function assigning scalar or tensor values over mesh entities, not an algebraic field such as $\mathbb{R}$ or $\mathbb{C}$.
Since our fields are often functions in finite element spaces or otherwise multi-valued across mesh entities (\cref{def:bgfes}), we must use care to model this as a discrete object on a mesh.
This starts with modeling vector spaces broken across a mesh.
\begin{definition}[Mesh Vector Space]
\label{def:MVS}
    Given a topological type $T$ and a mesh $(V, F)$, a \textbf{mesh vector space} is a sequence of vector spaces $R_{t}$ for all $t\in F$ with $\mathrm{Ty}(t)=T$.
\end{definition}
We note that \cref{def:MVS} is a more discrete notion of a fiber bundle (see \cite{introductiontosmoothmanifolds2012}) in differential geometry.
The key example of this is the tangent space defined per triangle on a discrete mesh, but we could consider many examples implied by \cref{grammar:pointwiseTensor} and \cref{grammar:tensorExpr1}.
We can now define our notion of fields:

\begin{definition}[Mesh Field]
\label{def:MF}
    Suppose $(V,F)$ is a mesh and $T$ is a topology type.
    Suppose we have a corresponding mesh vector space $R_{t}$.
    A \textbf{mesh field} is a collection of $\mathcal{C}^{\infty}$ functions, $h_{t}\colon\operatorname{Conv}(t)\to R_{t}$.
\end{definition}
A mesh field can be thought of as a discrete analogue to a function from a manifold to a bundle. 
A mesh field could model a function in a global finite element space, but it could also model a more generic object that we can sample when restricted to entities of a mesh (as in \cref{def:rof}).
For example, a mesh field could model a single linear Lagrange basis function or its gradient, which is discontinuous across triangles but $\mathcal{C}^{\infty}$ within them. 
Moreover, \cref{def:MF} accounts for situations such as a Python function (for example $(x,y)\mapsto xy$) defining a smooth function on the entire mesh.

While a field represents a single function or vector in a FEM like space, we will sometimes provide an entire basis or collection of fields on a mesh to build operators/SVAO.
We model this with the following definition.

\begin{definition}[Sectioned Mesh Field Sequence]
\label{def:SMFS}
    Suppose $(V,F)$ is a mesh, $f\colon L\to \mathcal{T}^{2}$ is a schema,  and $S_{V,F, f}$ is a section.
    Suppose $T$ is a topology type greater than or equal to any in $f(L)$.
    Suppose we have a corresponding mesh vector space $R_{t}$.
    A \textbf{sectioned mesh field sequence} is a sequence of $\mathcal{C}^{\infty}$ functions $g_{l,h,g,t}\colon \operatorname{Conv}(t)\to R_{t}$ indexed by $(l,h, g, t)\in L\times F^{3}$ with $(l,h, g,)\in S_{S, F, f}$, $\mathrm{Ty}(t)=T$, and $g\subseteq t$.
\end{definition}
We can think of these as a more computational counterpart to \cref{def:bgfes}.
Combining elements over $t$ reproduces our construction of global FEM spaces (\cref{def:gfes}) from local FEM spaces (\cref{def:bgfes}).
In this way, integration and storage topology types associated to an IPTEM (\cref{grammar:reduction}) implicitly  and automatically define the equivalence of DOFs used to make a FEM space (\cref{def:gfes}).
Returning to our linear Lagrange example, the restrictions of the basis functions to each incident triangle defines a section mesh field sequence.
If we had the linear Lagrange basis function $\phi_{v}$ for every vertex in the mesh, then $g_{\operatorname{vertexeval}, h, g, t} = \phi_{v}|_{t}$ for $t$ a triangle ($T=\operatorname{Simplex}(2)$) with $h=g=v\in t$.

\subsubsection{IPTEM as Global Computations}
\label{sec:globiptem}
We now reuse IPTEM as a specification. 
Following \eqref{eq:ptei} in \cref{sec:SCPTE}, we interpret IPTEM as operator parametrized by suitable mesh entities.
\begin{definition}[IPTEM on meshes]
\label{def:IPTEMONM}
Suppose we have an IPTEM with a  topological type $T_{I}$ for an integration domain, a topological type $T_{O}$ for output with $T_{I} \leq T_{O}$, $N$ topological types for field arguments, $fty_{i}$ and $M$ for chart arguments, $ty_{i}$
Suppose for $1\leq i \leq N$, we have a collection of mesh spaces, $R_{fty_{i}, i}$.
Suppose we have a finite dimensional vector space $V$.
An IPTEM is a function 
\begin{align*}
    I(c_{1},\dots, c_{M},f_{O}, f_{I}, f_{1},\dots, f_{N} ) \colon &\\\times_{i=1}^{M}\operatorname{Aff}(ty_{i})\times \operatorname{Aff}(T_{I})\times\operatorname{Aff}(T_{O})\times_{i=1}^{N+1}\operatorname{Aff}(fty_{i})
    \to&\\ \mathcal{F}(\times_{i=1}^{N}\mathcal{C}^{\infty}(\operatorname{Conv}(f_{i}), R_{f_{i}, i}), V)\\
\end{align*}

that takes the form
\begin{align*}
   (c_{1},\dots, c_{M},f_{O}, f_{I}, f_{1},\dots, f_{N} )& \mapsto (g_{1},\dots, g_{N})\\ \mapsto  &\int_{\operatorname{Conv}(f_{I})}F_{c_{1},\dots, c_{M},f_{O}, f_{1},\dots, f_{N}}(g_{0},\dots, g_{N})dx
\end{align*}
for some suitably typed $F$.
To understand this, we recall our discussion of \eqref{eq:ptei}: given some geometries, we take functions defined on paramterizations of various geometries and compute an integral on one of these geometries. Further, the integrand depends pointwise on the functions, the paramterizations, and derivatives of functions.
\end{definition}

This definition models the compilation of an IPTEM on a particular set of mesh entities modulo local reductions (see \cref{grammar:reduction} and \cref{grammar:pointwiseTensor}).
We observe that many good examples for this, such as the linear Lagrange DOF or Laplace operator, can be drawn from \cref{sec:examlples}.
Assembly mirrors topological enumeration (\cref{sec:TEP}) as assembly essentially interprets \cref{def:IPTEMONM} wherever possible just as we did for \cref{eq:ptei} in \cref{sec:SCPTE}.

\subsubsection{SVAO and Global Assembly}
\label{sec:definesvao}
We now have all the ingredients to define SVAOs, semantically, and as a sparse array.
Moreover, we argue that SVAOs are necessarily discretely meaningful, emerging from  well-defined discrete IPTEM operators and bases/global fields that are known to be well behaved in specific combinations of mesh entities.
In particular, mesh fields/bases combined with IPTEM allow us to easily see where computations are not multi-valued.
Thus, to produce all the possible well defined computations from an IPTEM (or a collection of IPTEMs with the compatible field arguments) and some input fields/bases, we simply pair IPTEMs with the data from fields in all possible ways that are well defined and not zero.

We now show how to do this mathematically to define assembly and a SVAO.
Our system implements these two definitions, so most users need to understand their meaning but do not need the implementation details; extending or debugging the system may require those details.
Suppose we have some IPTEMs $IS = \{I_{0},\dots, I_{Q}\}$. This allows us to define a schema $f_{IS}$ from their integration and storage topological types.
Each of these has $N$ field arguments and $M$ chart arguments\footnote{for simplicity, we don't represent this here, but this can actually be allowed to vary.}
We supply a mesh $(V, F)$.
Finally we provide $N$ input Section Mesh Field Sequences\footnote{the more general of the two options}, $g_{1},\dots, g_{N}$, with schema $f_{k}$, topological types $T_{k}$, sections $S_{V, F, f_{k}}$ for $k=1,\dots, N$.
We now have the ingredients for a call to assembly to produce a SVAO.
A SVAO is a sparse tensor with dimensions for
\begin{enumerate}
    \item The output section $S_{V, F, f_{IS}}$.
    \item For each $i=1,\dots, M$, a dimension for $F$ corresponding to a chart argument.
    \item For each $i=1,\dots,N$, a pair of dimensions for $S_{V, F, f_{k}}$ and $F$, representing a basis function and the domain from which the basis function was sampled. For mesh fields, we only include $F$.
\end{enumerate}
The SVAO stores all valid, in the sense of \cref{def:TVPTEA}, non-zero combinations of fields specified by the inputs, entities specified by the mesh, and IPTEMs.
Our assembly routine produces a SVAO from the call:
\begin{equation}
    S = \operatorname{assembly}((V, F), \{I_{0},\dots, I_{Q}\}, g_{1},\dots, g_{N})
\end{equation}
where
\begin{align}
    S_{(l_{0}, h_{0}, g_{0}), c_{1},\dots, c_{M}, ((l_{1}, h_{1}, g_{1}), t_{1}), \dots,  ((l_{N}, h_{N}, g_{N}), t_{N})} &\nonumber\\=  l_{0}(c_{1},\dots, c_{M},g_{1, ((l_{1}, h_{1}, g_{1}), t_{1} )},\dots, g_{N, ((l_{N}, h_{N}, g_{N}), t_{N} )})\label{eq:comp}\\\nonumber
\end{align}
if $g_{k}\subset t_{k}$ and $\mathrm{T}(t_{k})=T_{k}$ for all $k$ , $(l_{0}, h_{0}, g_{0})\in S_{V, F, f_{IS}}$, and $(l_{k}, h_{k}, g_{k})\in S_{V, F, f_{k}}$  for all $k$, and $( c_{1},\dots, c_{M}, h_{0}, g_{0}, t_{1},\dots, t_{N})$ are a topologically valid assignment (\cref{def:TVPTEA}) for $l_{0}$, which is one of the IPTEM.
We observe that each entry of this array is well defined/singly valued: the IPTEM is a single valued function on a given region and the input fields have been restricted to some region where they are $\mathcal{C}^{\infty}$.
On the other hand, we also compute all possible entries that could be well defined in this sense by using the validity definition: for each domain we must integrate on, we consider all possible ways we restrict fields to produce structurally non-zero well-defined values.
For our mesh field or section mesh fields on our conforming meshes, this is always the basis functions with supports that could intersect on the integration domain.

We consider an example to explain this.
Suppose we had an IPTEM for the Laplace operator, $l_{\Delta}$, as in \cref{lst:simplestExampleListing}.
Our schema would be  $f_{\Delta}(l_{\Delta}) = (\operatorname{Simplex(2)},\operatorname{Simplex(2)})$, which our system would automatically infer from the IPTEM code.
The section $S_{V, F, f_{\Delta}}$, which would be computed as part of assembly, would represent a Laplace operator per triangle. 
Let $g_{1}=g_{2}$ be section mesh field sequences corresponding to linear Lagrange basis functions, which are both sampled on triangles. 
Users would specify this with an element from \cref{sec:SEFEMOS}.
Then our SAVO assembly can be written
\begin{align*}
    S_{(l_{\Delta}, t_{0}, t_{0}), ((\operatorname{vertexeval}, v, v), t_{0}), ((\operatorname{vertexeval}, v^{\prime}, v^{\prime}), t_{0})} \\= l_{\Delta}(g_{\operatorname{vertexeval}, v,v,t_{0}}, g_{\operatorname{vertexeval}, v^{\prime},v^{\prime},t_{0}})
\end{align*}
for all triangles $t_{0}$ and vertices $v,v^{\prime}\in t_{0}$.
This is a typical abstraction\footnote{the reduction across triangles is accomplished via an interface in \cref{sec:oogo}} of the cotangent Laplacian assembly algorithm with just a constant factor more book keeping.
SVAO assembly arrives at this from the general definition in \eqref{eq:comp} via two elements of the definition.
First, via \cref{def:TVPTEA}, we sample the triangle defined basis functions on the same triangle where they are integrated.
Second, we get that $v,v^{\prime}\in t_{0}$ due to the definition of section mesh field sequences.

Reproducing assembly of a FEM matrix is not the only point though; our definition also reproduces operations such as interpolation (\cref{def:interp}), operations to mark boundaries, and more.
We recall interpolation in particular though because we recall that it is only partially defined due to multi-valued computation.
Our definition of assembly resolves this problem by computing all possible single-valued versions of a computation and then allowing users to resolve the problem in a way of their choosing via SVAO operations described in \cref{sec:oogo}.

\subsection{Implementation Aspects}
\label{sec:IAGF}
We describe code generation for assembly in \cref{sec:iasa}.
In particular, we describe interfaces to objects such as sectioned mesh field sequences that allow us to generate code.
Here, we wish to note two things to emphasize the parametric aspects of our system.

First, the design of SVAO and the implementation make plugging elements into assembly easy.
In particular, in \cref{sec:elemEasy}, we note that the infrastructure of the previous sections does most of the heavy lifting.
For example, schema to define global element numbering emerge naturally from our definition of IPTEMs with integration and storage topological types.
In our implementation, users need not think about the difference between an element and a global basis as our assembly routine recognizes the element and automatically does the wrapping via our infrastructure.
And our wrapping does the job of defining a sort of global tabulation function for a section mesh field sequence (\cref{lst:fieldInterface1}) that knows to generate code via \VA{} matrix inversion as in \cref{sec:SFIVFG}.

Second, our implementation heavily uses the infrastructure of \cref{sec:swrdoIPTEM} just as our implementation of elements does.
Both directly call \cref{lst:IptemCompilerArchitecture} for the local code.
We patch this local code through symbolic charts whose formula stand in for data extracted from mesh data structures at runtime.
See \cref{sec:AGC} for more information.

\subsection{Operations on Global Objects}
\label{sec:oogo}
With SVAO, we can now discuss the entire global interface.
In particular, we can now introduce a small number of additional operations that allow us to chain together assembly operations to express many FEM operations. 
The overall workflow is discussed in \cref{sec:OUW}.
All the objects and operations described here are discretely well defined so users can use them with SVAO to make discretely well defined pipelines.
All operations work discretely on SVAOs in a coherent easy to understand/NumPy-like manner.

We do need two more objects, sub-sections and sub-meshes, but these can be understood essentially as masks on the original objects.\footnote{though a sub-mesh must be a sub-poset so not any mask is valid, which is why internally we need a subset of a mesh that can be completed to a proper sub-mesh.}
With these, we have all our objects: mesh, IPTEM, SVAO, mesh field, and sectioned mesh field sequence, section, sub-mesh, sub-sections.
Our operations are four NumPy-like operations (reduction, slice, set-slice, and where), and seven simple specific operations (propagation, submersion, asMask, asInput, asDenseVector, asScipyMatrix, and asScalar).
Signatures for all operations are provided (in \cref{sec:appDGO}), but we describe the two categories roughly below.
Our eight objects and eleven operations constitute our user interface modulo all the tools to write IPTEM.
If users stick to this interface, code will remain discretely well defined as elements change; however, users can also just access underlying arrays if they really need to.

Our NumPy-like operations all utilize the idea that SVAO is still an array, but whose dimensions and entries have specific meaning.
(These operations are precisely enumerated in \cref{sec:GAAO}.)
By keeping the meanings of the entries and dimensions in mind, we believe this interface forces code that computes well defined quantities.
For example, to slice a SVAO, users must pair dimensions that are sections with sub-sections and dimensions that are mesh facets with sub-meshes.
We used these in our initial example, \cref{lst:simplestExampleListing}, to implement boundary conditions.
Dual to this, the where operation behaves like-NumPy where but the output object will be a sub-section for section dimensions, sub-meshes for facet dimensions, or even IPTEM for labels.
The where operation can be used to implement more sophisticated methods for identifying boundary entities or boundary DOFs.
The where operation combined with assembly is critical to extracting boundary DOFs to implement elimination methods as illustrated in \cref{sec:bebcerg}.
Lastly, reductions behave just like NumPy-reductions, but they can be specialized per IPTEM label of the output section, allowing complex default global reductions to be applied in many cases.
For reductions in particular, we note that the defaults inherited from IPTEM are nice as they allow users to not think about the many dimensions of SVAO if they just need to reduce away all dimensions but two section dimensions to get an operator as a matrix or vector or scalar.

Besides these NumPy operations, we provide a few more specific operations.
Four of these are simple: we offer the ability to use default reductions to attempt to convert a SVAO to a mesh field, dense vector, scipy sparse matrix, or scalar.
In other words, a short cut to apply a reduction and extract the data.
The defaults for global reductions are encoded into IPTEM (See \cref{grammar:reduction}).
For sub-sections and sub-meshes, we have three other operations starting with using the sub sets directly as a mask for array programming.
 Sub-meshes can always be propagated to a sub-section by simply finding all section elements associated to a sub-mesh (e.g., as in \cref{lst:simplestExampleListing}:\cref{line:prop}) while sub-sections can be submersed back into the mesh if a sub-section is an injection into the mesh.
 These two methods also support more robust boundary condition elimination as in \cref{sec:bebcerg}.
Precise definition for sub-objects and operations between them are provided in \cref{sec:subs}.

\
\section{Evaluation}
\label{sec:eval}
To evaluate our system, we construct several applications of interest to computer graphics where we can utilize a variety of elements: a biharmonic smoothing energy, a Hodge Decomposition, and a Stokes flow. 
For each application and relevant element, we validate the application via convergence plots and we also supply a simplistic analysis of some of the trade-offs between elements.
Thus, these applications demonstrate the potential utility of changing elements to computer graphics applications.
In some cases, these results are known, though they have never been achieved in a truly automated fashion and are not widely available to the graphics community.

To evaluate how our system achieves the above, we conduct three experiments to show that:
\begin{enumerate}
    \item Our system can specify FEs concisely compared to other systems, in large part due to our automatic capabilities.
    \item Our system can specify FEs in a malleable manner: the code required to implement new elements given the infrastructure of old elements is small, so the code to vary elements is small.
    \item Our system's automation does not cost us when it comes to the FE transformation theory - the code generated by our system require roughly the same work as compared to the transformation theory deduced by hand.
\end{enumerate}

Through the variety of applications, we hope to demonstrate that our system is sufficiently expressive to specify most elements that are available in the world.
Finally, we note that all of our applications provide evidence for the claim that our system is parametric over the definition of the element.
With the exception of the mixed method for the biharmonic applications, we did not need to change our program to change to another valid element for any of these applications.
For the biharmonic application, we did explore some changes to the weak formulation for the sake of exploring trade-offs, but we did not need to do so.
In other systems, this would not be the case due to the need to develop the element or the need to work out how interpolation or boundary conditions work for different elements.

\subsection{Applications}
\label{sec:evalAPPs}
\subsubsection{Biharmonic Energies}
\label{sec:bihmres}
\paragraph{Problem Setup}
Fourth-order problems are a central motivator for the development of scalar elements beyond the Lagrange element.
The model fourth-order problem is the biharmonic equation:
\begin{equation}
\label{eq:bihm2}
    \Delta^{2}u = f.
\end{equation}
The biharmonic equation emerges in many contexts, and we study several of them representing geometry processing and simulation concerns.
Our contexts include a variety of BCs, including Dirichlet conditions (also sometimes referred to as Dirichlet \& Neumann):
\begin{equation}
\label{eq:bc1}
    u |_{\partial\Omega} = h \mbox{ and } \frac{\partial u}{\partial n} = g
\end{equation}
for some given scalar functions $h,g$.
In simulation contexts, these are sometimes known as clamped plate BCs.
Simulation contexts also utilize a form of BC given by
\begin{equation}
\label{eq:bc2}
    \frac{\partial u}{\partial n} = g \mbox{ and } \frac{\partial \Delta u}{\partial n} = h.
\end{equation}
However, we also study the natural BCs, which occur in smoothing applications~\cite{naturalboundaryconditionsforsmoothingingeometryprocessing2018,asmoothnessenergywithoutboundarydistortionforcurvedsurfaces2020}:
\begin{equation}
\label{eq:bc3}
    n H_{u}n = 0 \mbox{ and } \frac{\partial\Delta u}{\partial n } + \frac{\partial u}{\partial n tn} = 0
\end{equation}
where $t$ is a vector tangent to the boundary.
In smoothing applications, we also see interpolation constraints:
\begin{equation}
\label{eq:bc4}
    u(x_{i}) = f_{i}
\end{equation}
where $x_{i}\in\Omega$ (typically vertices of the mesh) and $f_{i}\in\mathbb{R}$.

Finally, any of these BCs could be combined with a biharmonic eigenvalue problem:
\begin{equation}
    \Delta^{2}u = \lambda u.
\end{equation}
For example, the solution to the eigenvalue problem under the Dirichlet conditions provides the vibrational modes of a surface clamped at its boundaries. 

\paragraph{Operators}

Using our system, we implement the following operators to represent the biharmonic equation.
We use a Hessian-Hessian inner product for the equation
\begin{equation}
\label{eq:bi1}
    a_{H}(u,v) := \int_{\Omega} (H_{u}, H_{v}).
\end{equation}

Naturally, for the right hand side or the eigenvalue problem, we need a mass operator:
\begin{equation}
\label{eq:bi2}
    m(u,v) = \int_{\Omega}uv.
\end{equation}

To implement various BCs variationally (Neumann or Dirichlet via Nitsche's method~\cite{constructingnitschesmethodforvariationalproblems2024}), we utilize terms from the following operator:
\begin{equation}
\label{eq:bi3}
    a_{B}(u,v) := \int_{\partial\Omega} v \frac{\partial\Delta u}{\partial n} + \int_{\partial\Omega} \frac{\partial v}{\partial n}\Delta u + \int_{\partial\Omega} uv + \int_{\partial\Omega} \frac{\partial u}{\partial n}\frac{\partial v}{\partial n}.
\end{equation}

For natural BCs in smoothing, we do not need operators to enforce them. 
To enforce point constraints, we need linear operators that correspond to point evaluation at a fixed set of points $\{x_{i}\}_{i=1}^{N}$:
\begin{equation}
    P(u) = (u(x_{1}),\dots, u(x_{N}))\in\mathbb{R}^{N}.
\end{equation}
Also, to validate the BCs numerically\footnote{As a minor note, we used this to provide the first numerical validation of the BC offered in \cite{naturalboundaryconditionsforsmoothingingeometryprocessing2018}}, it is useful to have the following non-linear operator:
\begin{equation}
    e_{B}(u) = \int_{\partial\Omega} (nH_{u}n)^{2} + (\frac{\partial\Delta u}{\partial n } + \frac{\partial u}{\partial n tn})^{2}.
\end{equation}
Similarly, to check if an element is valid to use for a biharmonic problem on a given mesh (some unconditionally work, but others do not), it is useful to consider the patch test operator that operates on the interior edges $\cup e_{i} \subset \Omega$ with the notation that $K_{0, i}$ and $K_{1,i}$ are the two cells that include $e_{i}$:
\begin{equation}
    t(u,v) = \sum_{i} \int_{e_{i}} (\nabla u|_{K_{0,i}} - \nabla u|_{K_{1,i}})\cdot v.
\end{equation}
If $v$ is drawn from discontinuous piecewise linear functions defined the set of edges, then $t\equiv 0$ when $u$ is drawn from a given global finite element space is a necessary condition to use an element to solve a biharmonic problem.

Since we wish to compare to prior work, mainly~\cite{naturalboundaryconditionsforsmoothingingeometryprocessing2018,amixedfiniteelementmethodwithpiecewiselinearelementsforthebiharmonicequationonsurfaces2019}, we need operators to implement a mixed method for the biharmonic equation.
These come from the Laplace equation, except for smoothing, where we need a few additional operators, including an inner product on matrix-valued functions that represent Hessians and an integration of the divergence of a matrix against the gradient of a scalar function:
\begin{equation}
    a_{\operatorname{Div},\nabla}(\Lambda, u) = \int_{\Omega}\nabla\cdot\Lambda \cdot \nabla u.
\end{equation}

\paragraph{Elements}

We use the following elements on this problem: affine Lagrange, Morley, Hermite, Bell, Argyris.
For the purpose of later analysis, we provide code.
The Lagrange element in our system is:
\begin{lstlisting}[language=PythonPlus, caption={Affine Lagrange DOFs}\label{lst:llpexample},abovecaptionskip=5pt]
cty0, sty2 = ChartType(0), SimplexType(2)
@pointwiseEval(cty0, barycentricCoordinate=0.0)
def vertexEval(u: field(cty0, R)):
    return u(0)
LagrangeTri = EF.FE(2, Pure(sty2,1), [vertexEval])
\end{lstlisting}

The Morley element uses quadratic polynomials, the affine Lagrange DOFs, and evaluations of the gradient normal at midpoints. 
We note that the normal direction is computed with a global Hodge star on Euclidean space to ensure the value is consistent on edges examined from different cells:
\begin{lstlisting}[language=PythonPlus, caption={Morley DOFs}\label{lst:eval-mordofs},abovecaptionskip=5pt]
cty1 = ChartType(1) # On objects of dimension 1
@pointwiseEval(cty1, barycentricCoordinate=(0.5, 0.5)
def morleyDof(c: cty1, u: field(cty1, R, euclideanDerivative=True)):
    return hodge(u(1)).dot(c.E(c.tangentVector(0)))
Morley = EF.FE(2, Pure(sty2,2), [vertexEval, morleyDof])
\end{lstlisting}

Hermite DOFs are cubic and use Lagrange DOFs plus edge gradients and a center evaluation.
\begin{lstlisting}[language=PythonPlus, caption={Hermite DOFs}\label{lst:hermitedofs1},abovecaptionskip=5pt]
fld = field(cty0, R,  euclideanDerivative=True)
@pointwiseEval(cty0, barycentricCoordinate=0.0)
def xDervDof(u: fld):
    return u(1)[0]
@pointwiseEval(cty0, barycentricCoordinate=0.0)
def yDervDof(u: fld):
    return u(1)[1]
cty2 = ChartType(2)
@pointwiseEval(cty2, barycentricCoordinate=(1/3, 1/3, 1/3))
def centerDof(u: field(cty2, R)):
    return u(0)
HermiteDofs = EF.FE(2, Pure(sty2, 2), [vertexEval, xDervDof, yDervDof, centerDof])
\end{lstlisting}

The Argyris DOFs utilize the Hermite gradient DOFs plus the Morley DOFs and three additional Hessian DOFs per vertex:

\begin{lstlisting}[language=PythonPlus, caption={Argyris DOFs}\label{lst:argydofs},abovecaptionskip=5pt]
fld = field(cty0, R, euclideanDerivative=True)
@pointwiseEval(cty0, barycentricCoordinate=0.0)
def xxDervDof(u: fld):
    return u(2)[0,0]
@pointwiseEval(cty0, barycentricCoordinate=0.0)
def yyDervDof(u: fld):
    return u(2)[1,1]
@pointwiseEval(cty0, barycentricCoordinate=0.0)
def xyDervDof(u: fld):
    return u(2)[0,1]

ArgyrisDof = EF.FE(2, Pure(sty2, 5), [vertexEval, xDervDof, yDervDof, morleyDof, xxDervDof, yyDervDof, xyDervDof])
\end{lstlisting}

Finally, the Bell DOFs use the Argyris DOFs except the Morley DOFs.
Instead, the Bell space consists in quintic polynomials whose normal gradients are cubic polynomials on each edges.
To enforce this, we add Morley-like DOF that use the Legendre DOFs to the space:
\begin{lstlisting}[language=PythonPlus, caption={Bell DOFs}\label{lst:belldofs},abovecaptionskip=5pt]
@pointwiseIntegral(cty1)
def bellSpaceDof(c: cty1, 
                u: field(cty1, R, euclideanDerivative=True), 
                v: field(cty1, R) = legendreDofs(5).basisFunction(4)):
    return u(1).dot(c.normalVector(0)) * v(0)
dofsList = [vertexEval, xDervDof, yDervDof, xxDervDof, yyDervDof, xyDervDof]
BellDofs = EF.FE(2, BKernel(Pure(sty, 5), dofsList, dofsList + [bellSpaceDof],), dofsList)
\end{lstlisting}

The Legendre DOFs are just L2 duals of the monomials:
\begin{lstlisting}[language=PythonPlus, caption={Legendre DOFs}\label{lst:legdofs},abovecaptionskip=5pt]
def legendreDofs(deg: int):
    dofs = []
    for d in range(deg + 1):
        @pointwiseIntegral(cty1)
        def legendreDof(c:cty1, u: field(cty1, R)):
            return (c.x**d) * u(0)
        dofs.append(legendreDof)
    return EF.FE(1, Pure(Simplex(1), deg), dofs)
\end{lstlisting}

\begin{figure}
    \centering
    \includegraphics[width=1\linewidth]{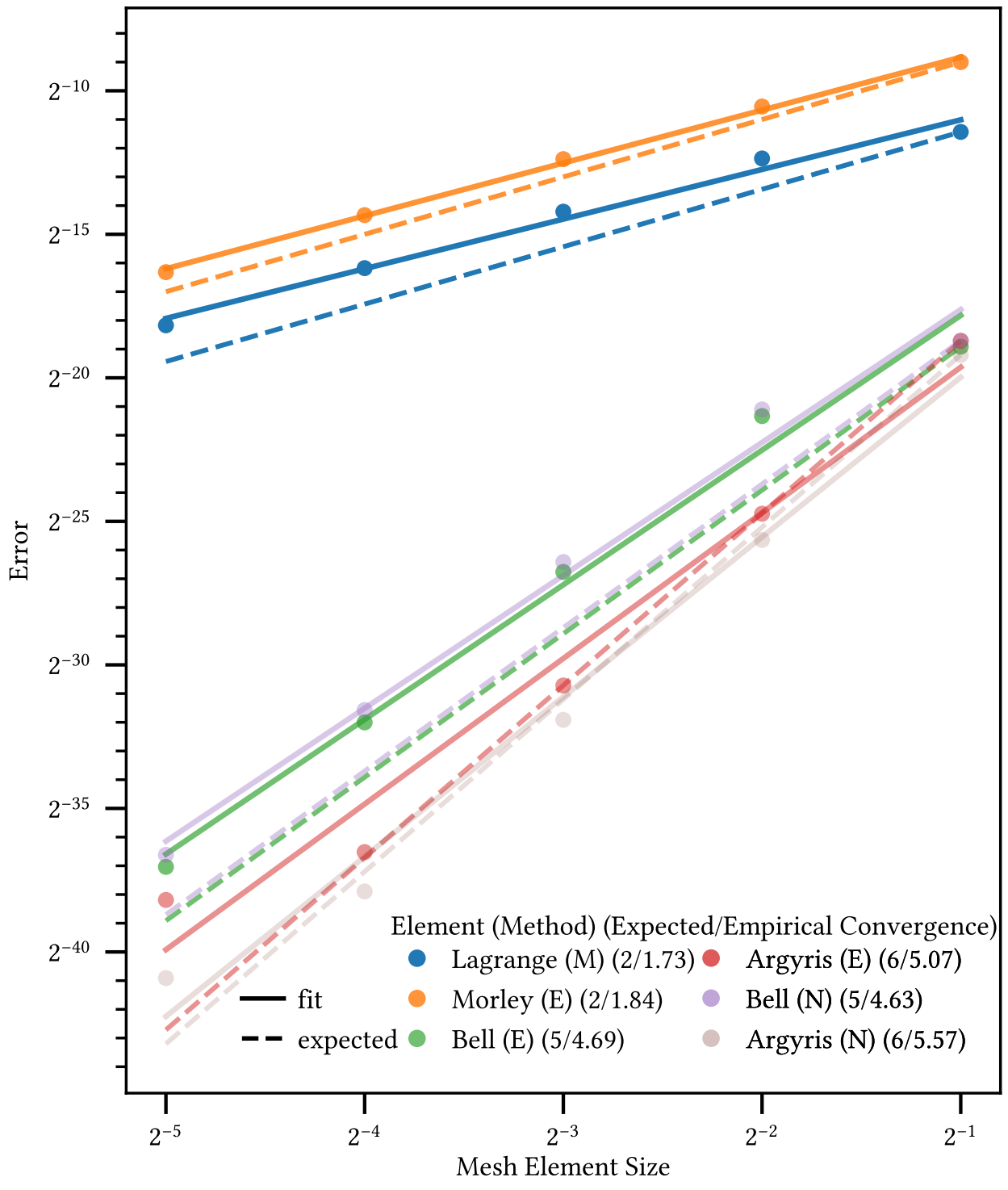}
    \caption[Convergence Data For Biharmonic Problems]{Convergence data for \cref{fig:covVisBiharmonic}. We enforce boundary conditions via either mixed methods, elimination methods, or Nitsche's method. Observe that we come close to the expected convergence rates in all cases.  For higher order elements, achieving exact convergence rates is difficult due to already low errors at initial sizes. Additionally, the parameter choices in Nitsche's method can impact observed convergence rates. }. 
    \label{fig:covPlotBiharmonic}
\end{figure}

\begin{figure}
    \centering
    \includegraphics[width=1\linewidth]{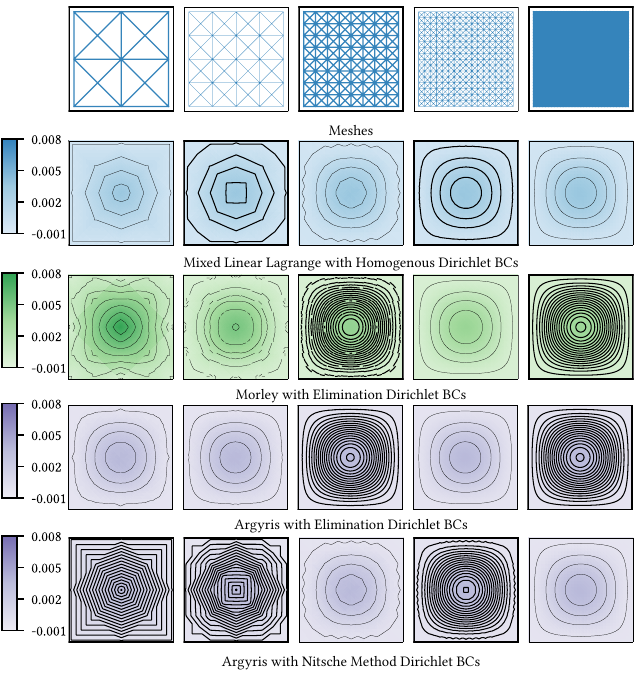}
    \caption[Solutions to Biharmonic Problems]{Solutions to the Biharmonic equation on the square with different elements, meshes, and BC enforcement methods.
    All are visualized on refined meshes according to polynomial degree.
    See \cref{fig:covPlotBiharmonic} for convergence data.
    These visually validate our convergence results and indicate trade offs between the elements.
    }
    \label{fig:covVisBiharmonic}
\end{figure}

\paragraph{Validation}
To validate our system, we provide a convergence study on a model problem: the biharmonic equation with homogeneous Dirichlet BCs on the unit square.
We use the analytical solution $u(x,y) = (x(x-1)y(y-1))^{2}$.
We compared the mixed method from~\cite{naturalboundaryconditionsforsmoothingingeometryprocessing2018,amixedfiniteelementmethodwithpiecewiselinearelementsforthebiharmonicequationonsurfaces2019} to an implementation using the operators in ~\eqref{eq:bi1} and ~\eqref{eq:bi2} with boundaries enforced by querying our system for the nodes corresponding to the Dirichlet BCs.
This amounts to a few lower level calls made automatically in the system to the last two terms of the operator in \eqref{eq:bi3}  (similar to \cref{lst:dgllisting}).
By comparing to other literature~\cite{c1finiteelementsfordune}, we have validated that the DOFs selected for the BCs are correct.
Since our particular variation of the Argyris and Bell elements only supports BCs on certain types of meshes (compare to \cite{optimalmultileveladaptivefemfortheargyriselement2022,essentialboundaryconditionswithstraightc1finiteelementsincurveddomains2010}), we also provide convergence data for a Nitsche method formulation~\cite{constructingnitschesmethodforvariationalproblems2024} that utilizes all of the terms in \eqref{eq:bi3}.
The results are depicted in \cref{fig:covVisBiharmonic} and \cref{fig:covPlotBiharmonic}.
The results validate the system, as convergence is in line with theoretical predictions.

\begin{figure}
    \centering
    \includegraphics[width=1\linewidth]{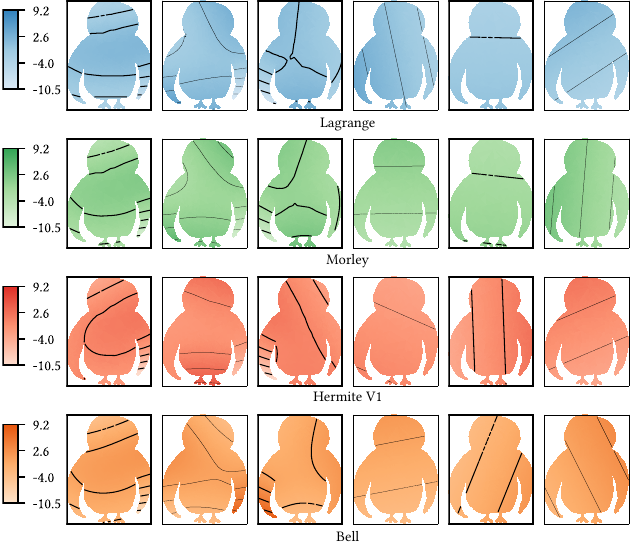}
    \caption[Eigenfunction of the Hessian Energy]{The first six eigenfunction of the Hessian energy operator with natural boundary conditions  \cite{naturalboundaryconditionsforsmoothingingeometryprocessing2018}.
    All pass rough visual muster, though Morley and Lagrange have kinks in the less refined belly regions of the third eigenvector.
    We also observe that the Hermite element
    may seem different from the rest, which is plausible, as Hermite elements fail the patch test.
    We do not include Argyris as it is visually indistinguishable from Bell. }
    \label{fig:eigenBiharmonic}
\end{figure}

To validate the system on a more realistic mesh, we provide the first six eigenvectors of the Hessian operator with natural BCs, allowing a visual comparison with \cite{naturalboundaryconditionsforsmoothingingeometryprocessing2018}.
To make the comparison possible here, we also provide the first six eigenvectors via the mixed method described in \cite{naturalboundaryconditionsforsmoothingingeometryprocessing2018}.
The results in \cref{fig:eigenBiharmonic} show that the first six eigenvectors have roughly the expected structure on a more complex mesh.
We also include Hermite elements in this example to show that they provide visually reasonable results, although they do not pass the patch test and do not converge on meshes used in our convergence validation.

\paragraph{Trade-Offs}
Our convergence results are the start of our discussion of trade-offs.
Examining \cref{fig:covVisBiharmonic}, we observe that Morley might improve the error, reaching the correct shape earlier than the Lagrange element, but the visual suffers due to discontinuity. 
Argyris in contrast reaches a visually pleasing and accurate solution much earlier.
However, we observe that this does not hold if we enforce boundary conditions variationally as opposed to an elimination method. 
Morley can always use elimination methods for Dirichlet BCs while Mixed Lagrange can only use homogeneous BCs and Argyris can only use elimination BCs on certain meshes. 
Since the visual benefits of Argyris can be harmed by boundary condition enforcement, we plan to seek alternatives with better boundary condition properties. 
Similarly, we ought to search for techniques to post-process discontinuous Morley (e.g. \cite{teosmooth, Mirzargar2015}) that produce continuous visually pleasing solutions with comparable error properties and boundary value preservation.

\begin{figure}
    \centering
    \includegraphics[width=\linewidth]{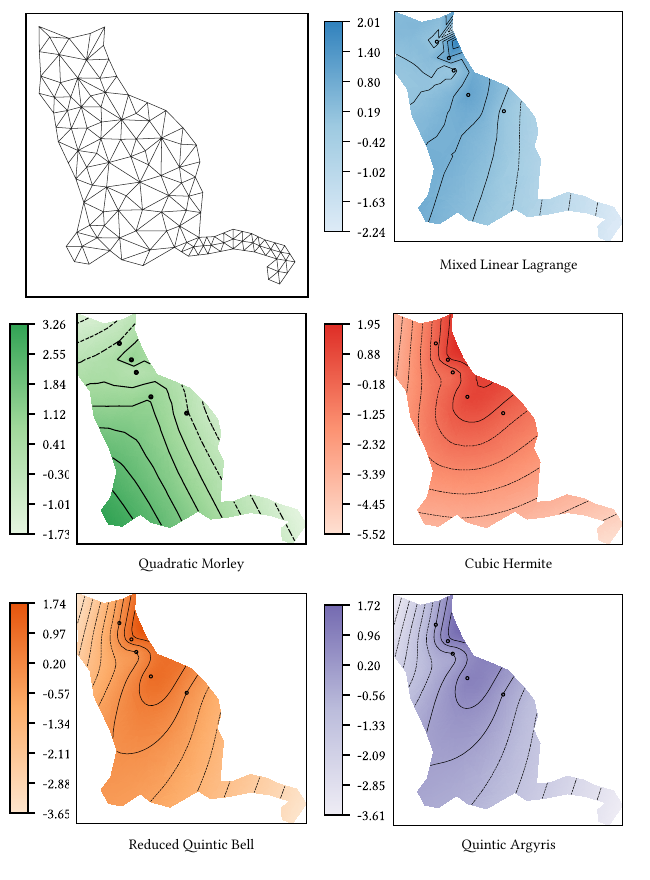}
    \caption[Biharmonic Interpolation Results]{Biharmonic interpolation on a low resolution mesh with different elements.
    All functions were rendered as affine on refined meshes of the appropriate polynomial degree (see \cref{fig:biharmonicInterpVisNote}).
    Observe that the head of the cat is improved by the cheaper quadratic Morley, though discontinuities are introduced. Hermite fixes the head of the cat, but the tail and bottom regions are less accurate compared to Bell. Argyris does not improve much on Bell for the additional cost. }
    \label{fig:biharmonicInterp}
\end{figure}

To further visually motivate our discussion of the trade-offs between elements for Biharmonic problems, we consider the interpolation problems from \cite{naturalboundaryconditionsforsmoothingingeometryprocessing2018}.
We replicate the mixed method described in \cite{naturalboundaryconditionsforsmoothingingeometryprocessing2018} and utilize a direct method via \eqref{eq:bi1} with four elements: Morley, Hermite, Bell, and Argyris. 
We apply all methods to a low-resolution mesh (210 elements).
We note that constructing a low-resolution mesh that worked on the mixed method was a struggle, as the mixed method fails when two edges of a triangle are on the boundary (``ears'').
In contrast, the Morley, Bell, and Argyris elements are valid on any conforming meshes for this problem.
The results are depicted in \cref{fig:biharmonicInterp}, where all are visualized by evaluating the solution at each vertex (so all solutions are still rendered as linear functions).
Visually, we see that distortions due to low poly count disappear as we use higher-order elements. 
What is striking about this result is that via various machine-independent metrics, several higher-order elements were cheaper: they used fewer DOFs per element and thus required smaller sparse matrices.
Of course, this is just one problem; the results are much more striking than the standard convergence results in \cref{fig:covVisBiharmonic}; there, the cost calculation might be more complex, especially if we explore variations of BCs.
There are many vicissitudes to explore, but to tease and conserve space, we summarize some trade-offs between elements for biharmonic problems in \cref{tab:harmonicTradeOff}.

\begin{figure}
    \centering
    \includegraphics[width=\linewidth]{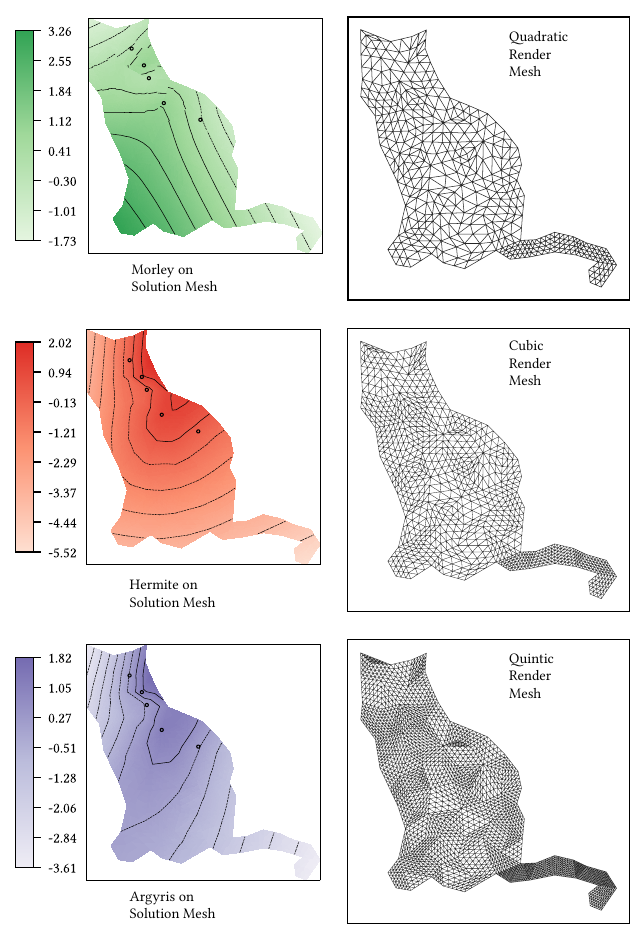}
    \caption[Biharmonic Interpolation Visualization Note]{Upsampled (higher order) render meshes contrasted with function solutions plotted on the mesh \emph{without} upsampling.
    Compared to~\cref{fig:biharmonicInterp} and we see that the visual benefits of higher order elements compared to the mixed method are partially realized without more expensive rendering technologies. }
    \label{fig:biharmonicInterpVisNote}
\end{figure}

\begin{table*}
\caption[Trade-Offs Between Elements]{A table of various trade-offs between elements. The first three rows give the number of DOFs used per triangle to solve certain types of problems. Then come rows indicating which formulations are supported (e.g., mixed methods, naive forms, IPO). The next three rows concern enforcement of boundary conditions. Transform NNZ is the number of non-zeros in the transform matrix.}
\label{tab:harmonicTradeOff}
\centering
\scriptsize
\begin{adjustbox}{max width=\textwidth}

 \begin{threeparttable}
\begin{tabularx}{\textwidth}{|p{0.9in}|X|X|X|X|X|X|X|}
\hline
\textbf{Trade Off/Element} & \textbf{Lagrange 1} & \textbf{Lagrange 2} & \textbf{Lagrange 3} & \textbf{Morley} & \textbf{Hermite V1} & \textbf{Bell} & \textbf{Argyris} \\
\hline
DOFs Per Tri & 3 & 6 & 10 & 6 & 10 & 18 & 21 \\
\hline
DOFs Per Tri BiLaplce & 6 & 12 & 10 & 6 & 10  & 18 & 21 \\
\hline
DOFs Per Tri Hessian & 15 & 30 & 10 & 6 & 10  & 18 & 21 \\
\hline
Supported Form\tnote{e} &
  Mixed &
  Mixed &
  IPM &
  Naive &
  Sometimes\tnote{a}  &
  All &
  All \\
\hline
Elimination BCs &
  Homogeneous\tnote{f} &
  Homogeneous\tnote{f} &
  Homogeneous\tnote{c} &
  Yes &
  Sometimes\tnote{b} &
  Sometimes\tnote{b} &
  Sometimes\tnote{b} \\
\hline
Supports Nitsche &
  No &
  No &
  Yes &
  Yes &
  Yes &
  Yes &
  Yes \\
\hline
Transform NNZ &
  3 &
  6 &
  10 &
  12 &
  16 &
  45 &
  63 \\
\hline
Continuity &
  $C^0$ &
  $C^0$ &
  $C^0$ &
  Complex\tnote{d} &
  $C^1$ at vertices &
  $C^1$ &
  $C^1$ \\
\hline
Ideal Convergence Rate &
  $h^{2}$ &
  Unknown &
  $h^4$ &
  $h^2$ &
  $h^4$ &
  $h^5$ &
  $h^6$ \\
\hline
Convergence Defects &
  Yes\tnote{g}   %
 & Unknown                        %
 & Unknown 
  &  None                     %
  &  Yes\tnote{h}  
  & None
  &  None                             
  
  \\ 
\hline
\end{tabularx}
\tiny
  \begin{tablenotes}[para]
        \item[a] \label{tn:a} All, but naive depends on the geometry.
        \item[b] \label{tn:b} Depends on geometry
        \item[c] \label{tn:c} Only $u=0$, but not $\frac{\partial u}{\partial n} = 0$
        \item[d] \label{tn:d} Minimal element that passes the patch test.
        \item[e] \label{tn:e} (Mixed, IPM (interior penalty method) Naive).
        \item[f] \label{tn:f} General case is unknown.
        \item[g] \label{tn:g} Fails on mesh with corners
        \item[h] \label{tn:h} Fails patch test on some meshes
        \end{tablenotes}

\end{threeparttable}
  \end{adjustbox}    
\end{table*}

\subsubsection{Hodge Decomposition}
\label{sec:feecres}
\paragraph{Problem Setup}
We assume the language of exterior calculus for this section (see \cite{exteriorcalculusingraphicscoursenotesforasiggraph2023course2023}) up to the Hodge Decomposition. For information on how exterior calculus primitives are expressed in our system, see \cref{sec:additional}.
\begin{definition}[Hodge Decomposition]
    Suppose $N$ is an $n$ manifold without boundary.
    Let $\omega$ be a $k$ form with $0\leq k\leq n$.
    The Hodge decomposition of $\omega$ is a triple of $k$ forms, $\alpha, \beta, \eta$, such that
    \begin{enumerate}
        \item $d\alpha = \delta\beta = (d\delta + \delta d)\eta = 0$,
        \item and $\omega = \alpha + \beta + \eta$.
    \end{enumerate}
    If $N$ has boundary, then the Hodge decomposition is a tuple of $5$ $k$ forms, $\alpha, \beta, \eta_{t}, \eta_{n}, \eta$ such that
    \begin{enumerate}
        \item $d\alpha = \delta\beta = (d\delta + \delta d)\eta =(d\delta + \delta d)\eta_{t} = (d\delta + \delta d)\eta_{n} =  0$, 
        \item and $\omega = \alpha + \beta + \eta_{t} + \eta_{n} + \eta$,
        \item and $\alpha |_{\partial\Omega} = \star\beta|_{\partial\Omega}  = \eta_{t}|_{\partial\Omega} =\star\eta_{n}|_{\partial\Omega} = 0$.
    \end{enumerate}

\end{definition}

Further variations on these decompositions are available when more complex varieties of BCs are used~\cite{3dhodgedecompositionsofedgeandfacebasedvectorfields2019}, but we stick to the standard variations.
Computing the Hodge decomposition has many applications, most notably as a primitive for solving many PDEs in fluids and electromagnetics~\cite{thehelmholtzhodgedecompositionasurvey2013}.
Additionally, the operators used to define the Hodge decomposition are at the core of vector field design problems~\cite{theabstracthodgediracoperatoranditsstablediscretization2016}.

\paragraph{Operators}

Traditionally, each Hodge decomposition amounts to the solution of two vector Poisson equations and two vector Poisson eigenvalue problems.
The Poisson equations allow us to compute the $\alpha$ and $\beta$ terms while the eigenvalue problems allow us to compute $\eta_{t},\eta_{n},\eta$ by projecting $\omega-\alpha-\beta$ to the null space of the two vector Poisson equations.
Further, we need two differential operators to set up the right hand side of these equations  ($d\omega, \delta \omega)$ and two of the same differential operators to process the results to form $\alpha$ and $\beta$.
A recent example of such an approach is~\cite{hodgedecompositionofvectorfieldsincartesiangrids2024}.
However, this involves a huge array of operators because for each choice of $k$ and $n$, a variety of differential operators might be involved.
We will avoid this complexity by simply specifying the operators involved using only $d$ and the inner product on forms.
We need only four operators that operate on pairs of either two $k$ forms, or $k$ and $k-1$ forms, or two $k-1$ forms:
\begin{equation}
    m(u,v) = \int_{N}<u,v>
\end{equation}

\begin{equation}
    m_{\star, d}(u,v) = \int_{N}<u,dv>
\end{equation}

\begin{equation}
    m_{d, \star}(u,v) = \int_{N}<du,v>
\end{equation}

\begin{equation}
    m_{d, d}(u,v) = \int_{N}<du,dv>.
\end{equation}

With the right choice of element, the BCs are Dirichlet or natural. 
Next, we need to use $d$ and $\delta$ pointwise to set up the right hand sides.
Since the first four are similar, we provide code for $m_{d,d}$:
\begin{lstlisting}[language=PythonPlus, caption={Example operator for FEEC}\label{lst:ddexanoke},abovecaptionskip=5pt]
def ddoperator(formDeg: int, cell: int):
    fld = field(ChartType(cell), Lambda(formDeg, cell))
    @pointwiseIntegral(ChartType(cell))
    def mdd(u: fld, v: fld):
        return antiSymmeterize(u(1)).dot(antiSymmeterize(v(1)))
    return mdd
\end{lstlisting}

We also provide an example with $\delta$:
\begin{lstlisting}[language=PythonPlus, caption={Implementation of a pointwise delta operator.}\label{lst:deltaExample},abovecaptionskip=5pt]
def deltaOperator(formDeg: int, cell: int):
    fld = field(ChartType(cell), Lambda(formDeg, cell))
    fldp = field(ChartType(cell), Lambda(formDeg - 1, cell))
    @pointwise(ChartType(cell))
    def delta(u: fld) -> fldp:
        return hodge(antiSymmetrize(D(hodge(u(0)))))
    return delta
\end{lstlisting}

\paragraph{Elements}

Just as with operators, there are elements for different choices of $k$ and $n$.
Also as with operators, these can be succinctly unified into a few choices, parametrized by dimension of the space/surface $N$, form degree $(k)$, polynomial degree $(r)$, Hodge duality $(\star)$, and trim $(-)$.
These elements are labeled $P_{r}\Lambda^{k}(N)$, $P_{r}^{-}\Lambda^{k}(\mathcal{T}N)$, $P_{r}^{\star}\Lambda^{k}(N)$, $P_{r}^{-\star}\Lambda^{k}(N)$.
Together these capture Lagrange elements, discontinuous Lagrange elements, Nedelec (type one and two) (face and edge) elements, Brezzi-Douglas-Marini elements, and Raviart Thomas elements.
Since we do not need the last two here, we will not discuss them here beyond providing their spaces, but they only require about 10-20 more lines of code.

We first describe these mathematically.
To describe the spaces, we need the Koszul operator $\kappa$ (see \cite{finiteelementexteriorcalculus2019}) and also the space of homogeneous polynomials (all terms have the same degree except $0$) $H_{r}$.
The latter gives rise to the space of homogeneous polynomial form spaces $H_{r}\Lambda^{k}$.
\begin{enumerate}
   \item $P_{r}\Lambda^{k}(\mathcal{T}N)$, the space of $r$ degree    polynomial valued $k$ forms on the tangent space of $N$.
    \item $P_{r}^{-}\Lambda^{k}(\mathcal{T}N) := P_{r-1}\Lambda^{k}(\mathcal{T}N) + \ker_{H_{r+1}\Lambda^{k-1}(\mathcal{T}N)}\kappa$, the degree $r-1$ $k$ forms plus the kernel of the Koszul derivative on the homogeneous polynomial degree $r+1$ $k-1$ forms. These are the so called trimmed spaces.
    \item $P_{r}^{\star}\Lambda^{k}(\mathcal{T}N) := \star P_{r}\Lambda^{n-k}(\mathcal{T}N)$, the space of $k$ forms represented via Hodge stars of $n-k$ forms. The idea of this spaces is to model normal components as opposed to tangential components.
    \item $P_{r}^{-\star}\Lambda^{k}(\mathcal{T}N) := \star P_{r}^{-}\Lambda^{n-k}(\mathcal{T}N)$, similar to the above but using trimmed spaces. 
\end{enumerate}

For each of these elements, the DOFs can be defined via mutual recursion.
The base case is the Lagrange elements, which we will discuss in the Stokes section.
We can define the DOFs on $k$ forms and simplices $f$ of dimension $d$ using forms of smaller degree defined on smaller simplices:
\begin{enumerate}
    \item $P_{r}\Lambda^{k}(\mathcal{T}f)$: for every face $f^{\prime}\subset f$,  let $\{\eta_{f^{\prime}, i}\}$ be a basis for $P^{-}_{r+k-\dim{f^{\prime}}}\Lambda^{\dim{f^{\prime}}-k}(\mathcal{T}f^{\prime})$. The DOFs are
    \begin{equation*}
        \phi_{f^{\prime},i}(u) = \int_{f^{\prime}} (u|_{f^{\prime}})\wedge \eta_{f^{\prime}, i}
    \end{equation*}
    for all $f^{\prime}\subset f$ and all valid $i$.
    \item $P^{-}_{r}\Lambda^{k}(\mathcal{T}f)$: for every face $f^{\prime}\subset f$,  let $\{\eta_{f^{\prime}, i}\}$ be a basis for $P_{r+k-\dim{f^{\prime}}-1}\Lambda^{\dim{f^{\prime}}-k}(\mathcal{T}f^{\prime})$. The DOFs are
    \begin{equation*}
        \phi_{f^{\prime},i}(u) = \int_{f^{\prime}} (u|_{f^{\prime}})\wedge \eta_{f^{\prime}, i}
    \end{equation*}
    for all $f^{\prime}\subset f$ and all valid $i$.
\end{enumerate}

These  DOFs capture the tangential components with forms;  DOFs for the latter two spaces would capture the orthogonal complements of the tangent space.
(Practically, the first two favor Dirichlet BCs on $u$, while the latter two favor Dirichlet BCs on $\star u$).
The trimmed spaces provide smaller, more economical versions that are harder to interpret; in particular, the trimmed spaces allow users to approximate k forms of different orders to the same order of accuracy~\cite{finiteelementexteriorcalculus2019}.

We provide code to capture all of these cases in stages.
First, we provide the DOFs to define the Koszul space.

\begin{lstlisting}[language=PythonPlus, caption={Implementation of Koszul  DOFs.}\label{lst:koszuldofs},abovecaptionskip=5pt, escapechar=|]
def koszulDofs(formDeg: int, degree: int, dim: int):
    evals = []
    dsf = filter(
        lambda x: sum(x) == degree + 1,
        product(*[range(degree + 2) for _ in range(dim)]),
    )
    for ds in dsfl:
        for k in range(int(binom(dim, formDegree - 1))):
            @P.pointwiseIntegrate(CellType(dim))
            def koszul_eval(c:  ChartType(dim), f: field(dim, Lambda(formDegree, dim)):
                x = c.P(c.x)
                xx = np.prod([c.x[j] ** ds[j] for j in (range(dim))])
                r = interiorEval(f(0), [0], [x])
                return (
                    r[(k,)] * xx |(See \cref{sec:uggappbad})|
            )   
        evals.append(koszul_eval)
    return evals
\end{lstlisting}

We define the Koszul DOFs to isolate the kernel of the Koszul operator.
We do this by testing the Koszul operator with every relevant monomial and form component. 
Then, we can define our spaces:
\begin{lstlisting}[language=PythonPlus, caption={Implementation of FEEC  spaces, utilizing Koszul DOFs for the Koszul space.}\label{lst:feecSpace},abovecaptionskip=5pt]
def feecSpace(formDeg: int, degree: int, dim: int, trim: bool):
    if not trim:
        return Pure(Simplex(dim), degree, Lambda(formDeg, dim))
dim))
    else:
        Prm1 = feecSpace(formDeg, max(degree - 1, 0), dim, trim)
        if formDegree  == 0 or formDegree == dim:
            return Prm1
        kdofs = koszulDofs(formDeg, degree, dim)
        return Prm1 + IKernel(kdofs, Pure(Simplex(dim), degree, minDegree=degree))
\end{lstlisting}

Finally, we can define our elements modulo a few lines of code to specify the base cases:
\begin{lstlisting}[language=PythonPlus, caption={Implementation of FEEC elements.}\label{lst:feecDofs},abovecaptionskip=5pt]
def feec(formDeg: int, degree: int, dim: int, trim: bool):
    if formDeg == 0:
        return langrangeDofs(dim, degree if not trim else degree - 1)
    assert dim >= formDeg >=0 and degree >=0 
    space = feecSpace(formDeg, degree, dim, hodge, trim)
    md = min(dim, formDegree + degree - 1)
    evals = []
    for j in range(formDegree, md + 1):
        ct = ChartType(j)
        field = field(ct, Lambda(formDegree, ct))
        recDegree = degree - j + formDegree - 1 if trim else degree - j + formDegree
        recDofs = feec(j, recDegree, j - formDegree, trim=not trim)
        for basisFunc in recDofs:
            @P.pointwiseIntegrate(ct, reduction=choice)
            def form_eval(u: field, v: recDofs.field() = basisFunc):
                return hodge(wedge([u(0), v(0)]))
            evals.append(form_eval)
    return EF.FE(dim, space, evals)
\end{lstlisting}

For our base case, we must also provide arbitrary order Lagrange DOFs.
We provide these in the Stokes section.

\paragraph{Validation}

\begin{figure}
    \centering
    \includegraphics[width=\linewidth]{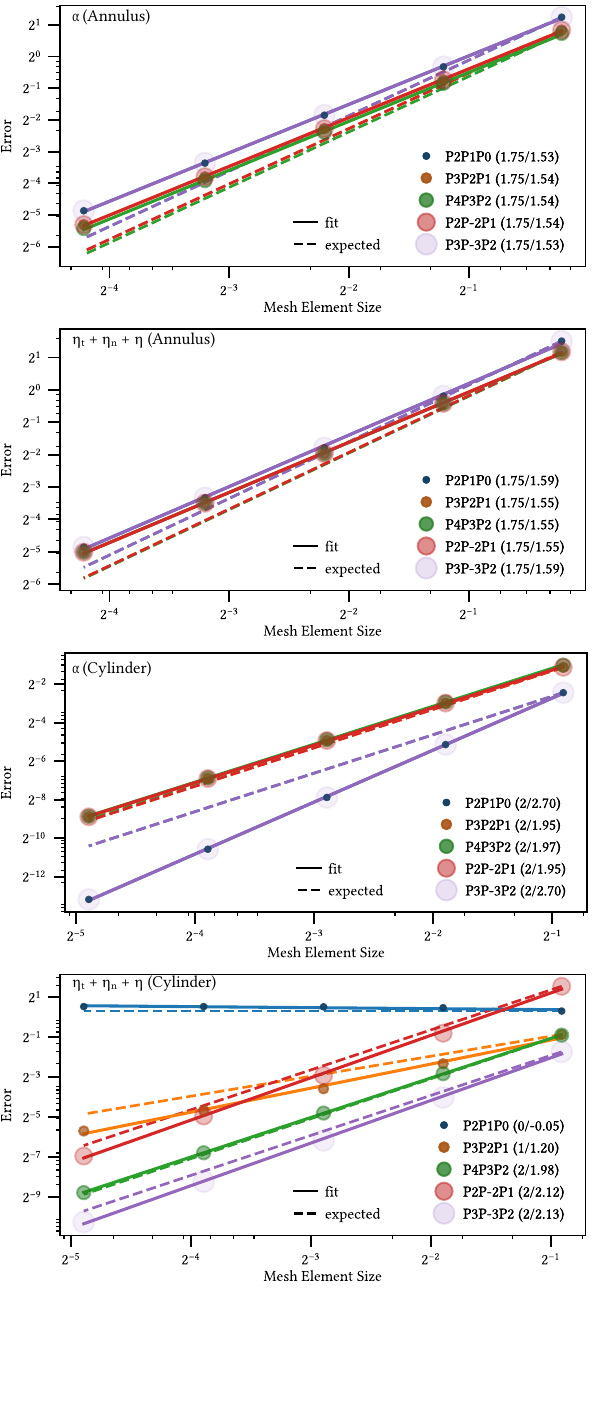}
    \caption{Convergence plots for five different elements on two different geometries for the $\alpha$ and total harmonic components of the Hodge Decomposition. We omit $\beta$ as it is similar to the total harmonic component. The theory of variational crimes indicates these are the correct results, validating our implementation and demonstrating the difficulty of a convergence advantage for analytical solutions via higher order exterior calculus methods.}
    \label{fig:hodgeConv}
\end{figure}

We use two problems to validate our solvers and provide convergence data.
We refer to the first as the annulus solution and the second as the cylinder solution.
See \cref{sec:ashd} for details.
For both cases, we must utilize the theory of variational crimes to temper the validation experiments; we should not expect convergence higher than second order because we are using a linear approximation of the geometry~\cite{geometricvariationalcrimeshilbertcomplexesfiniteelementexteriorcalculusandproblemsonhypersurfaces2012}.
For analytical data, we quickly achieve this if we consider convergence for $\alpha$, $\beta$, and the total harmonic part, $\eta_{t} + \eta_{n} + \eta$.
We provide convergence plots in \cref{fig:hodgeConv}.
For the individual harmonic components, our results differ by geometry.
For the annulus, we achieve results identical to the total harmonic component.
For the convergence in the cylinder scenario, we need a further proviso.
In particular, for $\eta_{t}$ and $\eta$, we achieve results comparable to the total harmonic component, but we do \textbf{not} achieve convergence for $\eta_{n}$.
Literature on convergence in the surface scenario of individual components of the algorithm is sparse, and most articles on the decomposition do not perform this analysis, sometimes reporting a difficulty in the convergence of the individual components of the harmonic part of the decomposition~\cite{boundaryawarehodgedecompositionsforpiecewiseconstantvectorfields2016,3dhodgedecompositionsofedgeandfacebasedvectorfields2019,hodgedecompositionofvectorfieldsincartesiangrids2024,topologypreservinghodgedecompositionintheeulerianrepresentation2024,discretemultiscalevectorfielddecomposition2003,hodgetypedecompositionsforpiecewiseconstantvectorfieldsonsimplicialsurfacesandsolidswithboundary2017}.
This might be a Babu\u{s}ka paradox type situation~\cite{gaussiancurvatureandbabuškasparadoxinthetheoryofplates2003}, but it might also be that the convergence in the theory applies to the PDEs but not the classic Hodge decomposition algorithms. 

\begin{figure}
    \centering
    \includegraphics[width=\linewidth]{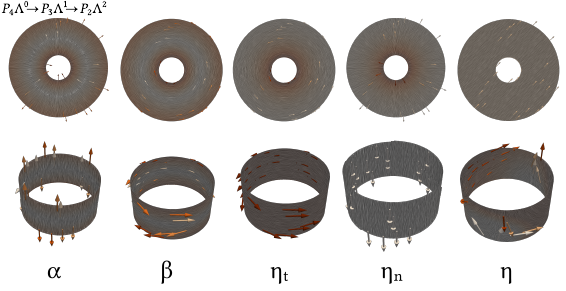}
    \caption{Higher order and resolution solutions for the Hodge decomposition in both scenarios. Visual results show the correct vector field topologies and no visual anomalies. }
    \label{fig:hodgeVis}
\end{figure}

\begin{figure}
    \centering
    \includegraphics[width=\linewidth]{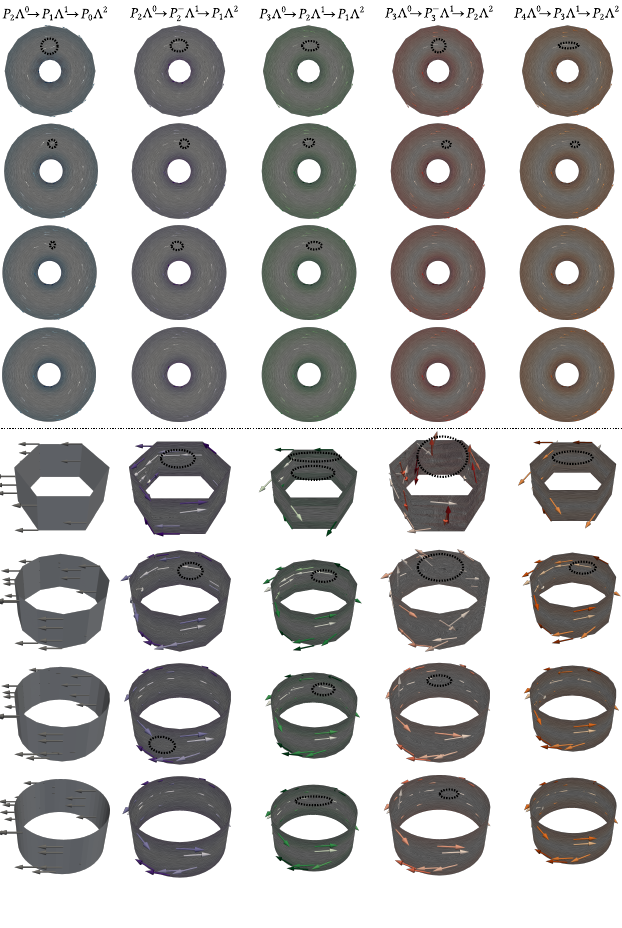}
    \caption{Visual results for $\beta$ produced by the convergence study. Except for the first column of the cylinders study where the method fails catastrophically, we have labeled regions visual anomalies with dashed lines. Though the pattern is not completely robust due to the $-$ elements, the last two components of the element complex control how coarse a mesh can be without obvious visual anomalies. }
    \label{fig:hodgeVolVis}
\end{figure}

We checked that each component has the correct angle with an interpolation of the true solution (zero to machine precision).
 These indicate that the elements and solvers are sound, but the steps to produce the five term Hodge decomposition (as found in many places in the literature) are not sound.
Despite this though, we do produce equally  visually satisfying solutions for all components (\cref{fig:hodgeVis}) that are indistinguishable from the true solutions. 
Further, the intermediate mesh resolutions of our convergence study provide a starting point for the discussion of trade-offs between elements.
We provide the results for $\beta$ for several mesh resolutions and elements in \cref{fig:hodgeVolVis}.
These results show that higher order elements provide advantages with respect to visual anomalies on low resolution meshes.

\paragraph{Trade Offs}

\begin{figure}
    \centering
    \includegraphics[width=1\linewidth]{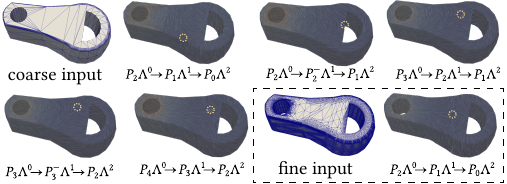}
    \caption[FEEC Zero Eigenvectors of Hodge Laplacian]{Zero eigenvectors of the Hodge Laplacian on a low-resolution mesh with a variety of elements and a comparison to a lower order element on a highly refined mesh (282 vertices vs. 29,000 vertices). Observe that the labeled critical point location drifts.}
    \label{fig:pretty}
\end{figure}

The trade-offs for higher order representations of forms have been noted throughout the literature~\cite{subdivisiondirectionalfields2020,edgesubdivisionschemesandtheconstructionofsmoothvectorfields2006,discretedifferentialformsforcomputationalmodeling2006,highorderdirectionalfields2022}, including applications such as vector field design, parameterization (fixed boundary, seamless), vector heat methods, and solutions to fluid equations.
In each of these, the story is roughly the same: higher-order elements provide smoother and more accurate solutions that benefit downstream applications at the expense of implementation pains and computational costs.
We can provide another data point by computing one of the zero eigenvectors of the Hodge Laplacian on a low resolution mesh.
We compare this to a higher resolution mesh and low order element scenario in ~\cref{fig:pretty}.
We follow~\cite{highorderdirectionalfields2022} in this comparison.
We note that not only are there visual distortions due to the lower order elements (jumps in color), but that the topology of the vector field is not accurately represented: the lowest orders do not capture the critical point in the center of the field and heavily distort the flow along the lines of the coarse geometry.

\subsubsection{Stokes Flow}
\paragraph{Problem Setup}
Let $d=2$ or $d=3$ and $\Omega\subset\mathbb{R}^{d}$.
Suppose we have a forcing function $f\colon [0, \infty)\times \Omega\to\mathbb{R}^{3}$ and a viscosity parameter $\nu\in\mathbb{R}$.
The Stokes equations seek a velocity field $u\colon [0, \infty)\times \Omega\to\mathbb{R}^{d}$ and a pressure field $p\colon [0, \infty)\times \Omega\to\mathbb{R}$ such that 
\begin{equation}
    u_{t} - \nu\Delta u + \nabla p = f
\end{equation}
and
\begin{equation}
    \nabla\cdot u = 0.
\end{equation}

We are primarily interested in no-slip BCs and other proscribed time-varying velocities:
\begin{equation}
u|_{\partial\Omega} = 0
\end{equation}
or 
\begin{equation}
    u|_{\partial\Omega} = w
\end{equation}
for some $w\colon  [0, \infty)\times \partial\Omega\to\mathbb{R}^{d}$.
We additionally enforce that the pressure is mean zero, i.e., $\int_{\Omega} p = 0$ for all time $t$.

We are primarily interested in the contrast between cases where $\nu \approx O(1)$ and $\nu \approx O(h)$ where $h$ is the characteristic edge length of the mesh.
The former resembles the case of a fluid like water while the latter resembles the case of a superfluid.

\paragraph{Operators and Time Stepping}

The standard weak formulation for stationary stokes is to find a $u$ and $p$ such that for all $v\colon\Omega\to\mathbb{R}^{d}$ and $q\colon\Omega\to\mathbb{R}$:
\begin{equation}
    F(u,p,q,v, f) = \int_{\Omega}\nabla u_{t}\cdot \nabla v - \int_{\Omega} p\nabla\cdot v + \int_{\Omega}q \nabla\cdot u - \int_{\Omega} fv
\end{equation}

We considered implicit Euler and Crank Nicholson time stepping.
For a time step size $dt$ and a $\theta\in [0,1]$, we solve for $u_{n+1}, p_{n+1}$ at time $t_{n+1}$ in terms of $u_{n}$ and $p_{n}$ at time $t_{n}$ via finding $u_{n+1}$ and $p_{n+1}$ such that
\begin{align*}
     \int_{\Omega} v(u_{n+1}-u_{n})/dt + &\\ \theta F(u_{n+1}, p_{n+1}, v, q, f(t_{n+1})) + (1-\theta) F(u_{n}, p_{n}, v, q, f(t_{n})) =0&\\
\end{align*}
holds for all $v,q$.
Since the resulting discrete system is a saddle point system, we must be careful with the solver. For all purposes in this section, we get away with PETSc' GMRES with LU (via MUMMPS) as a pre-conditioner~\cite{petscusersmanual2019,afullyasynchronousmultifrontalsolverusingdistributeddynamicscheduling2001,performanceandscalabilityoftheblocklowrankmultifrontalfactorizationonmulticorearchitectures2019}. 
\paragraph{Elements}

We follow the element choices offered in the experiments in \cite{pressureinducedlockinginmixedmethodsfortimedependentnavierstokesequations2019}, including various pairings of Lagrange elements (Taylor-Hood and Scott-Vogelius), mini elements, and Crouzeix-Raviart elements.
The Taylor-Hood (TH) element is simply the use of a $k$-degree vector Lagrange element for the velocity space and a $k-1$-degree scalar Lagrange element for the pressure space.
This requires a more general bit of code for the Lagrange element, which we provide here; this version allows for Lagrange elements on arbitrary Euclidean-like spaces and for discontinuous spaces:
\begin{lstlisting}[language=PythonPlus, caption={Tensor Lagrange Elements}\label{lst:lagfull},abovecaptionskip=5pt]
def lagrangeDofs(dim: int, degree: int, space = P.R, dg: bool = False):
    PSpace = Pure(Simplex(dim), degree, space)
    dofs = []
    for dimp in range(dim + 1):
        for point in equispacedBarycentric(dimp, degree, boundary=False):
            for vector in stdEuclideanBasis(space):
                @pointwiseEval(ChartType(dimp), barycentricCoordinate=point, storeAt=dimp if not dg else dim)
                def dof(u: field(ChartType(dimp), space)): return u(0).dot(vector)
                dofs.append(dof)
    return Dofs(PSpace, dofs)
\end{lstlisting}

Scott-Vogelius (SV) can be characterized as a Taylor-Hood element where the pressure space is discontinuous, but this characterization is facile due to the circumstances under which the element works. 
In particular, the SV element requires that there are no singular vertices (vertices incident to edges that all lie on two lines (or three planes)) and so the SV element is sometimes phrased as TH with discontinuous pressure on a barycentrically refined or macro triangle/tetrahedral meshes~\cite{thescottvogeliusfiniteelementsrevisited2019,areynoldsrobustpreconditionerforthescottvogeliusdiscretizationofthestationaryincompressiblenavierstokesequations2021,finiteelementapproximationoffluidstructureinteractionusingtaylorhoodandscottvogeliuselements2024,nonnestedmultigridsolversformixeddivergencefreescottvogeliusdiscretizations2008}.
In the future, we will model with these explicit macro elements, but for now we model this by just refining the mesh barycentrically. 
A bubble element is a Lagrange element where we include extra basis functions that are zero on the boundary of the geometry. 
The mini element on a $d$-dimensional space uses linear Lagrange for the pressure and the sum of linear Lagrange plus a $d$-degree bubble element for the velocity.
We can define a mini space similarly:
\begin{lstlisting}[language=PythonPlus, caption={Mini Elements}\label{lst:mini},abovecaptionskip=5pt]
def mini(dim: int, degree: int, bdegree: int, space = P.R, dg: bool = False):
    assert bdegree > degree and bdegree >= dim + 1
    idofs = []
    ndofs = []
    dofs = []
    for dimp in range(dim + 1):
        for point in equispacedBarycentric(dimp, bdegree, boundary=False):
            for vector in stdEuclideanBasis(space):
                @pointwiseEval(ChartType(dimp), barycentricCoordinate=point, storeAt=dimp if not dg else dim)
                def dof(u: field(ChartType(dimp), space)): return u(0).dot(vector)
                if dim == dimp:
                    idofs.append(dof)
                else:
                    ndofs.append(dof)
    PSpace =  Pure(Simplex(dim), degree, space) + IKernel(ndofs, Pure(Simplex(dim), bdegree, space))
    return Dofs(PSpace, lagrangeDofs(dim, degree, space).dofs + idofs)
\end{lstlisting}
Finally, we do not bother with the CR element as it is very similar to the Lagrange element.

\paragraph{Validation and Trade-Offs}

We set up two analytical problems to test our solver, which is sufficient to reveal a trade-off.
In particular, we follow ~\cite{pressureinducedlockinginmixedmethodsfortimedependentnavierstokesequations2019} to show that cheaper elements (fewer DOFs) tend to be non-robust to low viscosities, sometimes inducing locking behaviors similar to those found in elasticity simulation.

In two dimensions, we consider analytical velocities and pressures given by
\begin{equation}
    u(x,y,t) = (\cos(y), \sin(x)) (1+t)
\end{equation}
and
\begin{equation}
    p(x,y,t) = \sin(x+y) - (2\sin(1)-\sin(2)).
\end{equation}

In three dimensions, we consider
\begin{equation}
    u(x,y, z, t) = (\cos(y), \sin(z), cos(x)) (1+t)
\end{equation}
and
\begin{equation}
    p(x,y,z,t) = \sin(x+y+z) - 8 (1+2\cos(1) \sin(\frac{1}{2})^{4}.
\end{equation}

\begin{figure}
    \centering
    \includegraphics[width=\linewidth]{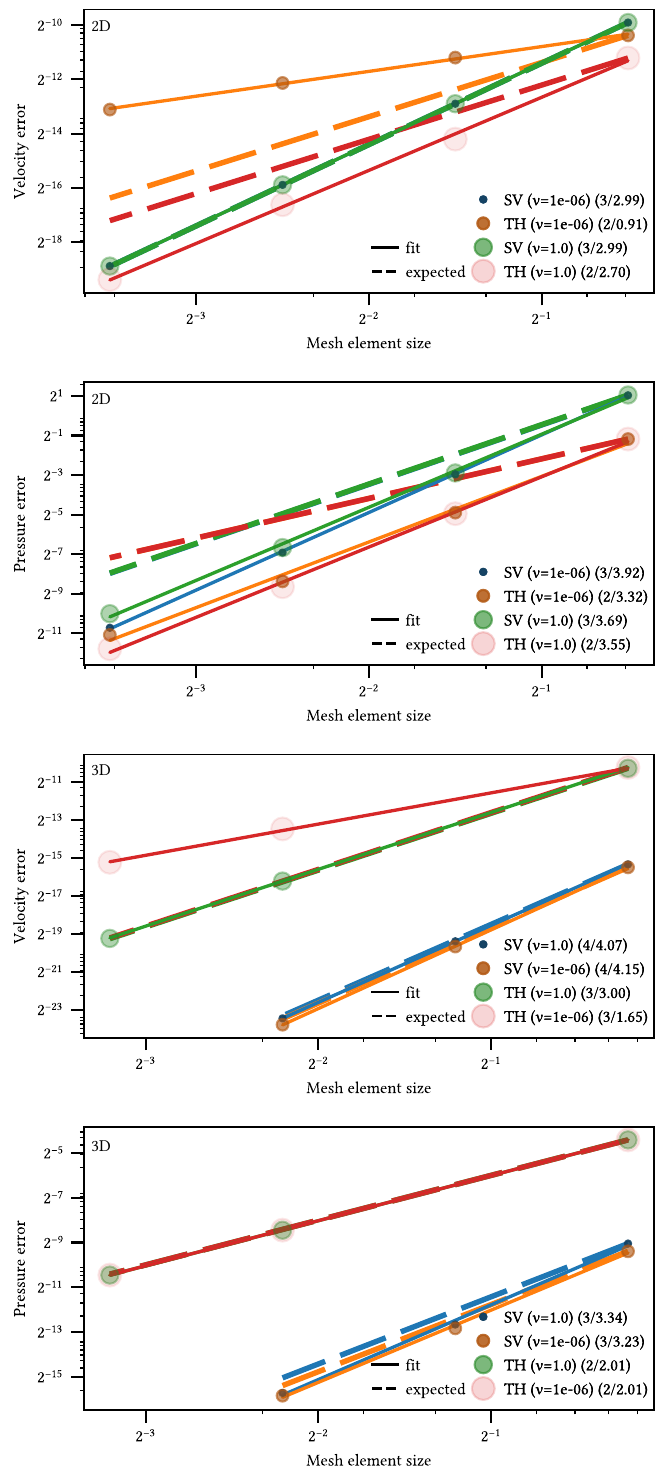}
    \caption{2D/3D Convergence results for velocity and pressure of a stokes simulation using two different elements and two different viscosities.  Note that the TH elements consistently underperformed  velocity convergence in the low viscosity regime whereas both methods are comparable in the normal regime. Also note that pressure convergence is not impacted.}
    \label{fig:stokesConv}
\end{figure}

We use these to construct a forcing function according to the Stokes equations and solve until time $0.01$ in ten time steps on the unit square or unit cube. 
We enforce Dirichlet conditions on the velocity while for the pressure we require that $\int p = 0$.
We do this for $\nu=1$ and $\nu=10^{-6}$.
We use the Taylor-Hood elements and the Scott-Vogelius elements at order $k=2$.
In 2D, we use Crank-Nicolson time stepping with $\theta=0.5$ whereas in 3D we used Euler.
Convergence results for velocities and pressures in 2D/3D are plotted in ~\cref{fig:stokesConv}.
We immediately observe that the convergence in velocity of the Taylor-Hood elements is massively impacted by lowering the viscosity while Scott-Vogelius is not impacted.
Interestingly, we observe that the pressure convergence is not impacted.
As it turns out in the analysis of~\cite{pressureinducedlockinginmixedmethodsfortimedependentnavierstokesequations2019}, as viscosity gets smaller, the error in velocity becomes dominated by the error in the gradient of pressure if an element is not discretely divergence free in a specific sense.
This theory is borne out in the velocity and pressure plots in~\cref{fig:stokesConv}, noting that convergence of a gradient is typically an order lower than convergence of the function.

Finally, to show that this difference can produce visually striking results, we set up a simulation on the Bob mesh.
We start from zero initial conditions.
We add inflow at the top of a head that looks like
\begin{equation}
   u(x,y,z,t)= (0, 0, -1) (1+10t)^{2}
\end{equation}
and an outflow that looks like

\begin{equation}
   u(x,y,z,t)= (0, 1, 0) (1+10t)^{2}.
\end{equation}
We assume a force due to gravity.
We compute 200 time steps starting at $t=0$ and ending at $t=0.2$.
We compare the mini element to the Scott-Vogelius element of order $k=3$. 
We note that the former uses 12 velocity DOFs per element while the latter uses 60.
We again contrast the behavior for $\nu=1$ vs $\nu=10^{-6}$.

\begin{figure}
    \centering
    \includegraphics[width=0.5\textwidth]{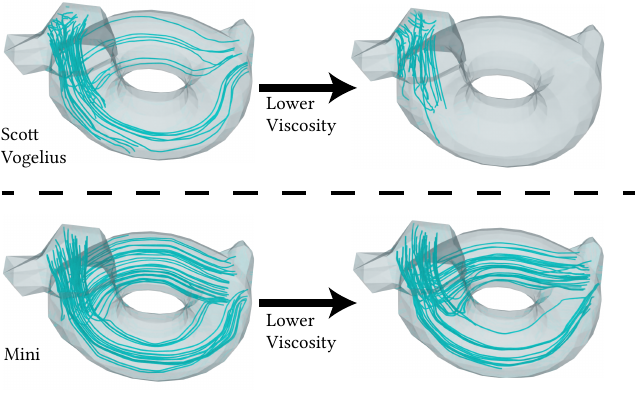}
    \caption[Visual Comparison of Stokes Elements in 3D for Viscosity Responsiveness]{Stream lines, seeded in the head, derived from the flow of our stokes simulation at the last time step with different elements and viscosities. We contrast the impact of changing viscosities ($\nu=1$ vs $\nu=10^{-6}$) on two different elements. The Scott Vogelius element correct accounts for the low viscosity whereas the mini element does not. In particular, the low viscosity at low pressure means that the fluid should move mainly due to gravity rather than inflow/outflow velocity, which is manifestly true for Scott-Vogelius but not for the Mini Element. Both produce plausible results for normal viscosity. }
    \label{fig:stokesPRetty}
\end{figure}

The results are seen in ~\cref{fig:stokesPRetty}.
When $\nu=1$, both simulations show reasonable results.
We observe that for Scott-Vogelius, the simulation changes dramatically with $\nu$, which is expected.
When $\nu$ is small, the fluid responds slowly to nearby changes in velocity so the inflow and outflows are slow to spread to the inner regions of the geometry.
For Scott-Vogelius, we can plainly see that away from the inflow/outflow, the fluid is mainly responding to gravity.
However, for the mini element, changing $\nu$ has little qualitative impact.
This is a convergence failure and a non-intuitive example of locking.

\subsection{Claim: Finite elements are specified concisely due to automation}
See~\cref{tab:element-SLOC-Counts}. In particular, we show that we are able to dramatically improve on the SOTA for specifying FEs within systems capable of assembling FE matrices.
We do not compare against systems like the symbolic finite elements library as not all of their specifications (e.g., Morley) can be used on multiple domains~\cite{symfemasymbolicfiniteelementdefinitionlibrary2021}.
Moreover, our improvements in lines of code come from two of our automations.
First, our efforts to automatically deduce transformations by simplifying formulas and symbolically inverting matrices represents non-trivial portions of code counts for many elements.
Not only does this portion represent a non-trivial number of lines, but these lines are hard to produce, requiring manual mathematical reasoning.
Second, our efforts to build a high-level language for IPTEMs, open in particular to the automation of topological enumeration, also dramatically reduce the lines of code for all elements.
This is best shown for Lagrange and FEEC elements: though the transformation is trivial, we still make the specification smaller by relying on topological enumeration and our high-level language for tensor expressions.

\begin{table}[]
\scriptsize
\begin{tabular}{|p{0.5\linewidth}|c|rl|}
\hline
Element         & \shortstack{\SYS{}\\SLOC} & \multicolumn{2}{p{0.20\linewidth}|}{FInAT/FIAT SLOC (Transform)} \\ \hline
Morley          & $\sim$10   & 84 & (31)              \\ \hline
Hermite         & $\sim$17   & 88 & (23)           \\ \hline
Argyris         & $\sim$31  & 221 & (117)            \\ \hline
Bell            & $\sim$44   & 138 & (63)           \\ \hline
( N\'ed\'ec (first kind)) + (DG Volume Lagrange) +( N\'ed\'ec (second kind)) &
  $\sim$46 &  534 & (0) \\ \hline
Lagrange        & $\sim$10         & 80  & (0)           \\ \hline

Mini & $\sim$17         & 28 &     (0)        \\ \hline
\end{tabular}
\caption[SLOC Results for Specifying Elements]{SLOC to specify all the elements used in our experiments and available in both \SYS{} and FInAT, currently the most comprehensive practical system for using different FEs. Note that for our system,  N\'ed\'ec (first kind), DG Volume Lagrange, and N\'ed\'ec (second kind) all share an implementation and so do BDM, so we group these together. If we added at most 10 to 20 lines of code we could also do Raviart-Thomas and Brezzi-Douglas-Marini, which combined require around 200 lines of code in FIAT. This and other variants on FEEC elements could make interesting additional case studies for \cref{tab:extra-SLOC-Counts1}. }
\label{tab:element-SLOC-Counts}
\end{table}

\subsubsection{Claim: The automation produces optimal implementations of the finite element transformation theory. }

In \cite{ageneralapproachtotransformingfiniteelements2018}, a key measure of the cost of using a more complex element is the number of non-zeros in the transformation.
Our automations and saved lines of code would not be as useful if the transformations were considerably more expensive.
We compare the number of symbolic non-zeros to assess the sparsity of the generated transformations. These counts do not measure compilation, assembly, or solve time; wall-clock comparisons remain necessary to assess runtime performance.
We do this in~\cref{tab:element-transform-sizes}, where we find that we have the number of non-zeros in matrices computed by hand in the literature.
Additionally, this shows that our inversion procedure does not default to a matrix inverse: \SYS{} enables the inverse to be computed more like a formula and with the optimal sparsity pattern visible at compile time.

\begin{table}[]
\scriptsize
\begin{tabular}{|p{0.5\linewidth}|c|}
\hline
Element         & Transformation NNZ  \\ \hline
Morley          & 12               \\ \hline
Hermite         & 16           \\ \hline
Argyris         & 81             \\ \hline
Bell            & 42         \\ \hline
(RT + N\'ed\'ec (first kind) + DG Volume Lagrange) + (BDM + N\'ed\'ec (second kind)) &
  $O(N)$ \\ \hline
Lagrange        & $O(N)$           \\ \hline

Mini & $O(N)$        \\ \hline
\end{tabular}
\caption[Analysis of DOF Transformations]{Transformation DOFs}
\label{tab:element-transform-sizes}
\end{table}

\subsection{Claim: Finite element specifications and usage are malleable}
To evaluate malleability, we ask: how many extra lines of code are required to change elements?
First, to literally add the element in the two simulations where the code changes,  we find that the number of lines of code is modest (\cref{tab:extra-SLOC-Counts1} and \cref{tab:extra-SLOC-Counts2}).
Similarly, in the case of the FEEC elements, we have a much more succinct implementation, especially for trimmed elements where we more directly translate the math than compared to other systems~\cite{algorithm839fiatanewparadigmforcomputingfiniteelementbasisfunctions2004}.
Second, we ask how much do programs change when we change elements?
The answer is that they do not unless an element requires a different formulation (e.g., the mixed Lagrange method in the biharmonic equation).
Compared to other systems, our advantage here lies in our construction of the front end; no other systems that we know of support interpolation into elements such as Hermite, Morley, or Argyris\footnote{See \url{https://www.firedrakeproject.org/interpolation.html}}, but this naturally emerges in our front end.
Similarly, \SYS{} supports direct enforcement of Dirichlet BCs when possible on these spaces, whereas that is typically a special case in other systems~\cite{c1finiteelementsfordune}.

\begin{table}[]
\scriptsize
\begin{tabular}{|p{0.5\linewidth}|c|}
\hline
Element         & Additional LOC  \\ \hline
Morley          & 4               \\ \hline
Hermite         & 11           \\ \hline
Argyris         & 11             \\ \hline
Bell            & 13         \\ \hline
\end{tabular}
\caption[LOC Delta for Biharmonic Elements]{Extra LOCs for biharmonic elements against affine Lagrange}
\label{tab:extra-SLOC-Counts1}
\end{table}

\begin{table}[]
\scriptsize
\begin{tabular}{|p{0.5\linewidth}|c|}
\hline
Element         & Additional LOC  \\ \hline
Mini          & 10               \\ \hline
CR         & 1           \\ \hline
\end{tabular}
\caption[LOC Delta for Stokes Elements]{Extra LOCs for Stokes elements against general Lagrange}
\label{tab:extra-SLOC-Counts2}
\end{table}

\section*{AI Use Disclosure}

The original implementation was written before our use of AI agents. We are now reimplementing the software with AI agents in preparation for public release.
We acknowledge the use of generative AI tools in the following ways: creating and reformatting figures, editing text, reformatting tables and listings, literature searches and citation checks, drafting and revising manuscript text, drafting and checking mathematical exposition, general latex debugging, producing scripts to organize data output from our system into a format suitable for bulk processing via Paraview or Matplotlib, producing scripts to control Paraview and Matplotlib on the output data in bulk, and producing scripts to perform bulk operations in Adobe Illustrator.

\begin{acks}
Theadora Fields Collin acknowledges the support of the NSF Graduate Research Fellowship Program.
This material is based upon work supported by the Department of Energy, National Nuclear Security Administration under Award Number DE-NA0004266.
Further, we graciously acknowledged the support of Intel and NSF PPoSS Grant CCF-2217064; DARPA PROWESS Award HR0011-23-C-0101, NSF SHF Grant CCF-2107244, DoE PSAAP Center DE-NA0003965, DoE SciDAC-5 DE-SC0023116, DARPA SBIR HR001123C0139, DoE DE-NA0004266.
Oded Stein  is generously supported by a gift from Adobe Inc and graciously acknowledges the support of the NSF award No. 2335493.
We acknowledge the support of the following individuals in developing these ideas and also in some cases editing this manuscript: Sofia Chandler-Freed, Eleanor Mast, Ridgway L. Scott, Gordon Kindlmann, David Ham, Robert Kirby (Barylor), Manya Bansal, Caleb Kim. Jaeyeon Won, Mit Kotak, Sina Nabizadeh, Christopher McNally, Jesse Michel, Willow Ahrens, Richard Sollee, Filippo Carloni, Francesco Peverelli, Robert L Van Heyningen, Jaime Peraire, Cuong Nguyen, and many more.
\end{acks}

\bibliographystyle{ACM-Reference-Format}
\bibliography{all}

\appendix
\crefalias{section}{appendix}
\crefalias{subsection}{appendix}
\crefalias{subsubsection}{appendix}

\section{Alternative Analysis of Finite Element Transformations}
\label{sec:AAFET}
We factor the generalized \VA{} matrix of~\cref{eq:transformVander} to separate local DOFs, coordinate changes, and the reference basis. This gives an alternative construction of the symbolic matrix in~\cref{sec:SFIVFG} and identifies structure that the inversion routine in~\cref{sec:SFSI} can exploit. With $T\colon K\to K^{\prime}$, the factorization is $\mathsf{C}=HFG$, and the basis transformation remains $\mathsf{P}=\mathsf{C}^{-T}$.

\begin{definition}[Dual Basis Cover]
\label{def:dualBasisCover}
Suppose $E=(K,V,P,\Sigma)$ is a finite element. Choose a basis for each tensor power of the tangent space used by the derivatives, and a basis of covectors on $V$. A derivative-evaluation functional has the form
\begin{equation}
    \gamma^K_{x,m,d,k}(p)
      =v_k\bigl((D^m p)(x)(w_{d,m})\bigr),
\end{equation}
where $w_{d,m}\in (T_xK)^{\otimes m}$ and $v_k\in V^{\star}$. A \textbf{dual basis cover} is a finite collection of these functionals whose restrictions span $P^{\star}$ and which includes all derivative and value components at each point and derivative order it uses. The collection may be redundant.
\end{definition}

For the affine charts used here, the chain rule maps a complete block of derivatives at a physical point to derivatives of the same order at its reference preimage. Corresponding covers can therefore be constructed from the evaluation or quadrature points used by the DOFs. Quadrature representations must be exact on the finite element space for the identities below to hold exactly. For non-affine maps, lower derivative orders may also be needed.

\begin{samepage}
\begin{theorem}[Factorization through dual basis covers]
\label{thm:pb2}
Assume the isomorphism condition of~\cref{thm:pb1}. Let $\{\gamma'_r\}_{r=1}^{m'}$ and $\{\gamma_s\}_{s=1}^{m}$ be physical and reference dual basis covers. Choose matrices $H\in\mathbb{R}^{|\Sigma'|\times m'}$, $F\in\mathbb{R}^{m'\times m}$, and $G\in\mathbb{R}^{m\times|\Sigma|}$ satisfying
\begin{align*}
    \phi'_i &= \sum_r H_{ir}\gamma'_r &&\text{on }P',\\
    T_{-\star}\gamma'_r &= \sum_s F_{rs}\gamma_s &&\text{on }P,\\
    \gamma_s &= \sum_j G_{sj}\phi_j &&\text{on }P.
\end{align*}
Then the matrices defined in~\cref{eq:transformVander,eq:erg} satisfy
\begin{equation}
\label{eq:coverFactorization}
    \mathsf{C}=HFG,\qquad \mathsf{P}=(HFG)^{-T}.
\end{equation}
\end{theorem}
\end{samepage}

\begin{proof}
The spanning properties of the covers and the fact that $\{\phi_j\}$ is a basis ensure that the three representations exist. Applying $T_{-\star}$ to the first identity and substituting the other two gives
\begin{equation*}
    T_{-\star}\phi'_i
      =\sum_{r,s,j}H_{ir}F_{rs}G_{sj}\phi_j.
\end{equation*}
Evaluate this identity on reference basis function $f_k$. Since $\phi_j(f_k)=\delta_{jk}$, the result is
\begin{equation*}
    \mathsf{C}_{ik}=(T_{-\star}\phi'_i)(f_k)=(HFG)_{ik}.
\end{equation*}
The relation $\mathsf{P}=\mathsf{C}^{-T}$ from~\cref{eq:erg} completes the proof.
\end{proof}

\paragraph*{$H$: physical DOF weights.}
Each row of $H$ combines physical derivative samples into one physical DOF. It contains evaluation or quadrature weights, contractions with normals or tangents, and any integration measure factors in the DOF specification. This is essentially a representation of a DOF in terms of samples that are easy to pushforward, but the coefficients dependent on the geometry. Thus $H$ exposes the local sampling pattern without constructing physical nodal basis functions.

\paragraph*{$F$: local coordinate changes.}
For corresponding complete derivative blocks under an affine map, $F$ is block diagonal by evaluation point and derivative order:  These blocks depend on the Jacobian, its inverse, and the tensor transformation of the field values. For scalar first derivatives in Cartesian coordinates, the block is $(DT)^{-T}$; scalar value blocks are identities. This component is the change of coordinates via pullbacks. 

\paragraph*{$G$: reference sampling and polynomial identities.}
The entries are directly computable as
\begin{equation}
\label{eq:coverReferenceSampling}
    G_{sj}=\gamma_s(f_j).
\end{equation}
For fixed reference points and component bases, $G$ is geometry-independent and can be precomputed once, using the reference basis already constructed in~\cref{lst:vanderOverall}. It records dependencies among the possibly redundant samples on $P$. A derivative sample can depend on several reference DOFs, so $G$ need not be sparse merely because each sample is local. Its zero pattern must come from evaluating or analyzing the reference basis.

\paragraph*{Consequences for sparsity and inversion.}
The product is read from right to left: $G$ samples the reference basis, $F$ converts those samples to physical components, and $H$ combines them into physical DOFs. A sufficient condition for $\mathsf{C}_{ij}=0$ is that $H_{ir}F_{rs}G_{sj}=0$ for every pair $(r,s)$. Thus the nonzero patterns of the factors give a conservative dependency pattern for $\mathsf{C}$: an upper bound on its possible nonzero entries, or equivalently a lower bound on its sparsity in the sense of guaranteed zeros. This is a useful structural guarantee that explains why those entries vanish. Running the construction and symbolically simplifying $HF$ and $HFG$ can reveal a more precise sparsity pattern by exposing cancellations that the factor patterns alone do not capture.

If an ordering gives a block lower-triangular $\mathsf{C}$ with invertible diagonal blocks, then $\mathsf{C}^{-1}$ is block lower triangular and $\mathsf{P}=\mathsf{C}^{-T}$ is block upper triangular. 

\paragraph*{An alternative construction algorithm.}
The factorization suggests the following organization of the computation:
\begin{enumerate}
    \item Construct corresponding covers from the DOF samples, completing derivative and value components. On the fixed reference element, tabulate $G$ and establish its zeros symbolically or from exact reference data, rather than from floating-point roundoff alone.
    \item Compile $H$ from the physical DOF specifications and build the local blocks of $F$ from the symbolic map. Reuse each transformed sample wherever it occurs in multiple DOFs.
    \item Assemble $\mathsf{C}=HFG$ by contracting these blocks with the precomputed reference data. Use the factor patterns to skip known zeros, then simplify the remaining entries and identify any valid block-triangular ordering.
    \item Solve $\mathsf{C}X=I$ symbolically, using the recursive block procedure of~\cref{sec:SFSI} where applicable. Return $X=\mathsf{P}^{T}$ to match~\cref{lst:vanderOverall}, or transpose it to obtain the paper's basis matrix $\mathsf{P}$.
\end{enumerate}
This constructs the same matrix as directly evaluating every physical DOF on every mapped reference basis function, but this directly exposes a useful conservative sparsity pattern.

A limitation of this organization is that it constructs the pullback blocks in $F$ independently of how the DOFs use them through $H$. Both factors depend on the same geometry, and the contractions in $H$ can simplify or cancel transformations in $F$. Constructing the complete blocks of $F$ can therefore introduce intermediate expressions that disappear in $HF$. Simplification in step~3 can recover these cancellations, but only after those expressions have been constructed. Our implemented route compiles the complete DOF expressions using the coordinate-independent tensor and basis machinery of~\cref{sec:TEBA,sec:TEFG} and our related Grassmann work~\cite{collin2026grassmann}, allowing the pullbacks to be simplified in the context of their use. The factorization remains useful for explaining and bounding sparsity, even when constructing its factors separately is not the best way to generate the formulas.

\section{Sample Change Of Basis Rules}
\label{sec:scofb}
We now sketch our implementation for \lstinline!changeOfBasis(basisSrc: Basis, basisDst: Basis, env: Optional[Env]) -> Tuple[Formula, Formula]:!.
We implement this with rules to convert every function to and from the standard basis:
\lstinline!toStandardBasis(basis: Basis, env: Optional[Env]) -> Tuple[Formula, Formula]:!.
A critical example of such a rule is the rule for converting a dual basis to the standard basis:
\begin{lstlisting}[ language=PythonPlus, caption={[Converting Bases]{A rule convert the dual basis of a space to the standard basis of that space.}}, label={lst:stdBasisVander},abovecaptionskip=5pt]
def dualToStandard(basis: Basis, env: Optional[Env]) -> Tuple[Formula, Formula]:
    space = spaceOfBasis(basis)
    stdBasis = standardBasis(space)
    if basis == stdBasis:
        return (eyeFormula(space), eyeFormula(space))
    else:
        primalOf = basis.primal
        (to, from) = toStandardBasis(primalOf, env)
        return (from.T, to.T)
\end{lstlisting}

This rule shows the virtue of our design: since we already have the inverse on hand, we can avoid recomputing or losing some clever way to do so. For example, a raw basis requires a recursive call to our representation function and an inverse:

\begin{lstlisting}[ language=PythonPlus, caption={[Converting Bases]{A rule convert the dual basis of a space to the standard basis of that space.}}, label={lst:stdBasisRaw},abovecaptionskip=5pt]
def rawToStandard(basis: Basis, env: Optional[Env]) -> Tuple[Formula, Formula]:
    space = spaceOfBasis(basis)
    stdBasis = standardBasis(space)
    # raw basis can't be std basis
    vectorsOfBasis = basis.vectors
    formulas = [representTensor(vec, stdBasis, env) for vec in vectorsOfBasis]
    # make a formula that produced a formula for np.hstack on the formulas
    from = hstackFormula(formulas)
    # make a formula that produces the inverse of the output of the previous formula:
    to = inverseFormula(from)
    retun (from, to)
\end{lstlisting}

Other rules such as image basis or primal bases can also require inverses or other symbolic additions in formulas.
Some rules like the tensor product though are quiet obvious and nice:
\begin{lstlisting}[ language=PythonPlus, caption={[Converting Bases]{A rule convert the tensor product of a space to the standard basis of that space.}}, label={lst:stdBasisProd},abovecaptionskip=5pt]
def prodToStandard(basis: Basis, env: Optional[Env]) -> Tuple[Formula, Formula]:
    space = spaceOfBasis(basis)
    stdBasis = standardBasis(space)
    if basis == stdBasis:
        return (eyeFormula(space), eyeFormula(space))
    else:
        productends = basis.bases
        tos = []
        froms = []
        for b in productends:
            (to, from) = toStandardBasis(b, env)
        tos.append(to)
        froms.append(from)
        # kronocker product of inverse is inverse of productands:
        return (kronFormula(tos), kronFormula(froms))
\end{lstlisting}

\section{Implementation Aspects of SVAO and assembly}
\label{sec:iasa}
We now sketch an interface for the field/basis input and an implementation for assembly.
We show assembly is parametric over elements in a practical way by sketching how to implement the section mesh field sequence with an element.

\subsubsection{(Sectioned) Mesh Field (Sequence) Interface}
A sketch of our interfaces is in \cref{lst:fieldInterface1}.
To satisfy the preconditions of both definitions (\cref{def:MF} and \cref{def:SMFS}), we supply the requites topological type and specification of the mesh vector space with a field type (\cref{line:corefild}) as well as a schema (\cref{line:corefild1}).
To facilitate quadrature degree estimation in assembly, the interface optionally supplies the degree of the field outputs as a function of the degree of the inputs (\cref{line:degfield}).
To satisfy the actual functions available in both definitions, we have a tabulation function (\cref{line:fieldInterface:tab2} and \cref{line:fieldInterface:tab1}).
The tabulation function takes care to understand because it operates symbolically i.e, it generates code that performs the functionality of the numerical version.
We present the numerical version first.
The tabulation function in numerical mode (\cref{line:fieldInterface:nfq}) consumes  positions in a chart (world and reference space), a chart given numerically, a number of requested derivatives, the specification of the mesh entity corresponding to the chart  as well as any mesh entities contained therein, and a desired basis for the output values.
The tabulation function can also consume global arrays or functions that might be needed to define the field or sequence of fields.
For a sectioned mesh field, the tabulation function returns a collection of outputs indexed to the parts of the section relating to the queried mesh entity (\cref{line:fieldInterface:bout}).
The numerical version essentially satisfies the practical existence of the $C^{\infty}$ functions available on some mesh entity.
The symbolic version of tabulate can be understood as a staging or symbolic execution~\cite{Futamura1999} of the numerical version: the symbolic version is simply the numerical one that runs on symbolic inputs to produce symbolic outputs.
For example, the array inputs and outputs to the numerical tabulate are replaced with formulas while the mesh entities will be replaced with symbols for mesh entities in the context that the mesh field is used, mainly assembly.

\begin{lstlisting}[language=PythonPlus, caption={Data provided by a global field at code generation time.}, label={{lst:fieldInterface1}},abovecaptionskip=5pt, captionpos=b,abovecaptionskip=5pt, escapechar=\%]
class CoreData: %\label{line:corefild}%
    field: fieldType
    topType: TopologicalType

    def degree(x: int | Tuple[int, ..]) -> Optional[int | Tuple[int, ..]]: ...  %\label{line:degfield}%

class NumericFieldQuery:  %\label{line:fieldInterface:nfq}%
    refPoints: NDArray
    worldPoints: NDArray
    chart: Tuple[NDArray, NDArray], 
    derivativse: int,
    meshEntity: MeshEntity,
    subMeshEntities: List[MeshEntities],
    basis: Optional[Basis] = None
    context: dict[str, Array | Any] = {}

class SymbolicFieldQuery: %\label{line:fieldInterface:symCon}%
    refPoints: Formula
    # as before but symbolic variables

class MeshField:
    # compile time data:
    core: CoreData
    basis: Optional[SectionedMeshField]
    # runtime data:
    sectionData: Optional[Tuple[Section, Array]]
    globalData: dict[str, Array | Any ]

    def tabulate(query: NumericFieldQuery | SymbolicFieldQuery) -> Tuple[NDArray | Formula, Basis]: ... %\label{line:fieldInterface:tab1}%

class BasisOut:  %\label{line:fieldInterface:bout}%
    commonFormulas: List[Tuple[str, Formula]]
    sectionedOut: dict[Tuple[Any, MeshEntity, MeshEntity], NDArray | Formula]
    basis: Basis

class SectionedMeshField:
    # compile time data:
    core: CoreData
    L: List[Any] 
    schema: dict[Any, Tuple[TopologicalType, TopologicalType]] %\label{line:corefild1} %
    
    def tabulate(query: NumericFieldQuery | SymbolicFieldQuery) -> BasisOut: ... %\label{line:fieldInterface:tab2}%
\end{lstlisting}

\subsubsection{Elements and Sectioned Mesh Field Sequences}
\label{sec:elemEasy}
To be usable, an element simply needs to implement a sectioned mesh field sequence, which is almost trivial to do automatically with our current infrastructure.
The primary topological type comes directly from the underlying domain of the element while the field and degree comes directly from the polynomial space.
The schema can be created by direct inspect of the DOFs/IPTEMs as they define $L$ and their integration and output types define the schema.
Finally, tabulate can be implemented by using the polynomial space to tabulate values on the reference space and then by using the Vandermonde matrix to transform those values to values for the DOFs on the given chart. 

\subsubsection{Assembly Code Generation}
\label{sec:AGC}
Our assembly function generates code for given a combination of mesh class\footnote{the topological types involved}, IPTEMs, and global fields (modulo runtime data).
The generated assembly function uses arrays associated to meshes and fields to build a tensor, the SVAO. 
Our assembly function caches the generated function and figures out how to wire runtime data (e.g. arrays of vertices or arrays for mesh relationships) into function arguments.
We sketch the code generation of the function, which reuses our existing infrastructure, mainly the compilation of IPTEM.
We sketch our code generation as five phases: preparation for multiple compilation, multiple compilation, analysis of multiple compilation, code generation, and meta data recording.

\begin{enumerate}
    \item \textbf{Preparation for IPTEM compilation:} Based on the IPTEM, we build a system of charts and fields that models the execution of parts of \eqref{eq:comp}. In particular, we determine the largest integration chart and generate all charts within that chart; this similar to the first stage of vandermonde matrix construction (~\cref{lst:vanderOverall}). Then for every field with a chart smaller than the largest integration chart, we introduce another symbolic chart to model neighboring geometries that a field could be sampled from. Each field input is then associated to at most two charts; we make up to two symbolic field inputs (\cref{lst:fieldInterfaceLocal}) that are placeholders to record how the fields from different charts are used.
    \item \textbf{IPTEM Compilation:} IPTEMs are paired with all the generated charts and fields to produce formulas. Local reductions are applied.  We make sure the integrals are compiled as loops and not evaluated. We just use the compiler of \cref{sec:swrdoIPTEM}, specifically \cref{lst:IptemCompilerArchitecture}.
    \item \textbf{Analysis of IPTEM Compilation} Based on the resulting formula, we can extract a wide variety of meta-data to generate our code. In particular, we can use the place holder fields to figure out how to query input fields i.e., at what points, geometries (one within the largest integration chart or one outside), and derivatives. We can also construct a schema for the output section and check if there are multiple values for any outputs. Finally, we can determine based on the used charts and fields that mesh/section queries/relationships are needed.
    \item \textbf{Code Generation:} The inner most loop of our generated code are the multiply compiled IPTEMs. The function that assembly runs is built out around from them. We sketch this structure in \cref{lst:template1}. First, these are lifted to operate on incoming groups of basis functions (\cref{line:t1} - \cref{line:t2}). Second, these are lifted to operate on particular geometric entities, which means place them within an outer loop over all integration charts and possibly inner loops from intersecting geometries on other charts (\cref{line:t3} - \cref{line:t1}). Within these loops, appropriate geometric information must be constructed to supply charts and then queries to fields must be created using the tabulate interface (\cref{line:f1,line:g1,line:f2,line:g2}). After the IPTEMs and within the geometric loops, the values of the IPTEMs must be added to the COO values and the compile times schema should be used to create appropriately ordered indices from queries to runtime meshes/sections (\cref{line:s1}-\cref{line:s2}). Finally, global arguments from meshes/fields/sections complete the function, modulo some metadata that allows meshes/fields to be unpacked for these (\cref{line:te}).
    \item \textbf{Meta Data:} Global reduction information is not used in the assembly function; instead, the global reduction is attached to the SVAO output as a default reduction. The output section is another global argument, but the schema metadata for the output section is used in the function that calls the generated code to construction a section if one is not present already.
\end{enumerate}

\begin{lstlisting}[language=PythonPlus, caption={[Global Code Generation Template]{A visual depiction of how the various intermediate outputs of our global code generation process are stitched togeather to produce a global kernel function.}},label={lst:template1},captionpos=b,abovecaptionskip=5pt, escapechar=\%]
def assemble(outSectionArrays,  
            meshTopologyArrays, 
            meshVertices,
            meshRelationships, 
            nTopCell,
            inputFieldSections,  inputFieldArrays, inputFieldOtherArgs,   % \label{line:te} %
            ) -> Tuple[Array, Array]: # BCOO
    coo_indices = []
    coo_vals = []
    for topCell in range(nTopCell): % \label{line:t3} %
        # Setup geometry for topCell and subcells % \label{line:f1} %
        # Setup fields at quadrature points on this cell % \label{line:g1} %
        for otherCell1 in ...:
            for otherCell2 in ...:
                # Setup fields at quadrature points  % \label{line:f2} %
                # Setup geometry for otherCells % \label{line:g2} %
                for dof0 in ...: % \label{line:t1} %
                    for dof1 in ...:
                        vals = # IPTEM quadrature loops % \label{line:t2} %
                        indices = ... % \label{line:s1} %
                        coo_vals.append(vals)
                        coo_indicies.append(indicies) % \label{line:s2} %
    # Return all blocks of values and indices
\end{lstlisting}

\section{Details on Global Operations}
\label{sec:appDGO}
\subsection{SVAO Operations}
\label{sec:GAAO}
We now outline the interface to each operation in turn.
For reductions, we support reduction over the non-specified axes and the reduction can be chosen up to indices from the specified axes. 
The defaults are inferred from the IPTEM's default reductions, which can be overridden in the global case (see ~\cref{grammar:reduction}).
Via the metadata for reductions and IPTEMs included, we support (when possible) automatic conversion to a global field (e.g., for interpolation) or automatic conversion to a scalar or dense vector or sparse matrix.
The interfaces are summarized in ~\cref{lst:red1}.
Slicing functionality is similar to slicing in NumPy except that dimensions are sliced with our global constructions: sub-sections, sub-meshes, list of IPTEM, or indices for local linear inputs.
Slicing can simply extract a slice, but the non-zeros of a slice can also be set.
As a convenience, we offer an option to set the diagonal over two axes, useful for boundary conditions.
Slicing is detailed in~\cref{lst:slice}.
Finally, we offer a where-functionality that allows us to construct subsections, sub-meshes or just lists of indices.
The interface is detailed in~\cref{lst:where}.
As a final addendum, we note that users can extract the output section and dimension information, which is how we expect users to get such information.
This is detailed in~\cref{lst:meta}.

\begin{lstlisting}[language=PythonPlus, caption={Reduction Interfaces},label={lst:red1},captionpos=b,abovecaptionskip=5pt]
# Reduction optionally using indices and/or the number of non-zeros in the reduction region.
Reducer = Callable[[NDArray], NDArray] | Callable[[int, NDArray], NDArray] | Callable[[NDArray, NDArray], NDArray] | Callable[[int, NDArray, NDArray], NDArray]

def reduce(ao: AssemblyOutput, axes: Optional[List[str]] = None, reducer: Optional[Reducer] = None) -> AssemblyOutput: ...
def asScalar(ao: AssemblyOutput) -> Optional[float]: ...
def asDenseVector(ao: AssemblyOutput) -> Optional[NDArray]: ...
def asScipySparseMatrix(ao : AssemblyOutput, format: str = "coo") -> AssemblyOutput: ...

\end{lstlisting}

\begin{lstlisting}[language=PythonPlus, caption={Slice Interfaces},label={lst:slice},captionpos=b,abovecaptionskip=5pt]

sliceIndex = List[IPTEM] | SubSection | SubMesh | NDArray | IPTEM | int

def slice(ao: AssemblyOutput, dimensions: List[str], subsets: List[sliceIndex]) -> AssemblyOutput

Setter = Zero | flat | Callable[[NDArray, NDArray, NDArray], NDArray] | AssemblyOutput 
# The second to last option takes slice indices, current values of the slice, and the non-zeros of the slice.
# The last option must have been sliced on the appropriate dimensions with the same subsets.

def setSlice(ao: AssemblyOutput, dimensions: List[str], subsets: List[sliceIndex], setter: Setter = Zero) -> AssemblyOutput: ...

def setDiagonal(ao: AssemblyOutput, dimension1: str, dimension2: str, subset1: sliceIndex, subset2: sliceIndex, val: float | NDArray) -> AssemblyOutput: ...

\end{lstlisting}

\begin{lstlisting}[language=PythonPlus, caption={Where Interfaces},label={lst:where},captionpos=b,abovecaptionskip=5pt]

sliceIndex = List[IPTEM] | SubSection | SubMesh | NDArray | IPTEM | int

FilterTy = allZero | allNonZero | Callable[[int, NDArray, NDArray], bool] 

def where(ao: AssemblyOutput, filter: Optional[FilterTy] = None) -> sliceIndex: ... 

\end{lstlisting}

\begin{lstlisting}[language=PythonPlus, caption={Metadata Interfaces},label={lst:meta},captionpos=b,abovecaptionskip=5pt]

def shape(ao: AssemblyOutput, dim: str) -> int: ...

def outSection(ao: AssemblyOutput) -> Section: ...


\end{lstlisting}

\subsection{Overall User Workflow}
\label{sec:OUW}
\begin{figure}
    \centering
    \includegraphics[width=0.75\linewidth]{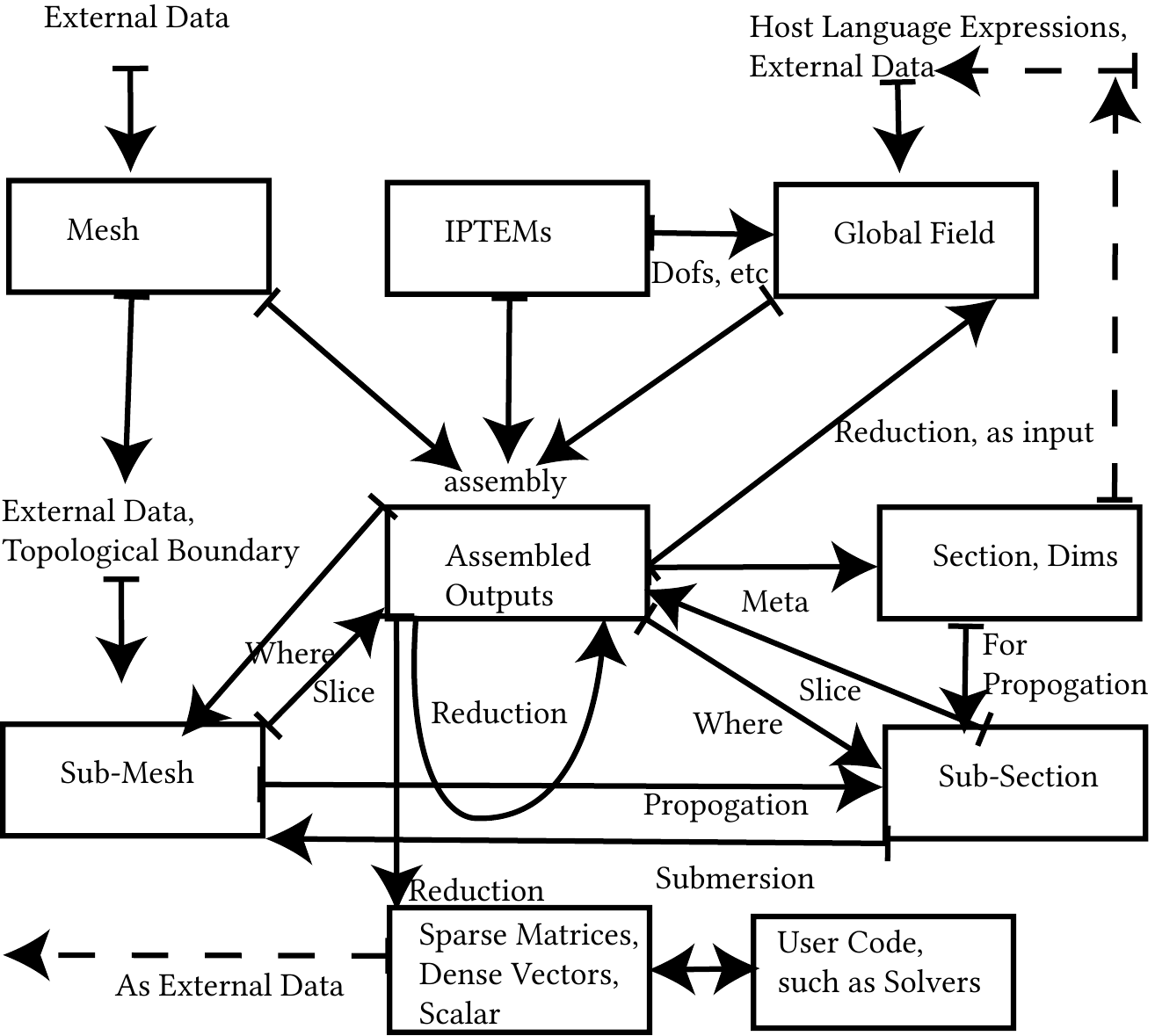}
    \caption{Overall Global Work Flow}
    \label{fig:workflow}
\end{figure}
We depict an overall user workflow in \cref{fig:workflow}.
As stated previously, the idea is to start with a few bespoke objects and then use IPTEMs and assembly to progressively build others that various post-processing operations convert to more objects to build or post-process further global objects.
In this version, we also suggest how this would interact with a full pipeline, involving solvers and other external user code.

\subsection{Sub-Meshes and Sub-Sections}
\label{sec:subs}
Meshes and sections have subsets, representing sub meshes (including boundaries) and subspaces.
These are simple to define and our main focus is on constructing subsets, as this is where our generality shines.

\begin{definition}[Sub-Mesh]
Given a mesh $(V, F)$, a sub-mesh is simply some other mesh $(V^{\prime}, F^{\prime})$ with $V^{\prime}\subset V$ and $F^{\prime}\subset F$.
\end{definition}

We note that this is differently from simply a subset of mesh:
\begin{definition}[Mesh Subset]
Given a mesh $(V, F)$, a mesh subset is simply a subset $F^{\prime}\subset F$.
\end{definition}
For sections, sub-sections are like mesh subsets:
\begin{definition}[Sub-Section]
    Suppose we have a mesh $(V, F)$ and a schema $f\colon L\to\mathcal{T}^{2}$, generating a section $S_{V,F, f}$.
    A sub-section is simply a subset of $S_{V, F, f}$.
\end{definition}

\begin{figure}
    \centering
    \includegraphics[width=0.75\linewidth]{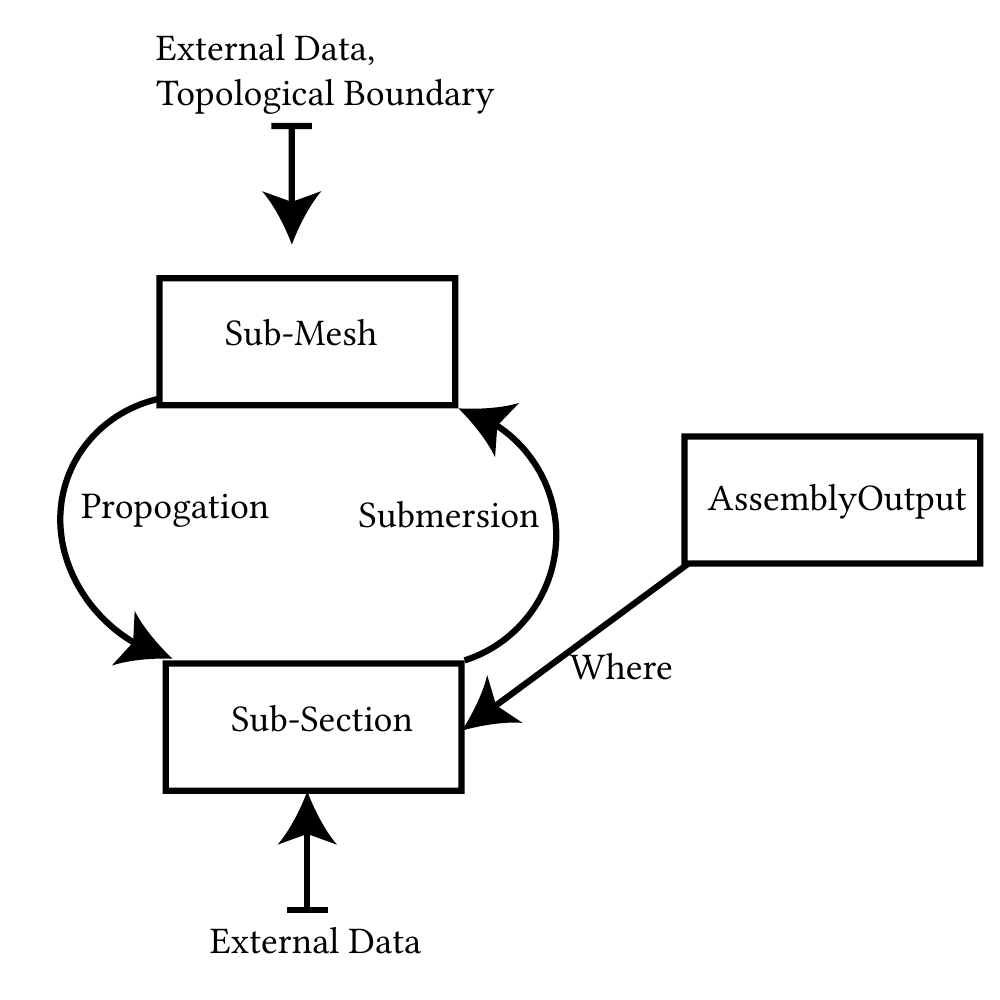}
    \caption{Mechanisms for constructing sub-meshes and sub-sections}
    \label{fig:submeshconstr}
\end{figure}

Sub-sections and sub-meshes can be constructed in a variety of ways, showcasing how various global aspects of the system come together.
We depict the situation in ~\cref{fig:submeshconstr}.
Sub-meshes and sub-sections can be constructed from external data or the special case of the topological boundary constructed from the relationships between mesh entities.
Critically, a sub-mesh $(V^{\prime}, F^{\prime})$  can be \textbf{propagated} to a sub-section:
\begin{definition}[Propagation]
    Given a sub-mesh $(V^{\prime}, F^{\prime})$ and a section $S_{V,F,f}$, the \textbf{propagation} of the sub-mesh to a sub-section is given by 
    \begin{equation*}
    \{(l, t1, t2)\colon (l, t1, t2)\in S_{V, F, f}, t1\in F^{\prime}\}.
\end{equation*}
\end{definition}

This simply extracts the portion of the section where the entities are on the sub-mesh; this captures the usual boundary basis functions of an FE space.
However, critically, we can also go the other way sometimes.
If we have a schema $f\colon L\to\mathcal{T}^{2}$ where $\pi_{1}\circ f$ is injective, then any section over $f$ can be converted into a sub-mesh. 
\begin{definition}[Submersion]
    Given a subset $S$ of a section $S_{V, F, f}$ where $\pi_{1}\circ f$ is injective, we can define a subset mesh via first defining a subset $F^{\prime}\subset F$  via:
    \begin{equation*}
        F^{\prime} = \{f\in F\colon \exists! (l,t_{1},t_{2})\in S_{V, F, f} \mbox{ s.t.} t_{1} = f\}.
    \end{equation*}
    and then we can define a sub-mesh $(V^{\prime}, F^{\prime\prime})$ as the smallest mesh such that  $F^{\prime}\subset F^{\prime\prime}$.
    We call this sub-mesh a \textbf{submersion}.
\end{definition}
This little bit of topological ingenuity allows all sorts of data to be transported to sub-meshes, such as using Lagrange elements to identify a boundary.
Finally, assembled outputs with a given section can be used with a predicate to filter a section or sub-section, via a where constructed to be discussed in~\cref{sec:GAAO}.
Additionally, we note that subsections and sub-meshes can be trivially combined with boolean operations, including and/or/complement, as they are simply masks.
We now proceed to discuss the assembled outputs, the runtime output of generated code, which can be used to create these sub-meshes via filtering or global fields using reductions.
We will see that these can also be sliced and filtered via sub-meshes and subsections, facilitating the construction of many algorithms, such as elimination of boundary DOFs.

\section{Permutations and FEM}
\label{sec:permutations}
By permutations, we mean permutations of vertices in a mesh.
Reasoning about permutations is a core issue in mesh data structures.
In the FEM literature, \cite{constructionofarbitraryorderfiniteelementdegreeoffreedommapsonpolygonalandpolyhedralcellmeshes2022} gave the most concise explanation for the impact of permutations of mesh data on the implementation of elements.
We summarize here and our solution is a programmatic version of their solution.

\paragraph{Core Problem} Suppose we have two triangles $T_{0}, T_{1}$, each parametrized in terms of a reference triangle $T$ via maps $F_{0}, F_{1}$.
The edge, $e=T_{0}\cap T_{1}$ can be parametrized in terms of a reference edge in either triangle: $e_{i} = F_{i}^{-1}(e)$.
Both reference edges admit paramterization $E_{i}\colon [0,1]\to e_{i}$, which leads to two different parameterizations of the $e$ via $G_{i} = F_{i}\circ E_{i}$.
Since our computations are not paramterization invariant, this choice matters and so can lead to inconsistent results.
For example, a DOF computed with an edge integral might depending on the choice of $G_{i}$.

\paragraph{Core Solution} Thankfully, there is a third paramterization of $e$: since our vertices are globally ordered, we can pick a $G\colon [0,1]\to e$.
This is a global orientation for $e$.
Further, based on the ordering of the vertices in $T_{i}$, we can compute $\sigma_{i}\colon [0,1]\to [0,1]$ so that $G_{i}\circ \sigma_{i} = G$.
These $\sigma_{i}$ are local orientations.
We need to adjust everything we do by $\sigma_{i}$ to ensure we get correct answers. 
Thankfully, we do have one useful identity to help adjust some calculations: $(G_{i}\circ \sigma_{i})^{\star} = \sigma_{i}^{\star}\circ G_{i}^{\star}$.
Thus, for some calculations, we can ignore the permutation and adjust later based on the local orientations.
Since there are finitely many options for local orientations,  this amounts to computing and applying a finite list of corrections to some calculations.

\paragraph{Outline} We divide the rest of this section into infrastructure for modeling permutations in our system and then the applications of actual corrections to ensure correct results.
Though this work is an appendix, we wish to stress that it was where most of the bugs occurred.

\subsection{Modeling Permutations}
We now sketch the infrastructure to model permutations in our system.

\paragraph{Permutations, Topological Types, and Charts}
We must characterize permutations and how they relate to charts to solve the above problem.
We first consider permutations of an object of a given topological type.
Our grammar for topological types (\cref{grammar:pointwiseTensorType}) limits the types of permutations we need to consider.
Given a $\operatorname{Simplex}(d)$, we obviously need to consider all permutations of the vertices $S^{d}$.
Given a $\operatorname{Product}(t_{0},\dots, t_{n})$, we can consider the symmetries of each $t_{0}$ as $S_{i}$.
We can also see that there are $S^{n}$ ways to arrange these.
This leads to a semi-direct product of $S^{n}$ with $\times_{i}S_{i}$.
One can check this reproduces the group $D_{4}$ as a semi-direct product of $S^{2}$ with $Z_{2}\times Z_{2}\cong Z_{4}$.
Thus, the inductive structure of our topological types induces a structure for the permutations.
Given a chart $(S_{1}, S_{2}, F)$ (\cref{def:chart1}), we add meta-data to encode if we wish to permute the vertices of $S_{1}$ in terms of an element of the permutation group of the associated topological type. 

\paragraph{Permutations and IPTEM Compilation}
Given an IPTEM, we can compute all the chart/topological types that it depends on.
This follow immediately from the grammar (\cref{grammar:pointwiseTensor} and \cref{grammar:reduction}).
Thus, when we compile IPTEMs, we can optionally supply a permutation for each chart input.

\paragraph{Permutations and Meshes}
Our meshes (\cref{def:meshtop}) store a global ordering on vertices $V$.
Given any $f,f^{\prime}\in F$ with $f\subset f^{\prime}$, we have an ordering of the vertices of $f$ in $f^{\prime}$ induced by the isomorphism from $f^{\prime}$ to the standard entity (\cref{def:meshtop}).
The orientation of $f$ in $f^{\prime}$ is the permutation of the vertices ordered according to $f^{\prime}$ that puts them into their global order.
Our meshes by default store the orientations of all entities with respect to the top level entities.
We can compute the others on demand.

\subsection{Applying Permutations}
We now sketch the infrastructure to apply permutations in our system.
\paragraph{Permutations and Elements}
We follow \cite{constructionofarbitraryorderfiniteelementdegreeoffreedommapsonpolygonalandpolyhedralcellmeshes2022} when we modify elements to deal with permutations.
When we tabulate basis function values or compute \VA{} matrices, we don't permute vertices.
These interfaces don't care about permutations and we assume the ordering induced by whatever charts are passed in.
Instead we separately apply a correction for permutations that allows those calculations to still occur symbolic once per reference element.
We use the infrastructure of \cref{sec:SRFE} and the above to compute the corrections.

Orientation corrections are computed once per element as a mapping from DOFs and potential permutations of entities in a reference element to rows of an adjustment matrix that adjust the DOFs in terms of other other DOFs.
To compute the corrections, for each DOF/IPTEM, we pre-compute how different combinations of permutations impact the DOF in terms of other DOF: following our infrastructure in \cref{sec:SRFE}, we compute (numerically) \VA{} matrices for every combination of permutations of charts of the reference element that a DOF depends on and extract the results.
We note that in cases where the topological type representing the domain of an element is not the top level element in a mesh, we must also consider permutations of this and that these must be propagated to the polynomial space.
This is accounted or because our field protocol (\cref{lst:fieldInterfaceLocal}) takes charts to define the geometries.
Since we are computing \VA{}, we note that this is close to computing pullback matrices on permutation adjustments, but that this logic does not hold when permutations of multiple mesh entities impact a DOF.
A better way to think about computing correction is this: we figure out how to write a basis function given certain permutation assumptions as a linear combination of other basis functions on the reference element.
This is the classic interpretation of \VA{} matrices. 

\paragraph{Permutations and Assembly}

Permutations impact our general assembly code generation (\cref{sec:AGC}) in two ways.
First, when we generate code to query elements, our implementation of \cref{lst:fieldInterface1} for elements must use the pre-computed permutation information to adjust sample basis function values.
Second, when we setup geometries in \cref{lst:template1}, we must actually setup reference geometries and geometries permuted according to global orientations.
The reference geometries and permutation meta-data go to sampled fields while the IPTEMs are assembled using the permuted geometries/charts.
In essence, when we sample fields, we expect to adjust them after doing reference calculations but when we actually compute an IPTEM we wish to do so in world space directly and so we just directly adjust the vertices that we patch in.

\section{IPTEM Default Arguments}
\label{sec:defaultArgs}
Our pointwise tensor expressions and integrated pointwise tensor expressions can have default arguments. 
Default arguments non-trivially aid the specification of elements, but they do requiring handling across multiple levels of our system.
The advantages can be seen through \cref{sec:bihmres} and \cref{sec:feecres}.
In both, element specifications are simplified via a recursive specification of basis functions wherein new DOF depend on primal basis functions, which we represent by allowing elements to provide access to the $k$th basis function and allowing this basis function to be supplied as a default argument.
In the case of \cref{sec:bihmres}, a true specification of the DOFs requires some way to use average characteristic length\footnote{Characteristic mesh length is the ratio of a triangle volume to permitter; the average of this quantity means that at an edge or vertex we average the characteristic length incident to an entity}: derivative DOFs must be weighted by this quantity in massive simulations as otherwise the condition number of the mass matrix varies too much with the mesh.
We handle this with default arguments too: we can expression this quantity via a few simple integrated pointwise tensor expressions that use each other as default arguments.
We find this to be quite elegant: a quantity that is typically builtin to other systems can emerge naturally.
On the other hand, this does require some plumbing.

\paragraph{Interface}
Our external interface is a simple modification to \cref{grammar:pointwiseTensor}: field arguments can have defaults.
The default can be another IPTEM (a zero degree field whose integration and output types allow the implementation of \cref{lst:fieldInterfaceLocal} given appropriate charts).
The default can also be anything that implements \cref{lst:fieldInterfaceLocal}.
For example, we extend elements to provide an implementation of \cref{lst:fieldInterfaceLocal} given a request for a $k$th basis function; internally, they simply generate the code/formulas they would to be sampled on a given chart in assembly and then extract the $k$th row from this.

\paragraph{IPTEM Compilation}
To handle default arguments, IPTEM compilation does not require much modification.
Moreover, the simplest approach is the best: we simply compile PTEM ignoring default arguments, treating them as unknown variables, and then query the fields for formulas to substitute in when we integrate. 
This does require some book keeping for quadrature points, bases, and charts.
This path works for reasoning about elements: generating Vandermonde matrices, reference space tabulations of field values, and generating permutation matrices.
Additionally, we can optionally postpone the query and substitution step so that global assembly can treat default arguments slightly differently.
\paragraph{Global Assembly}

Global assembly handles default arguments mostly with plumbing.

Global assembly must handle default arguments that appear in DOFs used to define a space as well in the IPTEM representing operators.
Thus, global assembly code generation must gather all the default arguments at the start of compilation.
Global assembly handles the two types of default arguments separately. 
For fields, global assembly just needs some additional bookkeeping to query the fields and substitute the code in before the necessary calculations. 
Thus global assembly just uses the IPTEM compilation path with additional bookkeeping.
For IPTEM arguments, global assembly acts recursively: global assembly generates code to assemble IPTEMs that are default arguments, applies the default reductions (including global ones), and then supplies the reduced SAVO for the IPTEMs as mesh fields.
To do this, global assembly code gen simply leaves the default arguments as non-linear input arguments and then at runtime global assembly provides default arguments assembled constructed by calling global assembly recursively.
By handling IPTEM defaults via a runtime and code gen separation, IPTEM defaults can cached between assembly calls.

\section{Additional IPTEM syntax in Python}
\label{sec:additional}
In our introduction to tensor expressions (\cref{grammar:tensorExpr1} and the examples (\cref{sec:examlples}), we slowly introduced a Python embedding of our language for IPTEMs. 
We now gather some of the details that might be unclear, in terms of semantics and implementation.
There are several topics like this that we gather here:  tracing, indexing operations, data inside charts, and exterior calculus.

\paragraph{Tracing} We utilize tracing~\cite{frostig2019compiling} to implement our DSL. Tracing, most notably utilized in Jax\cite{jax2018github} and PyTorch~\cite{10.1145/3620665.3640366}, is a technique to embed a DSL in a language.
In short, ordinary code is used to produce AST for a DSL by running the code on specialized tracing values whose operations are overloaded to produce AST instead of normal program values.
We utilize this technique with decorators such as \lstinline!@pointwise! or \lstinline!@pointwiseEval! or \lstinline!@pointwiseIntegral!.
These use Python's inspect functionality to find the field and chart types, construct appropriate field and chart tracing values, and run the function on these tracing values to extract an appropriate IPTEM expression.
To make our system more flexible, we could retrace under certain circumstances, such as different Euclidean dimensions, so that users could specialize IPTEMs to these circumstances.

\paragraph{(I)PTEM meta data such as reductions, integration, and storage}
To actually construct (I)PTEMs, we provide three decorators: \lstinline!@pointwise!, \lstinline!@pointwiseEval!, and \lstinline!@pointwiseIntegral!.
Consquently, these decorators must also supply metadata in \cref{grammar:reduction} and \cref{grammar:pointwiseTensor}.
We supply integration domains and barycentric coordinates directly to the decorator as required arguments. 
For reductions, we offer a variety of keyword arguments and further array of defaults.
In particular, reductions are assumed to be applied over charts by default unless overriden and reductions can be supplied via the builtin one or via python code conforming to the interface used in global reductions ().
Finally, storage can be overridden via a simple keyword argument to the decorator, \lstinline!storeAt=ChartType!.

\paragraph{PTEM: Chart interface}
Our extensions to the tensor expression language in \cref{grammar:pointwiseTensor} are reflected in a more Pythonic way in the implementation.
In particular, chart values expose methods, \lstinline!c.CT!,  \lstinline$c.T$, \lstinline$c.Rk$ for the tangent space, cotangent space, and Euclidean space.
As a convince, we also provide  \lstinline!c.P! and \lstinline!c.E! to project or embed between (co)tangent and (co)Euclidean spaces.
Thanks to tracing these can stand into both as conveniences.
The chart also exposes \lstinline!c.tangentVector!, \lstinline!c.cotangentVector!, \lstinline!c.normalVector!, and \lstinline!c.conormalVector!.
These match up to \cref{grammar:pointwiseTensor} by having the first chart value being \lstinline!c!.
Similarly, the spatial position, $\Xop$, is attached to any chart argument, via  \lstinline!c.x!.
Lastly, we support \lstinline!chartOf! as a utility for any tensor expression in an (I)PTEM that provides the unique chart associated to an expression if it exists; this avoids unnecessary chart arguments.

\paragraph{PTEM: Field interface}

Our expression of fields suffers a bit in the Python interface.
Via \cref{grammar:pointwiseTensor}, we have a field at the point $x$ via field access and we have derivatives.
With a field argument \lstinline!f!, we support field access with some number of derivatives via \lstinline!f(k)! with $k=0$ being the plain value.
We also unadorned  \lstinline!grad(f)! or \lstinline!D(f)! for a derivative as well as \lstinline!grad(e)! for general tensor expressions including  \lstinline!grad(f(k)!.

\paragraph{Indexing Operations and other overloads}
\label{sec:uggappbad}
Within our Pythonic embedding of the IPTEM language, we allow indexing of tensors to take on several meanings. 
We aim to balance a coordinate independent representation with convenience.
For example, for any vector in a space built purely from direct sums, tensor products, dual spaces, and Euclidean spaces, our system allows standard indexing because such spaces in oru system always operate in an inner product that is the dot Euclidean dot product.
On the other hand, for other spaces, we do not allow the standard Pythonic syntax for indexing of tensors.
Instead, for all other indexing, we only support non-standard indexing of the form \lstinline!a[(k,)]! for a fixed integer \lstinline!k!;
our system allows this as a hack to access the coefficient of some tensor in $k$th standard basis entry.
We would like to remove this in our system and only use it for constructions like the Koszul null space specification (\cref{lst:koszuldofs}) where we need DOFs that together zero a whole tensor.
We believe we can remove this by allowing more generalized DOFs that output tensors, but we decided that having this hack might be convenient if one really needs it. 
This would also simplify away the need for \lstinline$stdEuclideanBasis(space)$, an iterator over vectors forming the standard basis for a space built purely from direct sums, tensor products, dual spaces, and Euclidean spaces.

In addition to indexing, we overload the standard Pythonic binary operations as much as possible. We support \lstinline$+$ as sum, \lstinline$*$ as tensor product for spaces, vectors, and basis, \lstinline$|$ as direct sum for spaces, vectors, and basis. We also allow \lstinline$**$ for tensor power and \lstinline$^$ for direct sum power.

\paragraph{Exterior Calculus}
Since exterior calculus is fundamental to many types of physical modeling~\cite{exteriorcalculusingraphicscoursenotesforasiggraph2023course2023}, we provide users with helpers to make using it easy.
Since we use tracing to implement our DSL, these are nothing more than Python functions that construct tensor expressions that can be used in IPTEMs.
And these functions are quite simple since it is not hard to embed them.
First, we define \lstinline$Symmeterize(v)$, which implements tensor symmetrization for a tensor $v$ in a tensor product space.
Due to tracing, we can pattern match on the space of $v$ to construct the correct expression.
With this, we can define \lstinline$Lambda(formDeg, ct=None, space=None)$ as the $\Kernel$ of \lstinline$Symmeterize(v)$ where $v$ is assumed to be in either a given space or the cotangent space of a chart of a given chart or chart type.
Then, \lstinline$antiSymmeterize(v)$ is simply the projection to the kernel of the  symmetrization map determined by the space of \lstinline$v$
Like with \lstinline$Symmeterize$, we can define \lstinline$Wedge(v1,v2,...)$ via taking their tensor product, flattening the product with \lstinline$reshape$ and then applying \lstinline$antiSymmeterize$ to the result. 
For associativity of forms to hold, tracing must be used to find the involved form degrees and insert a correction factor (see \cite{linearalgebraviaexteriorproducts2010} for details).
We can also define \lstinline$interiorEval$ for values in a \lstinline$Lambda$ space of a dual space via $\Embed$ to expose the dual space, $\Eval$, and finally \lstinline$antiSymmeterize$ i.e $\Project$.
Finally, the Hodge start can be implemented by inverting the maps $h_{1} = x\mapsto (y\mapsto y\wedge x)$ and $h_{2} = x\mapsto (y \mapsto <x,y>dV)$ and then defining $\star x = h_{1}^{-1}(h_{2}^{-1}(x))$. The way to think of this is that $y\wedge \star x = <y,x>dV$ so given a particular $x$, we gain a vector $y\mapsto <y,x>dV$ and we wish to finds its image in $x\mapsto y\mapsto y\wedge x$.
Using $\Inverse$ and $\Abstract$ makes this possible up to two important details that tracing allows us to patch up: the standard inner product on $k$ forms induced by our language is the desired inner product up a scaling factor involving the factorials of the k forms (requiring us to detect the form degree and insert them) and the volume form must be specified in terms of the  corrected form inner product.
Again, we follow \cite{linearalgebraviaexteriorproducts2010} in this development of forms.

\section{Analytical Solutions for the Hodge Decomposition}
\label{sec:ashd}

First, we consider a flat annulus of outer radius $2$ and inner radius $1$.
We decompose the vector field:
\begin{equation}
    \omega(x,y) = (2 (x-y) + 1, 2(x+y)+1).
\end{equation}
The analytical form of the Hodge decomposition is known.
Set $vx(x,y) =  \frac{x}{||(x,y)||^{2}}$ and $vy(x,y) =  \frac{y}{||(x,y)||^{2}}$.
Then we can write the decomposition via
\begin{equation}
    d\alpha(x,y) = (2x- \frac{15}{\log{4}} vx(x,y), 2y- \frac{15}{\log{4}} vy(x,y))
\end{equation}
and
\begin{equation}
    d\beta(x,y) = \begin{bmatrix} 0 & -1\\ 1 & 0
    \end{bmatrix} d\alpha(x,y)
\end{equation}
and
\begin{equation}
    hn(x,y) = (\frac{15}{\log{4}} vx(x,y), \frac{15}{\log{4}} vy(x,y))
\end{equation}
and
\begin{equation}
    ht(x,y) = \begin{bmatrix} 0 & -1\\ 1 & 0
    \end{bmatrix} hn(x,y)
\end{equation}
and finally,
\begin{equation}
    \eta(x,y) = (1,1).
\end{equation}

Second, we consider a cylinder of radius $1$ and height $1$.
We further specify that this is a manifold with a metric inherited from $\mathbb{R}^{3}$.
To write this down with clarity, we use two bits of notation.
We utilize cylindrical coordinates, $(\theta, z)$, and Cartesian, $(x,y,z)$, simultaneously. 
We also utilize an orthonormal frame for the tangent space via
\begin{equation}
    tu(\theta, z) = (-\sin{\theta},\cos{\theta}, 0)
\end{equation}
and
\begin{equation}
    tv(\theta, z) = (0, 0, 1).
\end{equation}

Then the vector field 
\begin{equation}
\omega(\theta,z) = (\theta^{2} + \theta - z^{2}+2z)tu(\theta,z) -2 (z+z\theta)tv(\theta,z)
\end{equation}
has a decomposition as follows:

\begin{equation}
    d\alpha(\theta, z) = (0, 0, z-0.5).
\end{equation}
and
\begin{equation}
    d\beta(\theta, z) = (-2y(z-0.5), 2x(z-0.5), 0)
\end{equation}
and
\begin{equation}
    hn(\theta, z) = -tv(\theta, z)
\end{equation}
and
\begin{equation}
    ht(\theta, z) =(-4y, 4x, 0),
\end{equation}
and finally we can infer eta from the rest.

\section{Software Release}
\label{sec:softwareRelease}

Several components of our software stack are publicly available:
\begin{itemize}
    \item \texttt{polyarray}, an intermediate representation combining symbolic and numerical array computations: \url{https://github.com/wraith1995/polyarray}.
    \item PyAB (PyArrayBackend), an array intermediate representation with NumPy and PyTorch code generation: \url{https://github.com/wraith1995/PyArrayBackend}.
    \item Grassmann, a Python-embedded language for coordinate-free linear and multilinear algebra that compiles to array code: \url{https://github.com/wraith1995/Grassmann}.
\end{itemize}
We are working on preparing the remaining components of \SYS{} for public release.

\end{document}